\documentclass[final,3p,times]{elsarticle}
\usepackage{amssymb}
\usepackage{amsmath}

\journal{}

\begin{document}

\begin{frontmatter}



\title{The Allocation of Interests and The Maximization of Each Player's Benenfits:Transforming Strategic Games into Biform Games and Applications}

%%\author[label1,label2]{}

\author{Xiang Shuwen\textsuperscript{a,b}, Luo Enquan\textsuperscript{a}, Yang Yanlong\textsuperscript{b,*}}

\address[a]{College of Management, Guizhou University, Guiyang, 550025, Guizhou, China}
\address[b]{College of Mathematics and Statistics, Guizhou University, Guiyang, 550025, Guizhou, China}

%% Abstract
\begin{abstract}
As Aumann stated, cooperation and non-cooperation are different ways of viewing the same game, with the main difference being whether players can reach a binding cooperative agreement. In the real world, many games often coexist competition and cooperation. Based on the above reasons, we propose a method to transform strategic games into a biform game model, which retains the characteristics of cooperative games while considering the ultimate goal of players to maximize their own interests. Among them,the allocation function play a crucial role as a protocol on convention.  Furthermore, based on this game model, we analyze the impact of two different distribution methods, namely marginalism and egalitarianism, on the game results. As an application, we analyze how food producers seek maximum profits through cooperative pricing.
\end{abstract}

%% Keywords
\begin{keyword}
Biform game;Nash equilibrium; marginalism;egalitarianism;Quality improvement


\end{keyword}

\end{frontmatter}
\section{Introduction}
Game theory, a discipline that studies the strategic decision-making behavior of multiple agents in interactive scenarios, is broadly categorized into two major types based on whether the players collude or conspire cooperative game theory and non-cooperative game theory. Among these, the Nash equilibrium is the central solution concept in non-cooperative games, while cooperative games feature various solution concepts, primarily including the core, stable sets, the nucleolus, the Shapley value, and egalitarian solutions.

In the real world, including the vast majority of political and economic issues, decision-making often involves a coexistence of competition and cooperation. Competition is the objective whether in the natural world's survival of the fittest or in human society's dynastic shifts and civilizational progress, it undoubtedly serves as a catalyst and driving force. Cooperation, on the other hand, is the means. It facilitates resource sharing and mutual benefit, seeks common ground while preserving differences, reduces waste, and enables collective problem-solving to lower costs and improve efficiency. Ultimately, cooperation aims to enhance competitiveness on a broader or higher level. In the business realm, competition does not equate to a zero-sum game. Beyond dividing the pie, there is also the opportunity to make it bigger. Modern commerce often requires collaboration for mutual survival no one can succeed entirely on their own.

The cooperative behavior of players is motivated by the pursuit of gains. It entails forming a coalition to secure benefits through coordinated actions and sharing the resulting outcomes. As noted by renowned game theorist and Nobel laureate Maskin: "Cooperative theory appears to have the significant advantage of offering insights into how coalitions behave specifically, how members within a coalition bargain over which actions to take." \cite{maskin2016}. Although application scenarios abound, cooperative game theory has received far less attention and has been applied far less extensively than its non-cooperative counterpart. This situation largely stems from the absolute dominance of the rational economic man hypothesis. However, what cannot be ignored is one of the most essential characteristics of humanity sociality. Beyond the pursuit of self-interest, humans are also adept at communication and strategic assessment. Dr. Sun Yat-sen once said, "The principle of survival is competition for species, but for humankind, it is cooperation. Mankind thrives by following this principle and perishes by defying it." Since the Stone Age, humans have stood out among numerous species through cooperation, creating human civilization. In today's society, with an increasingly refined division of labor, cooperation is more conducive to individual specialization, and teamwork is essential for achieving steady and long-term progress.

Maskin \cite{maskin2016} pointed out the difficulties faced by cooperative game theory. Among these, beyond the rationality assumption, a significant challenge arises from the theory's failure to account for externalities related to profit distribution. A fundamental assumption in cooperative game theory is that the grand coalition comprising all players will form. This framework relies on a characteristic function that, by definition, assumes away externalities. In other words, it presumes that a coalition's payoff is not affected by the actions of any coalition outside of it. However, interactions between coalitions are crucial in economics, as seen in negotiations between trade unions and management, competition among firms, and trade between nations. Consequently, two-stage models were introduced in organizational economics; see \cite{grossman1986} and \cite{hart1990}. Brandenburger and Stuart \cite{brandenburger2007} formally proposed the biform game approach, which incorporates elements from both non-cooperative and cooperative game theory. In recent years, biform games have also been used to optimize strategies for multiple actors in supply chains, for example, \cite{feess2014,mahjoub2014,jia2023,zheng2023,zheng2024,liu2025}.
%in Feess and Thun (2014), Mahjoub and Hennet (2014), Jia et al. (2023), Zheng and Li (2023), Zheng et al. (2024), and Liu et al. (2025).

A game model requires three basic elements: a set of players, a set of strategies, and a payoff. Therefore, whether the players cooperate or not, a game with these three elements can be regarded as a strategic game. For a strategic game, it can be divided into two categories: non-cooperative games and cooperative games, depending on whether the players cooperate. When it is regarded as a cooperative game, the difficulty lies in establishing a suitable characteristic function. To be exact, for a certain coalition $S$, in addition to the members of the coalition choosing the strategy with the largest coalition revenue through cooperation, they must also face the question of what choices the players outside the coalition $S$ should make. Thus, the strategy choice of the players outside the coalition $S$will produce different valuations for the revenue of each coalition $S$. For example, Von Neumann and Morgenstern proposed the minimax representation method in their pioneering work on game theory \cite{von1944}, Myerson \cite{myerson1991} proposed the defensive equilibrium representation method, and Harsanyi \cite{harsanyi1963} proposed the rational threat representation method. These methods all answer this question from different perspectives. Because of the different values of the coalition $S$ revenue, we will further analyze their possible defects in the next section.

In this paper, we will propose a new research framework for any strategic game. It is a game model that takes into account both the cooperative and non-cooperative behaviors of the players. This framework overcomes the defects encountered in directly present the characteristic functions.To further verify the practical value of the proposed framework,  As applications, the choice of green technology in the market competition model will be discussed. It is evident from these examples that the new research framework can effectively break through the dilemmas embedded in them. 

This paper mainly includes two parts: the first part is to establish a research method of transforming from a strategic game (players can choose a cooperative game or a non-cooperative game) to a biform game; in the second part, we illustrate the method and sentencing game model proposed in the first part through two examples, and utilize this conversion method and biform model to analyze how food producers seek maximum profits through cooperative pricing.

The first part is mainly to establish a method that is not affected by the strategic choices of players outside the coalition, which not only retains the characteristics of a cooperative game but also takes into account the ultimate goal of players to maximize their own interests. In fact, this means the coexistence of cooperation and competition, so it should be a biform game, a biform game derived from the original strategic game, or to explore and establish a method of transforming a strategic game into a biform game.

Let's turn our attention to cooperative game theory, and specifically focous on the distribution of cooperation outcomes.The most important thing in cooperative game theory is the solution to the cooperative game, which is to answer the question: When we assume that the result of cooperation is the benefit of the grand coalition, how should the result of cooperation be distributed among the players in a reasonable way?  The rationality mentioned here includes fairness, the enthusiasm to encourage players to form alliances, and a certain degree of stability (the coalition is not easily destroyed). From a large body of literature, it is not difficult to find that the solutions of cooperative games are inseparable from the two logics of marginalism and egalitarianism. The Shapley value is a typical example of marginalism, while the equal division rule is a typical example of egalitarianism \cite{shapley1953,van2007,van2013,choudhury2022,borkotokey2023}. The Shapley value does not allocate anything to non-productive players (those who do not make a contribution), while the equal allocation rule does not consider the contribution rate of players to the coalition's revenue. However, neither of these two methods is completely consistent with the form of resource or outcome distribution in practical problems. When distributing the benefits of cooperation, we not only give incentives to players with high contributions, but also pay attention to the common sharing of the benefits of cooperation. What impact will marginalism and egalitarianism have on the final result in the distribution? How should our distribution method seek a balance between the two for specific problems? Or, what kind of distribution plan should we formulate in order to achieve the goal? These are the questions that the first part of this paper attempts to answer. This part will systematically study the impact of the distribution of coalition benefits based on the biform game derived from the strategic game, and explore a method to analyze how the distribution of coalition utility affects strategic behavior.

The second part applies the methods and models established in the first part to the domains of  food quality enhancement. It primarily addresses the following three issues: 1. A biform game analysis of the "anti-commons tragedy" in collaborative food safety supervision. The results indicate that effective collaboration requires not only stronger incentives for joint efforts but also an alignment between the distribution of cooperative benefits and the incurred costs. Specifically, if the benefits are shared equally, the collaboration costs must also be borne equally. 2. We utilize the aforementioned methods and game theory models to analyze the famous Chinese proverb, "One monk will get two buckets of water, two monks will share a load of water, three monks will have no water", demonstrating the impact of distribution methods based on marginalism and egalitarianism on game outcomes. Further elaborating, the egalitarian distribution approach is conducive to promoting cooperation, but it can lead to a lack of incentives for participating individuals or organizations, and even result in participants choosing to opt out; conversely, the marginalist distribution approach reflects the contributions of participants, but it is not conducive to cooperation, and the ultimate outcome still leads to a non-cooperative conclusion. 3. A game-theoretic analysis of green technology investment in a duopoly market, primarily based on the Bertrand model. By incorporating a green investment term, a model for green technology investment is constructed. Solving this biform game demonstrates how producers can achieve maximum profit through cooperative pricing, a process that simultaneously drives them to enhance their green technology investment levels. An analysis of green technology investment within a supply chain, building upon the duopoly market framework. The results reveal that when a supply chain engages in cost-sharing cooperation for green investment and the distribution of total profit is aligned with this cost-sharing mechanism, manufacturers are incentivized to increase their green technology investments. Furthermore, this elevated level of investment, in turn, boosts the total profit of the supply chain.
\section{The method of transforming strategic game into biform game}\label{Sec1}
\subsection{Problem Statement}

We first introduce some concepts and symbols.

Let $N = \left\{ {1, \cdots ,n} \right\}$ denote the set of $n$ players, and ${2^N}$ denote the set of all subsets of $N$, where each subset $S \in {2^N}$ represents a coalition formed by the players. A transferable utility cooperative game (or simply TU game) is denoted by $\left( {N,{\text{v}}} \right)$, where $v:{2^N} \to R$ is the characteristic function. When a participant $i \in S$ joins the coalition $S \subset N\backslash \left\{ {\text{i}} \right\}$, the increase in the coalition's payoff is called the marginal contribution of participant $i$ to the coalition $S$, denoted as  
\begin{equation*}
	m_i^v\left( S \right) = v\left( {S \cup \left\{ i \right\}} \right) - v\left( S \right).
\end{equation*}
Assume that the grand coalition $N$ is formed in such a way that players join the coalition sequentially. This joining order can be represented by a permutation $\pi: N \mapsto N$ of the players, and we denote the set of all permutations of  $N$ by $\pi \left( N \right)$. For each permutation $\pi  \in \pi \left( N \right)$, let $P\left( {\pi, i} \right) = \{ j \in N:\pi \left( j \right) < \pi \left( i \right)\} $ denote the set of players who join the coalition before participant $i$ in the order $\pi $.  The Shapley value (Shapley, 1953) is one of the most important solutions in cooperative games, denoted as $Sh\left( v \right) = (S{h_1}\left( v \right), \cdots, S{h_n}\left( v \right))$. It represents an allocation to all players, computed by assigning each participant their expected marginal contribution under the assumption that every joining order $\pi$ is equally likely with probability $\frac{1}{{n!}}$. Formally,  
\begin{equation*}
	\scalebox{0.95}{$
		S{h_i}\left( v \right) = \mathop \sum \nolimits_{\pi  \in \pi \left( N \right)} \frac{1}{{n!}}m_i^v\left( {P\left( {\pi ,i} \right)} \right) = \mathop \sum \nolimits_{S \subset N\backslash \left\{ i \right\}} \frac{{\left( {n - \left| S \right| - 1} \right)!\left( {\left| S \right|} \right)!}}{{n!}}m_i^v\left( S \right).$}
\end{equation*}

The biform game approach we aim to establish involves transforming a strategic game into a biform game, or in other words, deriving the corresponding biform game from a given strategic game.

Suppose the strategic game is $\left( {N, X,f} \right)$. Here $N = \left\{ {1, \cdots ,n} \right\}$ is the set of players, ${X_i}$ represents the strategy set of participant $i$, $X = \mathop \prod \nolimits_{i \in N} {X_i}$ is the strategy profile set, ${x_i} \in {X_i}$ represents the strategy selected by participant $i$, $\left( {{x_i},{x_{ - i}}} \right) \in {\text{X}}$ represents the strategy profile formed after each participant $i$ selects ${x_i}$, and ${x_{ - i}}$ represents the strategy selection component of all players except $i$. $f = \left( {{f_1}, \cdots ,{f_n}} \right)$ is a vector of payoff functions, where ${f_i}:X \to R$ is the payoff function of player $i$. In particular, when $N = \left\{ {1,2} \right\}$ and ${X_1} = {S_1}$ and ${X_2} = {S_2}$ are finite sets, then the two-person finite game $\left( {\left\{ {1,2} \right\},\left( {{S_i}} \right),\left( {{f_i}} \right)} \right)$ can be represented by a table, that is, the two payoff matrices are combined into one table.

When a strategic game is considered a non-cooperative game, we still use its notation, but assign the strategy set and payoff function to each participant, as, $\left( {N,\left( {{X_i}} \right),\left( {{f_i}} \right)} \right)$. In this scenario, each participant seeks to maximize their own payoff. A state of equilibrium, known as the Nash equilibrium the central solution concept in non-cooperative game theory is reached when no player can obtain a higher payoff by unilaterally changing their strategy.

We call $\left( {{x_i},{x_{ - i}}} \right) \in X$ a Nash equilibrium of $\left( {N,\left( {{X_i}} \right),\left( {{f_i}} \right)} \right)$ if for every $i \in N$:
\begin{equation*}{f_i}\left( {{x_i},{x_{ - i}}} \right) \geqslant {f_i}\left( {{y_i},{x_{ - i}}} \right), \forall {x_i} \in {X_i}.
\end{equation*}

If players can communicate and reach binding agreements to coordinate their strategic choices and redistribute gains (transferable utility) among themselves, we can also consider the strategic game $\left( {N, X, f} \right)$ a cooperative game. In fact, in strategic games, it's reasonable to consider the players' willingness to cooperate. This is both the original intention of scholars like Von Neumann and Morgenstern in developing characteristic functions for cooperative games and a key factor in the recent rise of cooperative evolution. When studying economic issues, the rationality assumption is undoubtedly a fundamental premise. However, rationality does not entirely negate the possibility that players can recognize the inefficiencies resulting from a refusal to cooperate. A key distinction between humans and other biological populations lies in humanity's superior ability to rely on the power of teamwork. From the perspective of human intelligence, if complete rationality entails the absolute pursuit of self-interest, then why not transcend simplistic rationality to pursue greater gains, especially when it is clear that refusing cooperation leads to mutual detriment? For example, in reality, price wars, as revealed by the Bertrand model, are common, but cases of fighting to the bitter end, resulting in mutual loss, are rare. Decisions made under conflicting interests involve not only overt and covert struggles but also compromises based on seeking common ground while preserving differences. Precisely because of this, businesses competing openly and covertly may also reach some form of cooperation. Even if this doesn't lead to cooperation, in many cases a tacit understanding will prevail. Competition is ubiquitous, and cooperation is also common. Given the social reality of the coexistence of competition and cooperation, biform games are undoubtedly the best choice when using game theory to analyze optimization and strategic choices among multiple players. Starting from the most basic strategic game theory, and considering the inherent nature of both competition and cooperation in practical problems, we no longer dwell on whether a game problem should be discussed as a non-cooperative or cooperative game. In other words, when players face a situation where there is both mutual benefit and self-interest, there is no need to arbitrarily force them to choose between cooperation and non-cooperation. This is the practical and theoretical value of biform games, and it is also the original intention of transforming strategic games into biform games.

The study of transforming strategic games into cooperative games began with the seminal work of von Neumann and Morgenstern \cite{von1944}. Although the only fully conflicting non-cooperative games at the time were zero-sum games, von Neumann and Morgenstern specifically discussed how to construct characteristic functions for cooperative games from strategic games. Since then, research on characteristic functions and cooperative game solutions has continued \cite{shi2012,zou2023,meng2013,parilina2020,parilina2024,kosian2020,liu2023a,liu2023b}.
% (see Shi Xiquan (2012), Zou et al. (2023), Meng et al. (2013), Parilina et al. (2020, 2023), Kosian and Petrosyan et al. (2020), Liu et al. (2023a, 2023b).

Von Neumann and Morgenstern \cite{von1944} defined the minimax characteristic function $v$ as follows:
\begin{equation}
	\begin{split}
		v(\emptyset) &= 0, \\
		v(S) &= \min_{x_{-S} \in -S} \max_{x_S \in S} \sum_{i \in S} f_i(x_S, x_{-S}); \quad S \ne \emptyset \in 2^N.
	\end{split}
	\label{2.1}
\end{equation}

Harsanyi \cite{harsanyi1963} proposed a rational threat characteristic function based on Nash's rational threat principle, which is defined as follows:
\begin{equation*}
	\scalebox{0.95}{$
		\begin{split}
			v(\emptyset) &= 0, \\
			v_H(S) &= \sum_{i \in S} f_i(x_S^*, x_{-S}^*) \quad \text{and} \quad v_H(-S) = \sum_{i \in -S} f_i(x_S^*, x_{-S}^*); \quad S \ne \emptyset \in 2^N.
		\end{split}$}
\end{equation*} 
Here $x_S^* \in {X_S}$, $x_{ - S}^* \in {X_{ - S}}$ satisfies
\begin{equation}
	\begin{split}
		x_S^* &\in \mathop{\text{argmax}}\limits_{x_S \in X_S} \left[ \sum_{i \in N} f_i(x_S, x_{-S}^*) - \sum_{i \in N\setminus S} f_i(x_S, x_{-S}^*) \right], \\
		x_{-S}^* &\in \mathop{\text{argmax}}\limits_{x_{-S} \in X_{-S}} \left[ \sum_{i \in N\setminus S} f_i(x_S^*, x_{-S}) - \sum_{i \in S} f_i(x_S^*, x_{-S}) \right].
	\end{split}
	\label{2.2}
\end{equation}  

Myerson \cite{myerson1991} proposed the defensive equilibrium characteristic function ${v_M}$, which is defined as follows:
\begin{equation*}
	\scalebox{0.95}{$
		\begin{split}
			v(\emptyset) &= 0, \\
			v_M(S) &= \sum_{i \in S} f_i(x_S^*, x_{-S}^*) \quad \text{and} \quad v_M(-S) = \sum_{i \in -S} f_i(x_S^*, x_{-S}^*); \quad S \ne \emptyset \in 2^N.
		\end{split}$}
\end{equation*}
Here $x_S^* \in {X_S}$, $x_{ - S}^* \in {X_{ - S}}$ satisfies:
\begin{equation}
	\begin{split}
		x_S^* &\in \mathop{\text{argmax}}\limits_{{x_S} \in {X_S}} \mathop \sum \nolimits_{i \in N} {f_i}\left( {{x_S},x_{ - S}^*} \right), \\
		x_{-S}^* &\in \mathop{\text{argmax}}\limits_{{x_{-S}} \in {X_{-S}}} \mathop \sum \nolimits_{i \in N\backslash S} {f_i}\left( {x_S^*,{x_{ - S}}} \right).
	\end{split}
	\label{2.3}
\end{equation}

The rational threat \eqref{2.2} and defensive equilibrium \eqref{2.3} characteristic functions differ from the von Neumann-Morgenstern minimax characteristic function. These three functions assign distinct values to each coalition, a divergence stemming from how they account for the externalities arising from the actions of players outside the coalition. The minimax characteristic function has at least two limitations in terms of the definition of each coalition's payoff: First, the antagonism of the strategy choices made by players outside the coalition $S$ is exaggerated. This can be seen from expression \eqref{2.1}, which assumes that players outside $S$ choose the strategy that minimizes the coalition's payoff $S$. In fact, in the two-person game ($\left\{ {1,2} \right\}, X, f$), the payment of participant 1 and participant 2 in the non-cooperative game clearly illustrates this problem. In this case, $X = {X_1} \times {X_1}$, the payoff functions of participant 1 and participant 2 are ${f_1}\left( {{x_1},{x_2}} \right)$ and ${f_2}\left( {{x_1},{x_2}} \right)$, respectively. According to the definition of the minimax characteristic function, the payoffs for participant 1 and 2 are the payoffs of their respective coalitions, ${f_1}\left( {{x_1},{x_2}} \right)$ and ${f_2}\left( {{x_1},{x_2}} \right)$. $v\left( {\left\{ 1 \right\}} \right) = \mathop {{\text{min}}}\limits_{{x_2} \in {X_2}} \mathop {{\text{max}}}\limits_{{x_1} \in {X_1}} {f_1}\left( {{x_1},{x_2}} \right)$and $v\left( {\left\{ 2 \right\}} \right) = \mathop {{\text{min}}}\limits_{{x_1} \in {X_1}} \mathop {{\text{max}}}\limits_{{x_2} \in {X_2}} {f_2}\left( {{x_1},{x_2}} \right)$. If the two-player game ($\left\{ {1,2} \right\},X,f)$ has a Nash equilibrium $\left( {x_1^*,x_2^*} \right)$, then
\begin{equation*}
	{f_1}\left( {x_1^*,x_2^*} \right) = \mathop {\max }\limits_{{x_1} \in {X_1}} {f_1}\left( {({x_1},x_2^*} \right) \geqslant \mathop {{\text{min}}}\limits_{{x_2} \in {X_2}} \mathop {{\text{max}}}\limits_{{x_1} \in {X_1}} {f_1}\left( {{x_1},{x_2}} \right) = v\left( {\left\{ 1 \right\}} \right).
\end{equation*}
Similarly, there are
\begin{equation*}
	{f_2}\left( {x_1^*,x_2^*} \right) \geqslant \mathop {{\text{min}}}\limits_{{x_1} \in {X_1}} \mathop {{\text{max}}}\limits_{{x_2} \in {X_2}} {f_2}\left( {{x_1},{x_2}} \right) = v\left( {\left\{ 2 \right\}} \right).
\end{equation*}
In non-zero-sum situations, we have ${f_1}\left( {x_1^*,x_2^*} \right) > v\left( {\left\{ 1 \right\}} \right)$ and ${f_2}\left( {x_1^*,x_2^*} \right) > v\left( {\left\{ 2 \right\}} \right)$. For example, the Battle of the Sexes game, the Hawk-Dove game, and the Wise Pigs game (also known as the free-rider game) all satisfy this property. Note that ${f_1}\left( {x_1^*,x_2^*} \right)$ and ${f_2}\left( {x_1^*,x_2^*} \right)$ are the payoffs under non-cooperation, while $v\left( {\left\{ 1 \right\}} \right)$ and $v\left( {\left\{ 2 \right\}} \right)$ are the payoffs under taking into account the possibility of cooperation. Considering the opponent's confrontation actually reduces the payoffs of the players, which seems logically unreasonable. Therefore, it is unreasonable to assume that players other than $S$ choose a confrontational strategy.

Second, for the characteristic functions ${v_H}$ and ${v_M}$, the point $\left( {x_S^*,x_{ - S}^*} \right)$ that satisfies \eqref{2.2} or \eqref{2.3} often has multiple values. Which value should be chosen as the value of the characteristic function? Answering this question is a difficult problem. Judging from the conditions satisfied by $\left( {x_S^*,x_{ - S}^*} \right)$, it is precisely the point that maximizes the respective payoffs, a point similar to a Nash equilibrium. Therefore, choosing among multiple values corresponding to $\left( {x_S^*,x_{ - S}^*} \right)$ is similar to the problem posed by the multiplicity of Nash equilibria. It is worth noting that the multiplicity of Nash equilibria remains a major issue in non-cooperative game theory. The selection and refinement of Nash equilibria, as well as evolutionary game theory, are all attempts to address this problem.

Furthermore, transforming strategic games into cooperative games focuses on cooperation between players, but fails to fully reflect each player's pursuit of self-interest. To overcome these limitations, we propose a biform game model derived from strategic games. This model assigns a coalition value to each strategy profile, regardless of the strategies chosen by players within or outside the coalition. Cooperation within the coalition manifests itself primarily through utility transfer and the potential generation of external benefits.

\subsection{Model Building}

For a given strategic game $\left( {N, X, f} \right)$. Let:
\begin{equation}
	u_f\left( {S,x} \right) = \mathop \sum \limits_{i \in S} {f_i}\left( x \right), \quad S \in N, x \in X. 
	\label{2.4}
\end{equation}                            
Similar to expressions \eqref{2.1}, \eqref{2.2} and \eqref{2.3}, ${u_f}\left( {S,x} \right)$ represents the payoff of coalition $S$ at the strategy profile $x$, referred to as the value of coalition $S$ at $x$, which is the total payoff of the players in coalition $S$.

As mentioned earlier, the characteristic function defined by \eqref{2.1}, \eqref{2.2} and \eqref{2.3} must be either the maximum or minimum value on ${X_S}$ or the corresponding ${X_{ - S}}$. By defining the coalition value according to \eqref{2.4}, there is no need to worry about whether to maximize or minimize it, as with $v$, ${v_H}$, and ${v_M}$. A way to circumvent this problem is to give, for each strategy profile $x \in X$, the value ${u_f}\left( {S,x} \right)$ of the coalition $S$ at $x$, then, ${u_f}\left( { \cdot,x} \right):{2^X} \to R$ is the characteristic function of the strategy profile $x$, or the coalition function associated with the strategy profile $x$. For each $x \in X$, ($N,{u_f}\left( { \cdot ,x} \right)$) constitutes a cooperative game.

According to the biform game model proposed by Brandenburger and Stuart \cite{brandenburger2007}, cooperative games with strategy profiles $x$ constitute biform games. Biform games first appeared in concrete examples (\cite{grossman1986} and Hart and Moore \cite{hart1990}), and the concept of biform games and their solutions was first introduced by Brandenburger and Stuart \cite{brandenburger2007}.

In order to make the model more general, in addition to considering utility transfer, we further consider the collaborative effect of cooperation, that is, considering the effect of "$1 + 1 > 2$" brought about by cooperation, a more general coalition function can be given:
\begin{equation*}
	{v_f}\left( {S,x} \right) = {u_f}\left( {S,x} \right) + \delta \left( {S,x} \right) = \mathop \sum \nolimits_{i \in S} {f_i}\left( x \right) + \delta \left( {S,x} \right), \quad S \subset N.
\end{equation*}
Here $\delta \left( {S,x} \right) \geqslant 0 $   
refers to the additional benefits generated by cooperation, beyond the cumulative gains of the players. in some cases, coalitions can bring additional benefits, such as collaborative effects, government subsidies, and cost savings after supply chain members form a coalition, as well as logistics and warehousing costs. Therefore,
\begin{equation}
	{v_f}\left( {S,x} \right) \geqslant \mathop \sum \nolimits_{i \in S} {f_i}\left( x \right), \quad S \subset N, x \in X.
\end{equation}  
We still call ${v_f}:{2^N} \times X \mapsto R$ the characteristic function with the strategy profile $x$, or the coalition function with the strategy profile $x$. In this case, ${u_f}$ becomes a special case of ${v_f}$, that is, the case where the additional benefit $\delta \left( { \cdot ,x} \right) = 0$.

Furthermore, considering that when players cooperate, in addition to the utility transfer and collaborative effects between players, another possibility is to restrict the scope of strategy choices through agreement and understandings. From the perspective of the strategy profile set $X$, this means adding constraints or limiting strategy choices to a subset ${X^C}$ of $X$. ${X^C}$ is called the collaboration set of the strategic game $\left( {N,X,f} \right)$, or simply the collaboration set. Based on the above analysis, the following definition is given.

\textbf{Definition 2.1} For a given strategic game $\left( {N, X, f} \right)$, ${v_f}:{2^N} \times X \mapsto R$ is the characteristic function with the strategy profile $x$, ${X^C} \subset X$ is the cooperation set, then $\left( {N,{v_f},{X^C}} \right)$ constitutes a biform game, which is called a biform game derived from $\left( {N,X,f} \right)$.

\textbf{Remark 2.1} (1) For every $x \in X$, $\left( {N,{v_f}\left( { \cdot ,x} \right)} \right)$ is the cooperative game with the strategy profile $x$, also called the cooperative game at $x$ derived from $\left( {N,X,f} \right)$.

(2) In many cases, we only consider the case where ${v_f} = {u_f}$, ${X^C} = X$, that is, the biform game $\left( {N,{u_f},{X^C}} \right)$.

(3) ${X^C} \subset X$ represents a mutual agreement among players to appropriately limit the range of strategic options, including government intervention in prices and production capacity under certain circumstances. A major role of the Organization of the Petroleum Exporting Countries (OPEC) is to restrict oil prices or production for its members. Recently, the state's efforts to combat involution in certain industries have also involved interventions in production capacity and prices.

For the biform game $\left( {N, {v_f}, X} \right)$ derived from the strategic game $\left( {N, X, f} \right)$, the first stage is a non-cooperative game, primarily describing the strategic choices of the players. This involves the interactive process of each participant selecting different strategies ${x_i}$ (forming a strategy profile $x$) to maximize their payoff. This also reflects the mutual influence of strategy choices as players pursue maximum payoff. The second stage is concerned with each cooperative game $\left( {N,{v_f}\left( { \cdot,x} \right)} \right)$, which reflects the formation of alliances among players and the distribution of cooperative gains among them. It determines the payoff value each participant receives when a strategy profile is realized, and also serves as the payoff function for the first-stage non-cooperative game. Transforming a strategic game into a biform game represents a novel analytical approach. It not only addresses the shortcomings of directly converting a strategic game into a cooperative game such as relying on max-min methods but also constitutes a new attempt to consider competition and cooperation simultaneously.

Before further analyzing the solution to the biform game $\left( {N, {v_f}, X} \right)$, let's briefly illustrate the model with an example game. This example is the famous "Tragedy of the Commons," first proposed by Hardin in 1968. It reveals the inherent logic of the misuse of common resources and is often used to illustrate the inevitability of environmental destruction. The model describes a group of herders who share a common pasture. Each herder's pursuit of maximum profit consistently puts them in a dilemma: the number of cattle exceeds the pasture's carrying capacity, resulting in a situation where the herders' profits are worse than if they had fewer cattle. This example belongs to the same type of game as the "Prisoner's Dilemma," which is also often used to diagnose various social issues, such as population growth and global warming.

Below, we use the classic two-person "tragedy of the commons" game to illustrate.

\textbf{Example 2.1} Two herders raise sheep on a common pasture. Each herder can choose between two strategies, denoted by C and NC. C represents cooperation and raising an appropriate number of sheep, while NC represents non-cooperation and overgrazing. Table \ref{21} shows the payoffs for the two herders under various scenarios.
\begin{table}[htbp]
	\centering
	\caption{Two-person ``tragedy of the commons''}
	\label{tab:tragedy_commons}
	\begin{tabular}{lcc}
		\hline
		& \multicolumn{2}{c}{Herder 1} \\
		\cline{2-3}
		Herder 2 & C & NC \\
		\hline
		C & 10, 10 & 0, 12 \\
		NC & 12, 0 & 5, 5 \\
		\hline
	\end{tabular}
	\label{21}
\end{table}
Let $N = \left\{ {1,2} \right\}$ denote the set of players (two herders), ${X_1} = {X_2} = \left\{ {C, NC} \right\}$ denote their strategy set, $X = {X_1} \times {X_2}$ denote the strategy profile set, ${f_1}\left( x \right),{f_2}\left( x \right)$ denote the payoff functions of herders 1 and 2 respectively, and each $f\left( x \right) = \left( {{f_1}\left( x \right),{f_2}\left( x \right)} \right)$ denote a payoff vector, where
\begin{equation*}
	\begin{split}
		f_1(C,C) &= f_2(C,C) = 10, \quad f_1(NC,NC)= f_2(NC,NC) = 5, \\
		f_1(C,NC) &= 0, f_2(C,NC) = 12, f_1(NC,C)= 12, f_2(NC,C) = 0.
	\end{split}
\end{equation*}
According to the above representation, the two-person "tragedy of the commons" is described as a strategic game $\left( {N, X, f} \right)$. Therefore, from $\left( {N,X,f} \right)$, we can derive the biform game $\left( {N,{u_f},X} \right)$, where ${u_f}:{2^N} \times X \to R$ is
\begin{equation*}
	\scalebox{0.95}{$
		\begin{split}
			u_f(\{1\},(C,C)) &= f_1(C,C) = 10, u_f(\{2\},(C,C))= f_2(C,C) = 10, \\
			u_f(\{1\},(C,NC)) &= f_1(C,NC) = 0, u_f(\{2\},(C,NC))= f_2(C,NC) = 12, \\
			u_f(\{1\},(NC,C)) &= f_1(NC,C) = 12, u_f(\{2\},(NC,C))= f_2(NC,C) = 0, \\
			u_f(\{1\},(NC,NC)) &= f_1(NC,NC) = 5, u_f(\{2\},(NC,NC))= f_2(NC,NC) = 5, \\
			u_f(\{1,2\},(C,C)) &= f_1(C,C) + f_2(C,C) = 20, \\
			u_f(\{1,2\},(C,NC)) &= f_1(C,NC) + f_2(C,NC) = 12, \\
			u_f(\{1,2\},(NC,C)) &= f_1(NC,C) + f_2(NC,C) = 12, \\
			u_f(\{1,2\},(NC,NC)) &= f_1(NC,NC) + f_2(NC,NC) = 10.
		\end{split}$ }
\end{equation*}
Here, ${u_f}$ does not involve the additional effects brought about by collaboration, that is, the case where $\delta  = 0$ and ${v_f} = {u_f}$.

\subsection{Distribution of Cooperative Outcomes and Biform Game Solutions}
Let$(N,v_f,X)$ be the biform game and $\left( {N,{v_f}\left( { \cdot ,x} \right)} \right)$ be the cooperative game derived from $\left( {N,{v_f},X} \right)$. For each player $i$, $S \subset N$ is a coalition and $i \notin S$, then the marginal contribution of participant $i$ to the coalition $S$:
\begin{equation*}
	m_i^{{v_f}}\left( {S,x} \right) = {v_f}\left( {S \cup \left\{ i \right\},x} \right) - {v_f}\left( {S,x} \right), S \subset N, x \in X.
\end{equation*}
Express ${v_f}\left( {S,x} \right)$ as a form with remainder, that is, ${v_f}\left( {S,x} \right) = {u_f}\left( {S,x} \right) + \delta \left( {S,x} \right)$. Let
\begin{equation*}
	\begin{split}
		m_i^{{u_f}}\left( {S,x} \right) &= {u_f}\left( {S \cup \left\{ i \right\},x} \right) - {u_f}\left( {S \cup \left\{ i \right\},x} \right), \\
		m_i^\delta \left( {S,x} \right) &= \delta\left( {S \cup \left\{ i \right\},x} \right) - \delta\left( {S \cup \left\{ i \right\},x} \right).
	\end{split}
\end{equation*}
Then $m_i^{{u_f}}\left( {S,x} \right) = {f_i}\left( x \right)$, so
\begin{equation*}
	\begin{split}
		m_i^{v_f}(S,x) &= v_f(S \cup \{i\},x) - v_f(S,x) \\
		&= u_f(S \cup \{i\},x) - u_f(S,x) + \delta(S \cup \{i\},x) - \delta(S,x) \\
		&= m_i^{u_f}(S,x) + m_i^\delta(S,x) = f_i(x) + m_i^\delta(S,x).
	\end{split}
\end{equation*}
For each $i\in N$ and $S\subset N$, if $i\in S$, set $m_i^{{v_f}}=0$. Then $m_i^{{v_f}}:{2^N} \times X \to R$ is called the marginal function of player $i \in S$ with respect to ${v_f}$. Let $m^{{v_f}}(S,X)=(m_1^{{v_f}},m_2^{{v_f}},...,m_n^{{v_f}}) $ . $m^{{v_f}}:{2^N} \times X \to R^N$ is called the marginal function respect to ${v_f}$.

\textbf{Definition 2.2}  Let $\left( {N,{v_f},X} \right)$ be a biform game derived from $\left( {N,X,f} \right)$. The marginal function $m^{{v_f}}$ is called payoff-dominant if for every $i \in N$ and $x,y \in X$, ${f_i}\left( x \right) > {f_i}\left( y \right)$, it follows that $m_i^{{u_f}}\left( {S,x} \right) > m_i^{{u_f}}\left( {S,y} \right)$ for all $S \subset N$.

\textbf{Remark 2.2} If for every $i \in N$ and $x,y \in X$, ${f_i}\left( x \right) > {f_i}\left( y \right)$, it follows that $m_i^\delta \left( {S,x} \right) > m_i^\delta \left( {S,y} \right)$, then $m_i^{{v_f}}$ is payment-dominated. This means that the larger the payoff to the participant, the larger the value of the additional benefit distributed. In particular, if $\delta \left( { \cdot ,x} \right) = c$, it is clear that $m_i^{{v_f}}$ is payment-dominated, and thus it is clear that $m^{{u_f}}$  is payment-dominated.

Before presenting the solution to the biform game, Let us start addressing a critical issue: the distribution of coalition payoffs during the cooperative phase. This determines the important properties of the biform game solution. When returning to the first stage of the non-cooperative game, the Nash equilibrium will be primarily influenced by the strategic choices of the players, whose goal is to maximize their respective payoffs, which are determined by the allocation function. In numerous practical problems, the allocation function is typically determined by either a formalized protocol-with clear definitions of the rights and obligations of all parties-or well-established conventions validated by long-term industry practice. This determination mechanism ensures the rationality and enforceability of distribution within a coalition.

Suppose the vector ${a^{{v_f}}}\left( x \right) = \left( {a_1^{{v_f}}\left( x \right), \cdots ,a_n^{{v_f}}\left( x \right)} \right)$ is the solution of the cooperative stage $\left( {N,{v_f}\left( { \cdot ,x} \right)} \right)$ of the cooperative game. Each $a_i^{{v_f}}\left( x \right)$ is the distribution of the grand coalition payoff ${v_f}\left( {N,x} \right)$ to the payoff of participant $i$. We call the vector-valued function ${a^{{v_f}}}:X \mapsto {R^n}$ an allocation function of ${v_f}$. The function $a_i^{{v_f}}\left( x \right)$ gives a new payoff to player $i$, which replaces the initial payoff value ${f_i}\left( x \right)$ in the strategic game. The non-cooperative game $\left( {N,\left( {{X_i}} \right),\left( {a_i^{{v_f}}} \right)} \right)$ constitutes the non-cooperative stage of the biform game $\left( {N,{v_f},X} \right)$. $\left( {N,\left( {{X_i}} \right),\left( {a_i^{{v_f}}} \right)} \right)$ is called the non-cooperative game derived from $\left( {N,{v_f},X} \right)$.

According to the definition of the solution of the biform game, the concept of the solution of the biform game $\left( {N, {v_f}, X} \right)$ derived from the strategic game is defined below.

\textbf{Definition 2.3} If the strategy profile $x \in X$ is a Nash equilibrium of the non-cooperative game $\left( {N,\left( {{X_i}} \right),\left( {a_i^{{v_f}}} \right)} \right)$ derived from the biform game, that is, $x$ satisfies 
\begin{equation*}
	a_i^{{v_f}}\left( {{x_i},{x_{ - i}}} \right) \geqslant a_i^{{v_f}}\left( {{y_i},{x_{ - i}}} \right), \forall {x_i} \in {X_i}, i \in N.
\end{equation*}
Then $x$ is called the solution of the biform game $\left( {N,{v_f},X} \right)$, also called the soution of the biform game $\left( {N,X,f} \right)$.

\textbf{Remark 2.3} According to the definition of the solution in the biform game, the allocation function $a_i^{{v_f}}$ generated in the cooperation stage is crucial. That is, the solution is determined by this allocation function.

Next, we will specifically discuss the impact of the distribution of cooperative results on the outcome of the game. Before we begin our discussion, we first introduce several concepts.

\textbf{Definition 2.4} The allocation rule $a^{v_f}$ is said to be egalitarian if, for any $x, y \in X$ such that $v_f(N, x) \geq v_f(N, y)$, it follows that $a_i^{v_f}(x) \geq a_i^{v_f}(y)$ for each $i$ and for all $x, y \in X$.

\textbf{Remark 2.4} If we replace ${v_f}\left( {N,x} \right) \geqslant {v_f}\left( {N,y} \right)$ with $\frac{{{v_f}\left( {N,x} \right)}}{n} \geqslant \frac{{{v_f}\left( {N,y} \right)}}{n}$, the meaning of "egalitarianism" becomes more obvious. It is not difficult to verify that if the solution of the cooperative game $\left( {N,{v_f}\left( { \cdot,x} \right)} \right)$ is directly decomposed using the equal decomposition \cite{brandenburger2007}, the resulting allocation function is clearly egalitarian. That is, if for each $i \in N$, the allocation function is $a_i^{{v_f}}\left( x \right) = \frac{{{v_f}\left( {N,x} \right)}}{n}$, then ${a^{{v_f}}}$ is egalitarian.

\textbf{Definition 2.5}  The allocation rule $a^{v_f}$ is said to be marginalist if, for any $x, y \in X$, $a_i^{v_f}(x) \leqslant a_i^{v_f}(y)$ for each $i \in N$ if and only if $f_i(x) \leqslant f_i(y)$ for each $i \in N$ and any $x, y \in X$. 

\textbf{Remark 2.5} (1) If the solution of a cooperative game  is defined by Shapley values, then
\begin{equation*}
	S{h_i}\left( {{v_f}\left( { \cdot ,x} \right)} \right) = \mathop \sum \nolimits_{S \subset N\backslash \left\{ i \right\}} \frac{{\left( {n - \left| S \right| - 1} \right)!\left( {\left| S \right|} \right)!}}{{n!}}m_i^{{v_f}}\left( {S,x} \right), \forall x \in X.
\end{equation*}
make
\begin{equation*}
	{a^{{v_f}}}\left( x \right) = sh\left( {{v_f}\left( { \cdot ,x} \right)} \right) = \left( {S{h_1}\left( {{v_f}\left( { \cdot ,x} \right)} \right), \cdots ,S{h_n}\left( {{v_f}\left( { \cdot ,x} \right)} \right)} \right), \forall x \in X.
\end{equation*}
We call ${a^{{v_f}}}$ the Shapley allocation function of ${v_f}$, abbreviated as ${a^{{v_f}}} = Sh\left( {{v_f}} \right)$. It is not difficult to verify that if the marginal function $m^{{v_f}}$ is payment-dominant, then the allocation function $Sh\left( {{v_f}} \right)$ is also marginalist. In fact, if ${f_i}\left( x \right) \leqslant {f_i}\left( y \right)$ holds for every $i \in N$ and any $x,y \in X$, it is equivalent to $m_i^{{v_f}}\left( {S,x} \right) \leqslant m_i^{{v_f}}\left( {S,y} \right)$ holding for every $S \subset N$, $i \in N$ and any $x,y \in X$, which is equivalent to $S{h_i}\left( {v\left( { \cdot ,x} \right)} \right) \leqslant S{h_i}\left( {v\left( { \cdot ,y} \right)} \right)\;$holding for every $i \in N$ and any $x,y \in X$ holds, that is, ${\text{if\;and\;only\;if\;}}a_i^{{v_f}}\left( x \right) = S{h_i}\left( {v\left( { \cdot ,x} \right)} \right) \leqslant S{h_i}\left( {v\left( { \cdot ,y} \right)} \right) = a_i^{{v_f}}\left( y \right)$ holding for every $i \in N$ and any $x,y \in X$. Hence, $Sh\left( {{v_f}} \right) = {a^{{v_f}}}$ is marginalist. 

(2)Distribution being marginalist means that within $v_f(S,x) = \mathop \sum \nolimits_{i \in S}f_i(x) + \delta(x) $ those whose contributions are greater shall be allocated a larger portion of the distribution. If there is no additional benefit, that is, the remainder $\delta  = 0$, ${v_f} = {u_f}$. In this case, ${a^{{u_f}}}$ is the Shapley allocation function of ${u_f}$, that is
\begin{equation*}
	{a^{{u_f}}}\left( x \right) = \left( {a_1^{{u_f}}\left( x \right), \cdots ,a_n^{{u_f}}\left( x \right)} \right) = \left( {S{h_1}\left( {{u_f}\left( { \cdot ,x} \right)} \right), \cdots ,S{h_n}\left( {{u_f}\left( { \cdot ,x} \right)} \right)} \right),
\end{equation*}
then ${a^{{u_f}}}$ is marginalist.

The allocation function is of vital importance in solving the biform game $(N,v_f,X)$, and how to determine it depends on the actual problems. We will analyze the profound impact of the two distribution methods of marginalism and egalitarianism on the solution of biform games.

\textbf{Proposition 2.1}  If the allocation function ${a^{{v_f}}}$ is marginalist, then a strategy profile $x$ is a solution of $\left( {N,{v_f},X} \right)$ if and only if $x$ is a Nash equilibrium of $\left( {N,\left( {{X_i}} \right),\left( {{f_i}} \right)} \right)$.

\textbf{Proof:} According to the definition of the solution of the biform game $\left( {N, {v_f}, X} \right)$, $x$ is a solution of $\left( {N, {v_f}, X} \right)$ if and only if $x$ is a Nash equilibrium of $\left( {N,\left( {{X_i}} \right),\left( {a_i^{{v_f}}} \right)} \right)$, that is,
\begin{equation*}
	a_i^{{v_f}}\left( {{x_i},{x_{ - i}}} \right) \geqslant a_i^{{v_f}}\left( {{y_i},{x_{ - i}}} \right), \forall i \in N, {y_i} \in {X_i}.
\end{equation*}
Noting that ${a^{{v_f}}}$ is marginalist, we have
\begin{equation*}
	{f_i}({x_i},{x_{ - i}}) \geqslant {f_i}({y_i},{x_{ - i}}), \forall i \in N, {y_i} \in {X_i}.
\end{equation*}
Therefore, $x$ is a Nash equilibrium of $\left( {N,\left( {{X_i}} \right),\left( {{f_i}} \right)} \right)$.

\textbf{Corollary 2.1}  For a given $\left( {N,{u_f},X} \right)$, if ${a^{{u_f}}}\left( x \right)$ is the Shapley allocation function of $\left( {N,{u_f}\left( { \cdot ,x} \right)} \right)$, then $x$ is a solution of $\left( {N,{u_f},X} \right)$ if and only if $x$ is a Nash equilibrium of $\left( {N,\left( {{X_i}} \right),\left( {u_f} \right)} \right)$.

Proposition 2.1 and Corollary 2.1 show that a marginalist distribution method cannot promote mutually beneficial cooperation. In this case, the solution to the biform game derived from the strategic game theory is still the non-cooperative solution to the game, namely, the Nash equilibrium of $\left( {N,\left( {{X_i}} \right),\left( {{f_i}} \right)} \right)$. This shows that if the cooperative outcomes are distributed strictly according to marginalist, the biform game derived from the strategic game will still be in a non-cooperative state.

\textbf{Proposition 2.2}  If the allocation function ${a^{{v_f}}}$ is egalitarian, then the strategy profile ${x^*} \in \mathop {argmax}\limits_{x \in X} {v_f}\left( {N,x} \right)$ is a Nash equilibrium of $\left( {N,\left( {{X_i}} \right),\left( {{f_i}} \right)} \right)$, and thus is also a solution of $\left( {N,{v_f},X} \right)$.

\textbf{Proof:} Noting that ${a^{{v_f}}}$ is egalitarian, it can be inferred from ${x^*} = \mathop {argmax}\limits_{x \in X} {v_f}\left( {N,x} \right)$ that for any $i \in N$ and $x \in X$, $a_i^{{v_f}}\left( {{x^*}} \right) \geqslant a_i^{{v_f}}\left( x \right)$ holds, that is, $a_i^{{v_f}}\left( {x_i^*,x_{ - i}^*} \right) \geqslant a_i^{{v_f}}\left( {{x_i},x_{ - i}^*} \right)$, $\forall {x_i} \in {X_i}$. From this we can see that ${x^*}$ is a Nash equilibrium of $\left( {N,\left( {{X_i}} \right),\left( {a_i^{{v_f}}} \right)} \right)$, and thus is also a solution of $\left( {N,{v_f},X} \right)$.

According to Proposition 2.2, the egalitarian distribution will maximize the grand coalition's benefit ${v_f}\left( {N,x} \right)$, that is, the maximum point of ${v_f}\left( {N,x} \right)$ becomes the solution to the biform game, which shows that the egalitarian distribution will promote cooperation and achieve the maximum overall benefit.

Next, we further discuss the case where there is no extra benefit, that is, the case where ${v_f} = {u_f}$.

Let $f\left( x \right) = \left( {{f_1}\left( x \right), \cdots ,{f_n}\left( x \right)} \right)$ be a payoff value in the strategic game $\left( {N,X,f} \right)$. If there does not exist $f\left( {x'} \right) = \left( {{f_1}\left( {x'} \right), \cdots ,{f_n}\left( {x'} \right)} \right)$, satisfying
\begin{equation*}
	\begin{split}
		f_i(x') &\leqslant f_i(x), \quad \forall i \in N, \\
		\text{and there exists at least one } j &\in N \text{ such that } f_j(x') < f_j(x);
	\end{split}
\end{equation*}
then $f\left( x \right) = \left( {{f_1}\left( x \right), \cdots ,{f_n}\left( x \right)} \right)$ is said to be the Pareto optimal payoff of $\left( {N,X,f} \right)$, or the strategy profile $x \in X$ is said to be Pareto optimal. In other words, if a Pareto optimal outcome is to be improved, the interests of at least one participant will inevitably be harmed. Usually, the Nash equilibrium is not Pareto optimal, which means that the benefits of some players can be improved.

For the case ${v_f} = {u_f}$,  ${x^*} \in \mathop {argmax}\limits_{x \in X} {u_f}\left( {N,x} \right) = \mathop {argmax}\limits_{x \in X} \mathop \sum \nolimits_{i \in N} {f_i}\left( x \right)$ is obviously Pareto optimal, and the following results can be obtained.

\textbf{Corollary 2.2}  If the allocation function $u_f$ is egalitarian, then the strategy profile ${x^*} \in \mathop {argmax}\limits_{x \in X} {u_f}\left( {N,x} \right)$ is a solution to $\;\left( {N,{u_f},X} \right)$, and $f\left( {{x^*}} \right) = \left( {{f_1}\left( {{x^*}} \right), \cdots ,{f_n}\left( {{x^*}} \right)} \right)$ is the Pareto optimal payoff of $\left( {N,X,f} \right)$.

The above results show that egalitarian distribution increas the willingness to cooperate and is conducive to achieving Pareto optimality. Marginalism, while emphasizing the contributions of players, is not conducive to cooperation, as the ultimate outcome still reverts to non-cooperation. These conclusions not only reveal the nature of the solutions to biform games but also offer a new perspective on the conceptual study of cooperative game solutions. It can also be argued that they offer a new perspective on the distribution of any type of cooperation and shared gains.

\subsection{Give examples to illustrate}

We continue to use the tragedy of the commons as an example and analyze it using a biform game. Unlike Example 2.1, for ease of discussion, we treat the number of sheep raised by the herders as a continuous variable.

\textbf{Example 2.2} On a common pasture, two herders raise sheep. Let ${q_1},{q_2}$ denote each herder's sheep stock, and ${\text{\;}}{\rho _1},{\text{\;}}{\rho _2}$ denote their profits from raising sheep. Given $M > 0$, $M$ denotes the carrying capacity of the pasture for sheep. That is, if the total sheep stock is ${q^ + } = {q_1} + {q_2} > M$, then each herder's profit ${\text{\;}}{\rho _i} < 0$. Therefore, we assume that the herders' sheep stock ${q_i} \in \left[ {0, M} \right]$. Let $\mu$ denote the slaughter rate per sheep on the pasture, which is the ratio of the total sheep stock ${q^+ }$. Assume that $\mu(M){p_0}={c_0}$, which means that if the total number of sheep raised exceeds the carrying capacity, there will be no profit at all, and $\mu \left( {{q^ + }} \right)$ is a second-order differentiable, decreasing, concave function with respect to ${q^ + }$. Let ${p_0}$ denote the unit price of mutton, and ${c_0}$ denote the cost of raising a sheep for each herder (including labor input).

The profit functions of herders 1 and 2 are:
\begin{equation*}
	\begin{split}
		\rho _1 = \mu \left( {{q^ + }} \right){p_0}{q_1} - {q_1}{c_0},\\
		\rho _2 = \mu \left( {{q^ + }} \right){p_0}{q_2} - {q_2}{c_0}.
	\end{split}
\end{equation*}

Let $X={X_1} \times {X_2} = \left[ {0,M} \right] \times \left[ {0,M} \right]$, $q = \left( {{q_1},{q_2}} \right)$, and $\rho  = \left( {{\rho _1},{\rho _2}} \right)$, then $\left( {\left\{ {1,2} \right\},X,\rho } \right)$ constitute a strategic game, where ${\rho _1}$ and ${\rho _2}$ are the payoff functions of herders 1 and 2. For ease of discussion, we assume that the number of sheep on hand is considered a continuous variable, the unit price ${p_0}$ is set to 1, and $0 \leqslant {c_0} \leqslant 1$ is considered a constant.Then the profit function is simplified as:
\begin{equation*}
	\begin{split}
		\rho _1 = \mu \left( {{q^ + }} \right){q_1} - {q_1}{c_0},\\
		\rho _2 = \mu \left( {{q^ + }} \right){q_2} - {q_2}{c_0}.
	\end{split}
\end{equation*}

If $\left( {\left\{ {1,2} \right\}, X,\rho } \right)$is regarded as a non-cooperative game, it can be expressed as  $\left( {\left\{ {1,2} \right\},\left( {{X_i}} \right),\left( {{\rho _i}} \right)} \right)$. The Nash equilibrium of this game is discussed below.

Calculate the partial derivatives of ${\rho _1},{\rho _2}$ with respect to ${q_1}$ and ${q_2}$ respectively, and set them to zero. Then,
\begin{equation*}
	\begin{split}
		\frac{\partial \rho_1}{\partial q_1} &= \mu'(q^+) q_1 + \mu(q^+) - c_0 = 0, \\
		\frac{\partial \rho_2}{\partial q_2} &= \mu'(q^+) q_2 + \mu(q^+) - c_0 = 0.
	\end{split}
\end{equation*}
It is not difficult to verify that the solution to the above equation satisfying  ${q_1}={q_2}$. So the above equations are equivalent to the following equation
\begin{equation}
	\frac{\mu'(q^+) q^+}{2} + \mu(q^+) - c_0 = 0.
\end{equation}
Noting the decreasing and concave nature of $\mu \left( {{q^ + }} \right)$, $\mu(0)>c_0$ and $\mu(M)=c_0$, we know that the above equation has a unique solution, denoted as ${q^{ + *}}$ which lies in $(0,M)$. Then, by ${q_1} = {q_2} = \frac{{{q^{ + *}}}}{2}$, we can obtain the Nash equilibrium of $\left( {\left\{ {1,2} \right\},\left( {{X_i}} \right),\left( {{\rho _i}} \right)} \right)$ as $\left( {\frac{{{q^{ + *}}}}{2},\frac{{{q^{ + *}}}}{2}} \right)$. At this point, the profits of herders 1 and 2 are both
\begin{equation*}
	\rho_1^* = \rho_2^* = \frac{\mu(q^{+*}) q^{+*}}{2} - q^{+*} c_0.
\end{equation*}
In addition, consider the total profit of the two herders, namely
\begin{equation*}
	\rho = \rho_1 + \rho_2 = \mu(q^+) q^+ - q^+ c_0.
\end{equation*}
Similar to the previous method, from the decreasing and concave properties of $\mu \left( {{q^ + }} \right)$, it can be seen that the unique maximum point of $\rho$ is the solution of the following equation:
\begin{equation*}
	\mu'(q^+) q^+ + \mu(q^+) - c_0 = 0.
\end{equation*}
Suppose the solution of the equation is ${q^{ + \# }}$. Then ${q^{ + \# }}$ is the maximum point of $\rho $.

Notice
\begin{equation*}
	\begin{split}
		\rho'(q^{+*}) &= \mu'(q^{+*}) q^{+*} + \mu(q^{+*}) - c_0 \\
		&= \frac{\mu'(q^{+*}) q^{+*}}{2} + \mu(q^{+*}) - c_0 + \frac{\mu'(q^{+*}) q^{+*}}{2} \\
		&= \frac{\mu'(q^{+*}) q^{+*}}{2} < 0.
	\end{split}
\end{equation*}
Since $\mu \left( {{q^ + }} \right)$ is a decreasing concave function and ${q^{ +\# }}$ is the maximum point of $\rho $, then $\rho '\left( {{q^{ + \#}}} \right) = 0$, and when ${q^ + } \geqslant {q^{ + \#}}$, then $\rho '\left( {{q^ + }} \right)<0$. Therefore, ${q^{ + \# }} < {q^{ + *}}$, and $\rho \left( {{q^{ + \# }}} \right) > \rho \left( {{q^{ + *}}} \right)$. 

This shows that the total profit corresponding to the Nash equilibrium $(\frac{{{q^{ + *}}}}{2},\frac{{{q^{ + *}}}}{2}$) is lower than the case of raising fewer sheep, that is, the total profit is lower than the case of $\left( {\frac{{{q^{ + \# }}}}{2},\frac{{{q^{ + \# }}}}{2}} \right)$.

A biform game is from the strategy game and the coalition function ${v_\rho }:{2^N} \times X \to R$ associated with the strategy profile $x$. Here we only need to consider the case $v_{\rho}(S,\cdot)=u_{\rho}(S,\cdot)=\sum_{i\in S}\rho_{i}(\cdot)$  where there is no extra benefit. The coalition function is as follows:
\begin{equation*}
	\begin{split}
		v_\rho(\emptyset, (q_1, q_2)) &= 0, \\
		v_\rho(\{1\}, (q_1, q_2)) &= u_\rho(\{1\}, (q_1, q_2)) = \rho_1 = \mu(q^+) q_1 - q_1 c_0, \\
		v_\rho(\{2\}, (q_1, q_2)) &= u_\rho(\{2\}, (q_1, q_2)) = \rho_2 = \mu(q^+) q_2 - q_2 c_0, \\
		v_\rho(\{1,2\}, (q_1, q_2)) &= u_\rho(\{1,2\}, (q_1, q_2)) = \rho_1 + \rho_2 = \mu(q^+) - q c_0.
	\end{split}
\end{equation*}
Therefore, the strategic game $\left( {\left\{ {1,2} \right\},X,\rho } \right)$is transformed into a biform game $\left( {\left\{ {1,2} \right\},{v_\rho },X} \right)$ for discussion.

Suppose the solution of the cooperative game $\left( {\left\{ {1,2} \right\},{u_\rho }\left( { \cdot ,x} \right)} \right)$, and suppose the allocation function is $$\left( {a_1^{{u_\rho }}\left( {{q_1},{q_2}} \right),a_2^{{u_\rho }}\left( {{q_1},{q_2}} \right)} \right).$$

If ${a^{{u_\rho }}}$ is marginalist, for example: $$\left( {a_1^{{u_\rho }}\left( {{q_1},{q_2}} \right),a_2^{{u_\rho }}\left( {{q_1},{q_2}} \right)} \right) = \left( {{\rho _1}\left( {{q_1},{q_2}} \right),{\rho _2}\left( {{q_1},{q_2}} \right)} \right),$$ then the Nash equilibrium of the non-cooperative game $\left( {\left\{ {1,2} \right\},\left( {{X_i}} \right),\left( {{\rho _i}} \right)} \right)$ is $\left( {\frac{{{q^{ + *}}}}{2},\frac{{{q^{ + *}}}}{2}} \right)$, by Corollary 2.1, $\left( {\frac{{{q^{ + *}}}}{2},\frac{{{q^{ + *}}}}{2}} \right)$ is a non-cooperative game, which is the solution of the biform game $\left( {\left\{ {1,2} \right\},{u_\rho },X} \right)$.

If ${a^{{u_\rho }}}$ is egalitarian, for example:
\begin{equation*}
	\scalebox{0.85}{$
		a_1^{{u_\rho }}\left( q \right) = a_2^{{u_\rho }}\left( q \right) = \frac{1}{2}{v_\rho }\left( {\left\{ {1,2} \right\},\left( {{q_1},{q_2}} \right)} \right) = \frac{1}{2}\left( {{\rho _1} + {\rho _2}} \right) = \frac{1}{2}\left[ {\mu \left( {{q^ + }} \right) - {q^ + }{c_0}} \right] = \frac{1}{2}\rho \left( {{q^ + }} \right).$}
\end{equation*}
From the previous analysis, let ${q^{ + \# }} \in \mathop {argmax}\limits_{{q^ + } \in X} \rho \left( {{q^ + }} \right)$ and $q_1^* + q_2^* = {q^{ + \# }}$, then by Corollary 2.2, we can see that $\left( {q_1^*,q_2^*} \right)$ is the Nash equilibrium of the non-cooperative game $\left( {\left\{ {1,2} \right\},\left( {{X_i}} \right),\left( {\frac{1}{2}\rho } \right)} \right)$, that is, the solution of the biform game $\left( {\left\{ {1,2} \right\},{u_\rho },X} \right)$, $\left( {q_1^*,q_2^*} \right)$ corresponds to the payoff that is Pareto optimal.

An analysis of the two-player "sheep grazing on common land" model as a biform game shows that two different allocation functions lead to two distinct outcomes. From the perspective of the total number of sheep raised, i.e., the inventory level, we have ${q^{ + \# }} < {q^{ + *}}$, while the profit is the opposite, with $\rho \left( {q_1^*,q_2^*} \right)\; > \rho \left( {\frac{{{q^{ + *}}}}{2},\frac{{{q^{ + *}}}}{2}} \right)$.

Let's return to the two-strategy scenario in Example 2.1. When the problem is viewed as a non-cooperative game, the Nash equilibrium is $\left( {NC, NC} \right)$, and the payoffs for the two herders are $\left( {5,5} \right)$, creating a social dilemma. When the problem is viewed as a biform game, if the allocation function is marginalist, the solution is still the Nash equilibrium $\left( {NC, NC} \right)$ of the non-cooperative game. If the allocation function is egalitarian, the solution is $\left( {C, C} \right)$, with the corresponding payoff $\left( {10,10} \right)$, which is Pareto optimal. Cooperation and egalitarian allocation allow players to escape the social dilemma of overgrazing in the "sheep grazing on common land" problem.

In this section, we use the analysis and examples of a biform game to demonstrate that marginalist allocation does not effectively promote cooperation. This is because the solution to the biform gam $\left( {\left\{ {1,2} \right\},{u_f},X} \right)$is equivalent to the solution to the strategic game as a non-cooperative game. Specifically, in the example of two-player "sheep grazing on common land", the solution to the biform game is the Nash equilibrium $\left(\frac{q^{+*}}{2},\frac{q^{+*}}{2}\right)$ of the non-cooperative game $\left( {\left\{ {1,2} \right\},\left( {{X_i}} \right),\left( {{\rho _i}} \right)} \right)$. This will cause the two shepherds to raise sheep beyond the capacity of the pasture, thus falling into a social deadlock similar to the "prisoner's dilemma." In contrast, an egalitarian distribution promotes cooperation. The Nash equilibrium in this case allows the two shepherds to reach an agreement and achieve the Pareto optimal solution $\left( {q_1^*,q_2^*} \right)$ by raising fewer sheep. Therefore, the distribution method determines the solution to the biform game. A reasonable distribution method can highlight the effectiveness of cooperation and overcome the social dilemma of inefficiency.

However, it is worth noting that egalitarian distribution can lead to a lack of motivation for participating individuals or organizations, and even make players choose to lie flat. In fact, from the solution of two-player "sheep grazing on common land", it can be seen that as long as $q_1^* + q_2^* = {q^{ + *}}$, the problem can be solved. This does not guarantee that any herder will have the motivation to raise more sheep, and will wait for another herder to raise more sheep to make up for the number of sheep he has fewer. The big pot meal that our country experienced during the Great Leap Forward was, to a certain extent, a reflection of this law. We have further discussed this issue in another article. The main content of this section can be found in the paper by Xiang et al. \cite{xiang2025}.

\section{Biform Game Analysis of Incentives for Multi-Departmental Regulatory Cooperation}
\subsection{Problem Statement}
Food safety is of vital importance to the national economy and people¡¯s livelihoods. Strengthening food safety supervision and improving regulatory efficiency are the responsibilities of the government and an important part of the modernization of the national governance system. Since the advent of the new era, the Party and government have attached great importance to food safety supervision. However, food safety incidents still occur from time to time. Looking into the root causes, the problems are complex and the reasons are diverse, but the lack of effective cooperative supervision is also one of the important reasons. China has long implemented a segmented regulatory system, with highly dispersed regulatory powers, which has led to regulatory fragmentation. Regulatory performance depends not only on the performance of individual departments but also on their ability to cooperate. Even after the institutional reform of the merger of three bureaus in 2018, many regulatory matters still rely on the cooperation between the State Administration for Market Regulation and other ministries and bureaus, and also depend on the cooperation between subordinate departments. From the local perspective, in addition to cooperation between departments, cross-regional cooperation between local governments is also indispensable.

Food safety issues are characterized by their distinct cross-boundary nature. This is reflected not only in the boundaries of knowledge and jurisdiction, but also more prominently in the complex and diverse range of stakeholders. The political interests of regulatory agencies, the economic interests of the regulated entities, and the social interests of the public are interwoven, creating significant tensions. There is a need to promote collaboration among multiple stakeholders, to build mutual trust in the process, and thereby to integrate their interests to improve regulatory effectiveness. Therefore, strengthening cooperative regulation is seen as an important measure to resolve the regulatory dilemma.

Under this circumstance, collaborative supervision in the field of food safety has become a research focus for academia, and scholars have made extensive explorations in promoting collaborative supervision. Many scholars have drawn nourishment from the perspective of holistic government and delved into the specific field of food safety supervision. Xu \cite{Bib30} proposed that in China, the forms of collaborative supervision of food safety include cooperation among government agencies, between government and non-governmental organizations, and among non-governmental organizations. However, cooperation among government agencies is the most favored, and unitary government supervision still dominates. Ma \cite{Bib31} argued that, considering the costs of integration, national conditions, and the complexity of food safety issues, a single-agency system may not be able to effectively solve all the problems of China's multi-departmental supervision. Yan and Nie \cite{Bib32} took relational contracts as the entry point and used the theory of transaction costs to explore the mechanism of the cooperation dilemma among China's food safety supervision departments. Yan \cite{Bib33} suggested that improvements in departmental cooperation should be made in terms of organizational structure, responsibility and incentive mechanisms, partnerships, and organizational culture.

From the perspective of institutional economics, food safety supervision is a public good, and government supervision essentially represents the exercise of public property rights. The government and relevant departments perform their supervisory duties as a means of using public power to maintain food safety and to intervene in and constrain the food market (see Chen \cite{Bib34}, Chen \cite{Bib35}, and Chen \cite{Bib36}). Public power originates from the delegation and authorization by the public. Given that public property rights are characterized by indivisibility, non-exclusivity of usage rights, and externality, the exercise of supervisory power over them is also subject to the "free-rider" phenomenon. Therefore, the public is reluctant to exercise this power individually and instead places it in the public domain, making it public power. Consequently, it is inevitable that the state or government acts on behalf of the public to exercise the public property right of food safety supervision. 

The research results of Parisi et al. \cite{Bib37} indicate that when public property rights are jointly owned by multiple departments with equal power status and no hierarchical relationships, the efforts each department makes to maximize its own interests are greater than those required to maximize collective interests. This insufficient cooperation among departments leads to an inadequate supply of public goods, resulting in the Tragedy of the Anticommons.The concept of the Tragedy of the Anticommons was first proposed by Heller \cite{Bib38} from the University of Michigan. Anticommons refers to a situation where a resource or property has many owners, each of whom has the formal or informal right to prevent others from using the resource, ultimately leading to no one having effective and substantive usage rights. The Tragedy of the Anticommons is in contrast to the Tragedy of the Commons,which describes the dilemma caused by the overconsumption of public resources. The Tragedy of the Anticommons reveals that the underutilization of public resources (power) can also lead to idleness and waste.

In the field of food quality and safety supervision, the Tragedy of the Anticommons is the root cause of the low efficiency of collaborative supervision. This is mainly due to the government granting public resources (power) to different authorities. Numerous regulatory departments, burdened with overlapping responsibilities, fail to perform their duties effectively and even shirk responsibility. In China, food quality and safety regulatory power was once divided among nearly 10 departments, including industry and commerce, health, drug supervision, quality supervision, agriculture, commerce, and customs. These departments were responsible for food raw materials, processing, production, distribution, and consumption, and each was granted certain regulatory powers. However, no single department had comprehensive regulatory responsibilities. With multiple authorities involved, each acting independently and even setting obstacles to pursue its own interests, collaboration became extremely difficult. Some regulatory responsibilities were lost in the process of mutual shirking, leading to the idleness and waste of public power and even regulatory failure.
In March 2018, the Central Committee of the Communist Party of China issued the "Plan for Deepening the Reform of Party and State Institutions." This plan integrated the responsibilities of the State Administration for Industry and Commerce, the General Administration of Quality Supervision, Inspection and Quarantine, the China Food and Drug Administration, the price supervision and anti-monopoly enforcement responsibilities of the National Development and Reform Commission, the anti-monopoly enforcement of business concentration of the Ministry of Commerce, and the Office of the State Council Anti-monopoly Commission to form the State Administration for Market Regulation. The State Administration for Market Regulation is responsible for the management and supervision of food safety. Its main responsibilities include formulating and implementing food safety policies, supervising food production and business activities, and issuing food safety standards and regulations to ensure food safety.

Does this mean that the much-criticized "nine dragons managing water" situation in the field of food safety has come to an end? The reality is not so simple. According to the "Food Safety Law of the People's Republic of China" (hereinafter referred to as the "Food Safety Law") (revised for the second time by the Standing Committee of the 13th National People's Congress on April 29, 2021) and other relevant laws, regulations, rules, and normative documents, in addition to the food department, departments such as agriculture, health, grain, environmental protection, public security, and industry and information technology still bear relevant functions of food safety supervision. That is to say, after March 2018, although the State Administration for Market Regulation is responsible for the supervision and management of food production and business activities, the multi-departmental supervision structure in China's food safety field has not been completely changed. Many areas still require coordinated cooperation and joint management by multiple departments. To strengthen the coordination and cooperation of food safety supervision departments, the State Council Food Safety Commission is responsible for coordinating the food safety supervision work of various departments. However, how to strengthen cooperative supervision and how to eliminate shirking and wrangling remain major challenges that food safety supervision must face.

According to the provisions of laws, regulations, and rules such as the Food Safety Law, as well as the specific practices of regulatory authorities, the methods of coordination and cooperation among regulatory authorities are diverse. These include seeking opinions (consultation), cooperation agreements (memoranda of understanding), joint decision-making (the Food Safety Law often uses expressions such as ¡°in conjunction with...¡± or ¡°organize and carry out...¡±), information notification, joint law enforcement, administrative assistance, case referral, joint meetings, and liaison officer meetings. To a certain extent, these activities have promoted communication and cooperation among regulatory authorities. However, they also suffer from inefficiencies and even ineffectiveness to varying degrees.

The lack of incentive mechanisms for cooperation is a significant flaw in collaborative supervision. In practice, some regulatory authorities, in order to better fulfill their duties, have voluntarily taken some coordination and cooperation actions, such as signing cooperation agreements or memoranda of understanding, and establishing joint meeting or liaison officer meeting systems. However, such voluntary cooperation often lacks legal binding force. Cooperation agreements (memoranda) may be terminated at any time due to one party¡¯s refusal to cooperate, and joint meetings may not be held regularly due to the non-cooperation of any party. In other words, whether the purpose of voluntary cooperation can be achieved largely depends on whether the cooperating parties are honest, trustworthy, and committed to fulfilling their obligations. This, in turn, is closely related to whether the regulatory authorities can benefit from cooperation and the extent of such benefits. Under the existing system, there is an imbalance in power and resource allocation among food safety regulatory authorities. Some departments always give more and gain less in the cooperation process (such as information notification), while some departments may be pure beneficiaries. In the absence of specific incentive mechanisms, the enthusiasm of the former for voluntary cooperation will inevitably be affected. Since 2018, the supervision and management of food production and business operations have been the responsibility of the Market Supervision and Administration Bureau, which has led to centralized and unified management of regulatory work. However, local Market Supervision and Administration Bureaus, which operate under the management of local governments, need to strengthen cooperative supervision and overcome the constraints of local protectionism.
Coordination and cooperation among regulatory authorities cannot merely remain at the level of principled provisions and slogan-like advocacy; they require scientific and meticulous institutional design. In addition to institutional constraints, incentive mechanisms are equally important for stimulating the proactivity and enthusiasm of regulatory authorities in coordination and cooperation. Drawing from international experience, an effective coordination and cooperation mechanism must have clear cooperation goals and a performance evaluation system. In the performance evaluation system, not only should there be assessments of the regulatory effectiveness of individual departments, but evaluations of inter-departmental coordination and cooperation are also essential. Moreover, the weight assigned to these evaluations is crucial for promoting multi-departmental coordination and cooperation. Establishing a scientific and rational performance evaluation system is of great significance for improving regulatory efficiency. On the one hand, performance assessments ensure that regulatory authorities fulfill their departmental responsibilities, which is consistent with the current evaluation mechanisms. On the other hand, by making inter-departmental coordination and cooperation an important part of the evaluation, regulatory authorities are encouraged to actively participate in collaborative supervision. Only in this way can a positive situation be created where departments strive to maximize their own interests by focusing on their main tasks while actively coordinating and cooperating in joint supervision. Therefore, the coexistence of competition and cooperation is the norm in multi-departmental supervision.

This section applies the biform game model established in Section \ref{Sec1} to the cooperative supervision of food quality and safety. It examines the gains and costs associated with departments' inputs into cooperative supervision, as well as the distribution of cooperative gains, to investigate the strategic choices of participating departments. Specifically, it conducts an in-depth analysis of the competition and cooperation among regulatory authorities through the lens of the biform game. Focusing on the example of cooperative supervision involving three departments, it constructs a basic strategic game model. The strategic game is then treated as a two-stage biform game. By selecting appropriate cooperative gain distribution functions and analyzing the corresponding solutions of the biform game, it explores the strategic behaviors of regulatory authorities. Additionally, it offers suggestions and insights for the development of incentive mechanisms and the promotion of cooperative supervision.
\subsection{Problem Description}
In one area of food safety supervision, three supervisory departments need to cooperate. These departments may be agencies or departments within a single region involved in this area of supervision, or they may be cross-regional agencies or departments. Denote the set of supervisory departments as $N=\big\{1,2,3\big\}$. For each supervisory department $i \in N$, there are two strategic choices: to participate in collaborative supervision or not to participate in collaborative supervision, denoted as "$p$" and "$np$", respectively. The set of strategies for a supervisory department is represented by $S_i=\big\{p,np\big\}$. For the collaborative supervision task, the higher-level department or local government provides corresponding inputs or performance recognition. These inputs and performance are regarded as the utilities of the supervisory departments.

If local governments or higher-level authorities only conduct an overall assessment of the work without evaluating individual departments, the total utility cannot be disaggregated among the participating departments; instead, it must be split equally among the three. Under this rule, should a single department take on and complete the entire cooperative-supervision task, it would receive only one-third of the total benefit. Because the cost of acting alone exceeds this share, the department¡¯s net utility would be negative. If two departments work together, each still receives one-third of the total benefit; although collaboration lowers the aggregate cost, the benefit accruing to each participant remains below its individual cost, so their utilities are still negative. Only when all three departments participate does each again receive one-third of the total, but the cost falls further, so that each department¡¯s benefit exceeds its individual cost and its net utility becomes positive.

To construct the game model, we make the following assumptions:

{\bf A1}. The payoff for completing the entire collaborative supervision task is $R$.

{\bf A2}. The cost for a single department to complete the collaborative supervision task alone is $C$, satisfying $R>C$ and $\frac{R}{2}<C$. In the absence of cooperation between departments, if two departments participate and complete the collaborative supervision task, the cost $C$ is equally shared between the two departments. If three departments participate and complete the task, the cost $C$ is equally shared among the three departments.

{\bf A3}. Considering the efficiency improvement due to the synergistic effect, if two departments collaborate to complete the collaborative supervision task, the combined cost for the two departments is $rC<C$; if three departments collaborate to complete the task, the combined cost for the three departments is $qC<rC<C$.

{\bf A4}. Regarding the collaborative supervision task, the utility of a supervisory department is the payoff minus the cost.

To maximize their own utility, there is competition among the departments. Meanwhile, in regulatory fields involving multiple departments, cooperation is needed to improve efficiency. Therefore, competition and cooperation coexist. The above problem can be analyzed using the dual-type game discussed in Section \ref{Sec1}.
\subsection{Non-Cooperative Scenarios in Strategy Games}

Let ${S=\prod_{i\in N}S_{i}}$ be the set of strategy profiles. For each $i \in N$, the utility of the supervisory department is as follows:
\begin{align}
	\nonumber
	\begin{split}
	\begin{aligned}
		& \pi_{{i}}({np,np,np})=0, \\
		& \pi_{1}({p,np,np})=\frac{1}{3}{R-C,}\ \pi_{2}({p,np,np})=\frac{1}{3}{R,}\ \pi_{3}({p,np,np})=\frac{1}{3}{R,} \\
		& \pi_{1}({np,p,np})=\frac{1}{3}{R,}\ \pi_{2}({np,p,np})=\frac{1}{3}{R-C,}\ \pi_{3}({np,p,np})=\frac{1}{3}{R,} \\
		& \pi_{1}({np,np,p})=\frac{1}{3}{R,}\ \pi_{2}({np,np,p})=\frac{1}{3}{R,}\ \pi_{3}({np,np,p})=\frac{1}{3}{R-C,} \\
		& \pi_{1}({np,p,p})=\frac{1}{3}{R,}\ \pi_{2}({np,p,p})=\frac{1}{3}{R-\frac{1}{2}C,}\ \pi_{3}({np,p,p})=\frac{1}{3}{R-\frac{1}{2}C,} \\
		& \pi_{1}({p},{n}{p},{p})=\frac{1}{3}{R}-\frac{1}{2}{C},\ \pi_{2}({p},{n}{p},{p})=\frac{1}{3}{R},\ \pi_{3}({p},{n}{p},{p})=\frac{1}{3}{R}-\frac{1}{2}{C}, \\
		& \pi_{1}({p},{p},{n}{p})=\frac{1}{3}{R}-\frac{1}{2}{C},\ \pi_{2}({p},{p},{n}{p})=\frac{1}{3}{R}-\frac{1}{2}{C},\ \pi_{3}({p},{p},{n}{p})=\frac{1}{3}{R},\\
		&\pi_{i}({p,p,p})=\frac{1}{3}{R-\frac{1}{3}C.}
	\end{aligned}
	\end{split}
\end{align}

Therefore, the cooperative regulation problem can be described as a strategic game with ${(N,(S_{i}),(\pi_{i})),\ \pi_{i}:S\to R}$ as the payoff functions of the supervisory department $i$.

To take into account the participation level or the probability of participation of the supervisory departments, we directly discuss the mixed game over $(N,(S_{i}),(\pi_{i}))$. For any $i \in N$, let
\begin{align}
	\nonumber
	\Delta_i=\big\{u_i=(x_i;1-x_i);\ x_i\in [0,1]\big\}
\end{align}
and ${\Delta=\prod_{i\in N}\Delta_{i}}$. Then $\Delta_i$ represents the set of mixed strategies for $i$, and  $\Delta$ represents the set of mixed strategy profiles for the game. Here,  $x_i$ denotes the degree or probability of department  participating in cooperative regulation, and correspondingly, $1-x_i$ denotes the degree or probability of non-participation. The payoff function for the mixed game is
\begin{align}
	\nonumber
	\begin{aligned}
		f_1(u_1,u_2,u_3)&= x_{1}x_{2}x_{3}\pi_{1}(p,p,p)+x_{1}x_{2}(1-x_{3})\pi_{1}(p,p,np)+x_{1}(1-x_{2})x_{3}\pi_{1}(p,np,p)\\
		&+(1-x_{1})x_{2}x_{3}\pi_{1}(np,p,p)+x_1(1-x_2)(1-x_3)\pi_1(p,np,np)\\
		&+(1-x_{1})x_{2}(1-x_{3})\pi_{1}(np,p,np)+(1-x_1)(1-x_2)x_3\pi_1(np,np,p)\\
		& +(1-x_1)(1-x_2)(1-x_3)\pi_1(np,np,np) \\
		& =\left(\frac{1}{3}R-\frac{1}{3}C\right)x_1x_2x_3-\left(\frac{1}{3}R-\frac{1}{2}C\right)x_1x_2-\left(\frac{1}{3}R-\frac{1}{2}C\right)x_1x_3 \\
		& -\frac{1}{3}Rx_2x_3+\left(\frac{1}{3}R-C\right)x_1+\frac{1}{3}R(x_2+x_3).
	\end{aligned}
\end{align}

Similarly, for any $(u_1,u_2,u_3)\in \Delta$, we can obtain
\begin{align}
	\nonumber
	\begin{aligned}
 f_2(u_1,u_2,u_3)&=\left(\frac{1}{3}R-\frac{1}{3}C\right)x_1x_2x_3-\left(\frac{1}{3}R-\frac{1}{2}C\right)x_1x_2-\left(\frac{1}{3}R-\frac{1}{2}C\right)x_2x_3 \\
		& -\frac{1}{3}Rx_{1}x_{3}+\frac{1}{3}Rx_{1}+\left(\frac{1}{3}R-C\right)x_{2}+\frac{1}{3}Rx_{3}, \\
 f_3(u_1,u_2,u_3)&=\left(\frac{1}{3}R-\frac{1}{3}C\right)x_{1}x_{2}x_{3}-\frac{1}{3}Rx_{1}x_{2}-\left(\frac{1}{3}R-\frac{1}{2}C\right)x_{2}x_{3} \\
		& -\left(\frac{1}{3}R-\frac{1}{2}C\right)x_1x_3+\frac{1}{3}R(x_1+x_2)+\left(\frac{1}{3}R-C\right)x_3.
	\end{aligned}
\end{align}

Since $u_i=(x_i,1-x_i)$, then for any $(x_i,1-x_i) \in \Delta_i$, the aforementioned payoff function can be expressed as
\begin{align}\label{H1}
	\begin{aligned}
		 f_1(x_1,x_2,x_3)&=\left(\frac{1}{3}R-\frac{1}{3}C\right)x_{1}x_{2}x_{3}-\left(\frac{1}{3}R-\frac{1}{2}C\right)x_{1}x_{2}-\left(\frac{1}{3}R-\frac{1}{2}C\right)x_{1}x_{3} \\
		& -\frac{1}{3}Rx_2x_3+\left(\frac{1}{3}R-C\right)x_1+\frac{1}{3}R(x_2+x_3), \\
	 f_2(x_1,x_2,x_3)&=\left(\frac{1}{3}R-\frac{1}{3}C\right)x_1x_2x_3-\left(\frac{1}{3}R-\frac{1}{2}C\right)x_1x_2-\left(\frac{1}{3}R-\frac{1}{2}C\right)x_2x_3 \\
		& -\frac{1}{3}Rx_{1}x_{3}+\frac{1}{3}Rx_{1}+\left(\frac{1}{3}R-C\right)x_{2}+\frac{1}{3}Rx_{3}, \\
 f_3(x_1,x_2,x_3)&=\left(\frac{1}{3}R-\frac{1}{3}C\right)x_1x_2x_3-\frac{1}{3}Rx_1x_2-\left(\frac{1}{3}R-\frac{1}{2}C\right)x_2x_3 \\
		& -\left(\frac{1}{3}R-\frac{1}{2}C\right)x_1x_3+\frac{1}{3}R(x_1+x_2)+\left(\frac{1}{3}R-C\right)x_3.
	\end{aligned}
\end{align}
 
 Eq. \eqref{H1} constitutes a strategic game, denoted by $(N,(\Delta_i),(f_i))$. If we regard Eq. \eqref{H1} as a non-cooperative game, we discuss the Nash equilibrium of this game. By taking the partial derivatives of  $f_1,f_2,f_3$ with respect to $x_1,x_2,x_3$, respectively, we can obtain
 \begin{align}
 	\nonumber
 	\begin{gathered}
 		\frac{\partial{f}_{1}}{\partial{x}_{1}}=\left(\frac{1}{3}{R}-\frac{1}{3}{C}\right){x}_{2}{x}_{3}-\left(\frac{1}{3}{R}-\frac{1}{2}{C}\right){x}_{2}-\left(\frac{1}{3}{R}-\frac{1}{2}{C}\right){x}_{3}+\left(\frac{1}{3}{R}-{C}\right), \\
 		\frac{\partial{f}_{2}}{\partial{x}_{2}}=\left(\frac{1}{3}{R}-\frac{1}{3}{C}\right){x}_1{x}_3-\left(\frac{1}{3}{R}-\frac{1}{2}{C}\right){x}_1-\left(\frac{1}{3}{R}-\frac{1}{2}{C}\right){x}_3+\left(\frac{1}{3}{R}-{C}\right), \\
 		\frac{\partial{f}_{3}}{\partial{x}_{3}}=\left(\frac{1}{3}{R}-\frac{1}{3}{C}\right){x}_1{x}_2-\left(\frac{1}{3}{R}-\frac{1}{2}{C}\right){x}_2-\left(\frac{1}{3}{R}-\frac{1}{2}{C}\right){x}_1+\left(\frac{1}{3}{R}-{C}\right).
 	\end{gathered}
 \end{align}

It is noted that
\begin{align}
	\nonumber
	\begin{aligned}
		\frac{\partial^{\prime}{f_1}}{\partial{~x_1}}&=\left(\frac{1}{3}{R}-\frac{1}{3}{C}\right){x_2}{x_3}-\left(\frac{1}{3}{R}-\frac{1}{3}{C}\right){x_2}-\left(\frac{1}{3}{R}-\frac{1}{3}{C}\right){x_3}+\frac{1}{6}{C}{x_2}+\frac{1}{6}{C}{x_3}+\left(\frac{1}{3}{R}-{C}\right) \\
	&	\leq-\left(\frac{1}{3}{R}-\frac{1}{3}{C}\right)[{x}_{2}+(1-{x}_{2}){x}_{3}]+\frac{1}{6}{C}+\left(\frac{1}{3}{R}-{C}\right) \\
	&	\leq\frac{1}{6}{C}+\left(\frac{1}{3}{R}-{C}\right)=\frac{1}{3}{R}-\frac{5}{6}{C}<\frac{1}{3}{R}-\frac{5}{12}{C}<0.
	\end{aligned}
\end{align}

Similarly, we can obtain $\frac{\partial^{\prime}{f_2}}{\partial{~x_2}} <0$ and $\frac{\partial^{\prime}{f_3}}{\partial{~x_3}} <0$. From this, it can be seen that ${f_{1}(x_{1},x_{2},x_{3}),f_{2}(x_{1},x_{2},x_{3}),f_{3}(x_{1},x_{2},x_{3})}$ are decreasing functions with respect to $x_1,x_2,x_3$, respectively. Therefore,  $(0,0,0)$ constitute the unique Nash equilibrium of the game (2.2.1). In this case, each department, aiming to maximize its own utility, will inevitably choose not to participate in cooperative regulation. This conclusion reflects the ¡°tragedy of the anti-commons¡± in food safety cooperative regulation, that is to say, when performing the function of cooperative regulation, it often falls into such a deadlock or dilemma, namely, the regulatory departments will often choose inaction in cooperative regulation.

It is worth noting that in the model given by Eq. \eqref{H1}, the total utility obtained by the three departments through joint cooperative regulation is higher than the total utility obtained by inaction, that is
\begin{align}
	\nonumber
	f_1(1,1,1)+f_2(1,1,1)+f_3(1,1,1)=\frac{1}{3}(R-C)>0=f_1(0,0,0)+f_2(0,0,0)+f_3(0,0,0).
\end{align}

Furthermore, denote
\begin{align}
	\nonumber
	{g(x_1,x_2,x_3)=f_1(x_1,x_2,x_3)+f_2(x_1,x_2,x_3)+f_3(x_1,x_2,x_3).}
\end{align}

Then,
\begin{align}
	\nonumber
	\begin{aligned}
		\frac{\partial g}{\partial x_{1}}&=\left(\frac{1}{3}R-\frac{1}{3}C\right)x_{2}x_{3}-\left(\frac{1}{3}R-\frac{1}{2}C\right)x_{2}-\left(\frac{1}{3}R-\frac{1}{2}C\right)x_{3}+\left(\frac{1}{3}R-C\right) \\
		&+\left(\frac{1}{3}R-\frac{1}{3}C\right)x_2x_3-\left(\frac{1}{3}R-\frac{1}{2}C\right)x_2-\frac{1}{3}Rx_3+\frac{1}{3}R \\
		&+\left(\frac{1}{3}R-\frac{1}{3}C\right)x_2x_3-\frac{1}{3}Rx_2-\left(\frac{1}{3}R-\frac{1}{2}C\right)x_3+\frac{1}{3}R \\
		&=(R-C)x_{2}x_{3}-(R-C)x_{2}-(R-C)x_{3}+(R-C).
	\end{aligned}
\end{align}

It is not difficult to deduce that when $x_{2} \neq 1$ and $x_{3} \neq 1$ hold, $\frac{\partial g}{\partial x_{1}} > 0$ follows. Similarly, we can obtain $\frac{\partial g}{\partial x_2} \geq 0$ and $\frac{\partial g}{\partial x_3} \geq 0$. When $x_1 \neq 1$ and $x_3 \neq 1$, $\frac{\partial g}{\partial x_2} > 0$ holds. When $x_1 \neq 1$ and $x_2 \neq 1$, $\frac{\partial g}{\partial x_3} > 0$ holds. Therefore, $g(x_1, x_2, x_3)$ are increasing with respect to $x_1, x_2, and x_3$, respectively. Thus, for any $x_1, x_2, x_3 \neq 1$, we have
\begin{align}
	\nonumber
	\begin{aligned}
	g(1,1,1) & =f_{1}(1,1,1)+f_{2}(1,1,1)+f_{3}(1,1,1) \\
	& >g(x_{1},x_{2},x_{3})=f_{1}(x_{1},x_{2},x_{3})+f_{2}(x_{1},x_{2},x_{3})+f_{3}(x_{1},x_{2},x_{3}).
\end{aligned}
\end{align}

This means that when the three regulatory departments fully collaborate to complete the cooperative regulation, the total utility obtained is higher than that achieved by any other alternative. At this point, $(0,0,0)$ becomes a Nash equilibrium, indicating that the non-participation of the three regulatory departments in cooperative regulation creates a social dilemma. To break out of this dilemma, cooperation among regulatory departments is essential to achieve the maximum total utility, with each department's payoff being Pareto optimal. However, regulatory departments inevitably have their own interests and will pursue the maximization of their own utilities. In other words, even with cooperation, competition still exists. Therefore, this issue can be analyzed using a biform game. Following the model constructed in the previous section, the strategic game \eqref{H1} can be transformed into a biform game for analysis.
\subsection{Solution of the Biform Game}
Section \ref{Sec1} of this chapter discusses the transformation of strategic-form games into biform games for analysis. We will apply this approach to discuss the issue of cooperative regulation, establishing a biform game model derived from the strategic-form game \eqref{H1}.

First, we present the biform game in pure strategies, namely the biform game $(N, v_\pi, S)$ derived from $(N, S, \pi)$, with the characteristic function $v: 2^N \times S \to R$ as follows:
\begin{align}
	\nonumber
	\begin{aligned}
		&{v}_{f}(\emptyset,(s_{1},s_{2},s_{3}))=0,\ v_{f}(\{i\},(s_{1},s_{2},s_{3})))=\pi_{i}((s_{1},s_{2},s_{3})),\ s_{i}\in\{p,np\},i\in\{1,2,3\},\\
		& v_{f}\left(\{i,j\},(np,np,np)\right)=0,i\neq j\ {and}\ i\in\{1,2,3\}, \\
		& v_{f}\left(\{i,j\},(p,p,p)\right)=2\left(\frac{1}{3}R-\frac{r}{3}C\right),i\neq j\ {and}\ i\in\{1,2,3\},\\
		& v_{f}\left(\{1,2\},(p,p,np)\right)=2\left(\frac{1}{3}R-\frac{r}{2}C\right),v_{f}\left(\{1,3\},(p,p,np)\right)=\left(\frac{1}{3}R-\frac{1}{2}C\right)+\frac{1}{3}R, \\
		& v_{f}\left(\{2,3\},(p,p,np)\right)=\left(\frac{1}{3}R-\frac{1}{2}C\right)+\frac{1}{3}R, \\
		& v_{f}\left(\{1,2\},(p,np,p)\right)=\left(\frac{1}{3}R-\frac{1}{2}C\right)+\frac{1}{3}R,v_{f}\left(\{1,3\},(p,np,p)\right)=2\left(\frac{1}{3}R-\frac{r}{2}C\right), \\
		& v_{f}\left(\{2,3\},(p,np,p)\right)=\left(\frac{1}{3}R-\frac{1}{2}C\right)+\frac{1}{3}R, \\
		& v_{f}\left(\{1,2\},(np,p,p)\right)=\left(\frac{1}{3}R-\frac{1}{2}C\right)+\frac{1}{3}R,v_{f}\left(\{1,3\},(np,p,p)\right)=\left(\frac{1}{3}R-\frac{1}{2}C\right)+\frac{1}{3}R, \\
		& v_{f}\left(\{2,3\},(np,p,p)\right)=2\left(\frac{1}{3}R-\frac{r}{2}C\right), \\
		& v_{f}\left(\{1,2\},(np,p,p)\right)=\left(\frac{1}{3}R-\frac{1}{2}C\right)+\frac{1}{3}R,\quad v_{f}\left(\{1,3\},(np,p,p)\right)=\left(\frac{1}{3}R-\frac{1}{2}C\right)+\frac{1}{3}R, \\
		& v_{f}\left(\{2,3\},(np,p,p)\right)=2\left(\frac{1}{3}R-\frac{r}{2}C\right), \\
		& v_{f}\left(\{1,2\},(p,np,np)\right)=\left(\frac{1}{3}R-C\right)+\frac{1}{3}R,v_{f}\left(\{1,3\},(p,np,np)\right)=\left(\frac{1}{3}R-C\right)+\frac{1}{3}R, \\
		& v_{f}\left(\{2,3\},(p,np,np)\right)=\frac{2}{3}R, \\
		& v_{f}\left(\{1,2\},(np,p,np)\right)=\left(\frac{1}{3}R-C\right)+\frac{1}{3}R,v_{f}\left(\{1,3\},(np,p,np)\right)\frac{2}{3}R,\\
		 & v_{f}\left(\{2,3\},(np,p,np)\right)=\left({\frac{1}{3}}R-C\right)+{\frac{1}{3}}R, \\
		& v_{f}\left(\{1,2\},(np,np,p)\right)=\frac{2}{3}R,v_{f}\left(\{1,3\},(np,np,p)\right)=\left(\frac{1}{3}R-C\right)+\frac{1}{3}R, \\
		& v_{f}\left(\{2,3\},(np,np,p)\right)=\left(\frac{1}{3}R-C\right)+\frac{1}{3}R, \\
		& v_{f}\left(\{1,2\},(np,np,p)\right)=\frac{2}{3}R,v_{f}\left(\{1,3\},(np,np,p)\right)=\left(\frac{1}{3}R-C\right)+\frac{1}{3}R, \\
		& v_{f}\left(\{2,3\},(np,np,p)\right)=\left(\frac{1}{3}R-C\right)+\frac{1}{3}R.
	\end{aligned}
\end{align}

Next, we present the biform game in mixed strategies, namely the biform game $(N, v_f, \Delta)$ derived from $(N, \Delta,f)$, with the characteristic function $v_f : 2^N \times \Delta \to R$, as follows:
\begin{align}
	\nonumber
	\begin{aligned}
		& v_{f}(\emptyset,(x_{1},x_{2},x_{3}))=0, \\
		& v_{f}(\{i\},(x_{1},x_{2},x_{3}))=f_{i}(x_{1},x_{2},x_{3})=u_{f}\left(\{i\},f_{1}(x_{1},x_{2},x_{3})\right),i\in\{1,2,3\},
	\end{aligned}
\end{align}
\begin{align}
	\nonumber
\begin{split}
	v_f\left( \{1,2\}, (x_1, x_2, x_3) \right)
	&= x_1x_2x_3\left[2\left(\frac{1}{3}R - \frac{r}{3}C\right)\right] + x_1x_2(1 - x_3)\left[2\left(\frac{1}{3}R - \frac{r}{2}C\right)\right]\\
	&+ x_1(1 - x_2)x_3\left[\left(\frac{1}{3}R - \frac{1}{2}C\right) + \frac{1}{3}R\right] + (1 - x_1)x_2x_3\left[\frac{1}{3}R + \left(\frac{1}{3}R - \frac{1}{2}C\right)\right]\\
	&+ x_1(1 - x_2)(1 - x_3)\left[\left(\frac{1}{3}R - C\right) + \frac{1}{3}R\right] + (1 - x_1)x_2(1 - x_3)\left[\frac{1}{3}R + \left(\frac{1}{3}R - C\right)\right]\\&+ (1 - x_1)(1 - x_2)x_3\left[\frac{1}{3}R + \frac{1}{3}R\right]\\
	&= \left[\frac{2}{3}R - \left(1 - \frac{r}{3}\right)C\right]x_1x_2x_3 - \left[\left(\frac{2}{3}R - (2 - r)C\right)\right]x_1x_2 - \left(\frac{2}{3}R - \frac{1}{2}C\right)x_1x_3\\
	&- \left(\frac{2}{3}R - \frac{1}{2}C\right)x_2x_3 + \left(\frac{2}{3}R - C\right)x_1 + \left(\frac{2}{3}R - C\right)x_2 + \frac{2}{3}Rx_3,
\end{split}
\end{align}
\begin{align}
	\nonumber
	\begin{split}
	v_f\left( \{1,3\}, (x_1, x_2, x_3) \right)
	&= x_1x_2x_3\left[2\left(\frac{1}{3}R - \frac{r}{3}C\right)\right] + x_1x_2(1 - x_3)\left[\left(\frac{1}{3}R - \frac{1}{2}C\right) + \frac{1}{3}R\right]\\
	&+ x_1(1 - x_2)x_3\left[2\left(\frac{1}{3}R - \frac{r}{2}C\right)\right] + (1 - x_1)x_2x_3\left[\left(\frac{1}{3}R - \frac{1}{2}C\right) + \frac{1}{3}R\right]\\
	&+ x_1(1 - x_2)(1 - x_3)\left[\left(\frac{1}{3}R - C\right) + \frac{1}{3}R\right] + (1 - x_1)x_2(1 - x_3)\left[\frac{1}{3}R + \frac{1}{3}R\right]\\
	&+ (1 - x_1)(1 - x_2)x_3\left[\frac{1}{3}R + \left(\frac{1}{3}R - C\right)\right]\\
	&= \left[\frac{2}{3}R - \left(1 - \frac{r}{3}\right)C\right]x_1x_2x_3 - \left(\frac{2}{3}R - \frac{1}{2}C\right)x_1x_2 - \left[\left(\frac{2}{3}R - (2 - r)C\right)\right]x_1x_3\\
	&- \left(\frac{2}{3}R - \frac{1}{2}C\right)x_2x_3 + \left(\frac{2}{3}R - C\right)x_1 + \frac{2}{3}Rx_2 + \left(\frac{2}{3}R - C\right)x_3,
	\end{split}
\end{align}
\begin{align}
	\nonumber
	\begin{split}
	v_f\left( \{2,3\}, (x_1, x_2, x_3) \right)
	&= x_1x_2x_3\left[2\left(\frac{1}{3}R - \frac{r}{3}C\right)\right] + x_1x_2(1 - x_3)\left[\left(\frac{1}{3}R - \frac{1}{2}C\right) + \frac{1}{3}R\right]\\&
	+ x_1(1 - x_2)x_3\left[\left(\frac{1}{3}R - \frac{1}{2}C\right) + \frac{1}{3}R\right] + (1 - x_1)x_2x_3\left[2\left(\frac{1}{3}R - \frac{r}{2}C\right)\right]\\
	&+ x_1(1 - x_2)(1 - x_3)\left[\frac{1}{3}R + \frac{1}{3}R\right] + (1 - x_1)x_2(1 - x_3)\left[\left(\frac{1}{3}R - C\right) + \frac{1}{3}R\right]\\
	&+ (1 - x_1)(1 - x_2)x_3\left[\frac{1}{3}R + \left(\frac{1}{3}R - C\right)\right]\\
	&= \left[\frac{2}{3}R - \left(1 - \frac{r}{3}\right)C\right]x_1x_2x_3 - \left(\frac{2}{3}R - \frac{1}{2}C\right)x_1x_2 - \left(\frac{2}{3}R - \frac{1}{2}C\right)x_2x_3\\
	&- \left[\left(\frac{2}{3}R - (2 - r)C\right)\right]x_1x_3 + \frac{2}{3}Rx_1 + \left(\frac{2}{3}R - C\right)x_2 + \left(\frac{2}{3}R - C\right)x_3,
	\end{split}
\end{align}
\begin{align}
	\nonumber
	\begin{split}
	v_f\left( \{1,2,3\}, (x_1, x_2, x_3) \right)
	&= x_1x_2x_3(R - qC) + x_1x_2(1 - x_3)\left[\left(\frac{2}{3}R - rC\right) + \frac{1}{3}R\right]\\
	&+ x_1(1 - x_2)x_3\left[\left(\frac{2}{3}R - rC\right) + \frac{1}{3}R\right] + (1 - x_1)x_2x_3\left[\left(\frac{2}{3}R - rC\right) + \frac{1}{3}R\right]\\
	&+ x_1(1 - x_2)(1 - x_3)\left[\left(\frac{1}{3}R - C\right) + \frac{2}{3}R\right] + (1 - x_1)x_2(1 - x_3)\left[\frac{2}{3}R + \left(\frac{1}{3}R - C\right)\right]\\
	&+ (1 - x_1)(1 - x_2)x_3\left[\frac{2}{3}R + \left(\frac{1}{3}R - C\right)\right]\\
	&= (R - 3C + 3rC - qC)x_1x_2x_3 - (R + rC - 2C)(x_1x_2 + x_1x_3 + x_2x_3) + (R - C)(x_1 + x_2 + x_3).
	\end{split}
\end{align}

Note that
\begin{align}
	\nonumber
	\begin{split}
	v_f\left( \{1,2\}, (x_1, x_2, x_3) \right) &- \left[f_1(x_1, x_2, x_3) + f_2(x_1, x_2, x_3)\right]\\
	&= \left[\frac{2}{3}R - \left(1 - \frac{r}{3}\right)C\right]x_1x_2x_3 - \left[\left(\frac{2}{3}R - (2 - r)C\right)\right]x_1x_2 - \left(\frac{2}{3}R - \frac{1}{2}C\right)x_1x_3\\
	&- \left(\frac{2}{3}R - \frac{1}{2}C\right)x_2x_3 + \left(\frac{2}{3}R - C\right)x_1 + \left(\frac{2}{3}R - C\right)x_2 + \frac{2}{3}Rx_3\\
	&- - \left[\left(\frac{2}{3}R - \frac{2}{3}C\right)x_1x_2x_3 - \left(\frac{2}{3}R - C\right)x_1x_2 - \left(\frac{2}{3}R - \frac{1}{2}C\right)x_1x_3
	\right.\\
	&\left.
	- \left(\frac{2}{3}R - \frac{1}{2}C\right)x_2x_3 + \left(\frac{2}{3}R - C\right)x_1 + \left(\frac{2}{3}R - C\right)x_2 + \frac{2}{3}Rx_3\right]\\
	&= -\frac{1 - r}{3}Cx_1x_2x_3 + (1 - r)Cx_1x_2 \geq -\frac{1 - r}{3}Cx_1x_2 + (3 - r)Cx_1x_2\\
	&\geq \frac{2(1 - r)}{3}Cx_1x_2 > 0.
	\end{split}
\end{align}

Then,

\[
v_f\left( \{1,2\}, (x_1, x_2, x_3) \right) - u_f\left( \{1,2\}, (x_1, x_2, x_3) \right)
= v_f\left( \{1,2\}, (x_1, x_2, x_3) \right) - \left[f_1(x_1, x_2, x_3) + f_2(x_1, x_2, x_3)\right] > 0.
\]

Similarly, we can obtain

\[
v_f\left( \{1,3\}, (x_1, x_2, x_3) \right) - u_f\left( \{1,3\}, (x_1, x_2, x_3) \right) > 0,
\]

\[
v_f\left( \{2,3\}, (x_1, x_2, x_3) \right) - u_f\left( \{2,3\}, (x_1, x_2, x_3) \right) > 0.
\]

Additionally,
\begin{align}
	\nonumber
	\begin{split}
		v_f\left( \{1,2,3\}, (x_1, x_2, x_3) \right) &- \left[v_f\left( \{1,2\}, (x_1, x_2, x_3) \right) + v_f\left( \{3\}, (x_1, x_2, x_3) \right)\right]\\
		&= (r - q)Cx_1x_2x_3 + (R - C)x_1x_2x_3 - (1 - r)Cx_1x_2x_3\\
		&- (R + rC - 2C)(x_1x_2 + x_1x_3 + x_2x_3)\\
		&- \left\{\left[\frac{2}{3}R - \left(1 - \frac{r}{3}\right)C\right]x_1x_2x_3 - \left[\left(\frac{2}{3}R - (2 - r)C\right)\right]x_1x_2 - \left(\frac{2}{3}R - \frac{1}{2}C\right)x_1x_3
		\right.\\
		&\left.
		+ \left(\frac{1}{3}R - \frac{1}{3}C\right)x_1x_2x_3 - \frac{1}{3}Rx_1x_2 - \left(\frac{1}{3}R - \frac{1}{2}C\right)x_2x_3\right\}\\
		&= -\left(\frac{5}{3} - \frac{8r}{3} + q\right)Cx_1x_2x_3 + 2(2 - r)Cx_1x_2 + (1 - r)Cx_1x_3 + (1 - r)Cx_2x_3\\
		&\geq \left[\frac{4}{3}(1 - r) + (3 - q)\right]Cx_1x_2x_3 \geq 0.
	\end{split}
\end{align}

That is,

\[
v_f\left( \{1,2,3\}, (x_1, x_2, x_3) \right) \geq v_f\left( \{1,2\}, (x_1, x_2, x_3) \right) + v_f\left( \{3\}, (x_1, x_2, x_3) \right).
\]

Similarly, we can obtain

\[
v_f\left( \{1,2,3\}, (x_1, x_2, x_3) \right) \geq v_f\left( \{1,3\}, (x_1, x_2, x_3) \right) + v_f\left( \{2\}, (x_1, x_2, x_3) \right),
\]

\[
v_f\left( \{1,2,3\}, (x_1, x_2, x_3) \right) \geq v_f\left( \{1\}, (x_1, x_2, x_3) \right) + v_f\left( \{2,3\}, (x_1, x_2, x_3) \right).
\]

Then,

\[
v_f\left( \{1,2,3\}, (x_1, x_2, x_3) \right) \geq v_f\left( \{1\}, (x_1, x_2, x_3) \right) + v_f\left( \{2\}, (x_1, x_2, x_3) \right) + v_f\left( \{3\}, (x_1, x_2, x_3) \right).
\]

From Section \ref{Sec1}, regarding the cooperative game
$(\{1,2\}, v_f(\bullet, (x_1, x_2, x_3)))$, the allocation function will play a crucial role. If the solution of the cooperative game is the Shapley value, denote the allocation function as

\[
a^{v_f}(x_1, x_2, x_3) = \left( Sh_1^{v_f}(x_1, x_2, x_3), Sh_2^{v_f}(x_1, x_2, x_3), Sh_3^{v_f}(x_1, x_2, x_3) \right).
\]

Then, we obtain
\begin{align}
	\nonumber
	\begin{split}
	Sh_1^{v_f}(x_1, x_2, x_3) &= \frac{1}{6}\Big\{\left[v_f\left( \{1,2,3\}, (x_1, x_2, x_3) \right) - v_f\left( \{1,2\}, (x_1, x_2, x_3) \right)\right]\\
	&
	+ \left[v_f\left( \{1,3\}, (x_1, x_2, x_3) \right) - v_f\left( \{1\}, (x_1, x_2, x_3) \right)\right]
	+ \left[v_f\left( \{2,3\}, (x_1, x_2, x_3) \right) - v_f\left( \{2\}, (x_1, x_2, x_3) \right)\right]\Big\}\\
	&= \frac{1}{3}(R - 3C + 3rC - qC)x_1x_2x_3 - \frac{1}{3}\left(R - 3C + \frac{3r}{2}C\right)x_1x_2\\
	&- \frac{1}{3}\left(R - 3C + \frac{3r}{2}C\right)x_1x_3 - \frac{1}{3}Rx_2x_3 + \frac{1}{3}(R - 3C)x_1 + \frac{1}{3}Rx_2 + \frac{1}{3}Rx_3.
	\end{split}
\end{align}

Taking the partial derivative of $Sh_1^{v_f}$ with respect to $x_1$, we obtain
\begin{align}
	\nonumber
	\begin{split}
	\frac{\partial Sh_1^{v_f}}{\partial x_1} &= \frac{1}{3}\left[(R - 3C + 3rC - qC)x_2x_3 - \left(R - 3C + \frac{3r}{2}C\right)(x_2 + x_3) + (R - 3C)\right]\\
	&= \frac{1}{3}\left[(R - 3C)x_2x_3 - (R - 3C)(x_2 + x_3) + (R - 3C)\right.\\
	&\left. + \frac{3r}{2}C(2x_2x_3 - x_2 - x_3) - qCx_2x_3\right]\\
	&= \left(\frac{1}{3}R - C\right)(1 - x_2)(1 - x_3) - \frac{r}{2}C\left[(1 - x_2)x_3 + (1 - x_3)x_2\right] - \frac{1}{3}qCx_2x_3 < 0.
	\end{split}
\end{align}

Similarly, we can obtain
\[
\frac{\partial Sh_{2}^{v_{f}}}{\partial x_{2}} < 0 \quad \text{and} \quad \frac{\partial Sh_{3}^{v_{f}}}{\partial x_{3}} < 0.
\]
Therefore, if the solution of the biform game in the cooperative stage is the Shapley value, then the solution of the biform game is the Nash equilibrium of Eq. \eqref{H1} in the non-cooperative case, that is, the solution of the biform game is $(0,0,0)$. In fact, it is not difficult to verify that the allocation function
\[
\left( Sh_{1}^{v_{f}}(x_{1},x_{2},x_{3}),\; Sh_{2}^{v_{f}}(x_{1},x_{2},x_{3}),\; Sh_{3}^{v_{f}}(x_{1},x_{2},x_{3}) \right)
\]
satisfies the property of marginalism. Because
\[
Sh_{i}^{v_{f}}(x_{i},x_{-i}) \geq Sh_{i}^{v_{f}}(y_{i},x_{-i})
\]
holds for every $i \in \{1,2,3\}$ and for any $(x_{1},x_{2},x_{3}), (y_{1},y_{2},y_{3}) \in \Delta$, if and only if $x_{i} \geq y_{i}$ holds for every $i \in \{1,2,3\}$ and for any $(x_{1},x_{2},x_{3}), (y_{1},y_{2},y_{3}) \in \Delta$, and thus if and only if
\[
f_{1}(x_{i},x_{-i}) \leq f_{1}(y_{i},x_{-i})
\]
holds for every $i \in \{1,2,3\}$ and for any $(x_{1},x_{2},x_{3}), (y_{1},y_{2},y_{3}) \in \Delta$.

Since $(0,0,0)$ is the solution to the biform game, we arrive at the following conclusion: In the case of engaging in cooperation, if the distribution of cooperative benefits is improper, each regulatory body will still choose to fully refrain from participating in cooperative regulation. From the above analysis, it is evident that if the allocation function is based on marginalism, it cannot overcome the social dilemma of regulatory bodies' inaction in cooperative regulation.

If the cooperative benefits are distributed equally, that is, the solution in the cooperative game stage is egalitarian, then the dilemma of the ``tragedy of the commons'' can be resolved. Below, the corresponding results are presented.

For the biform game $(N,v_{f},\mathrm{\Delta})$ derived from
$(N,\mathrm{\Delta},f)$ and the cooperative game
$(N,v_{f}( \bullet ,\left( x_{1},x_{2},x_{3} \right))$ with strategy
profile $x = \left( x_{1},x_{2},x_{3} \right)$, let the allocation
function be defined as follows: for any $(x_{i},1 - x_{i}) \in {\Delta}_{i}$ and $i \in N$, there is
\begin{align}\label{H2}
	\begin{split}
	a_{1}^{v_{f}}\left( x_{1},x_{2},x_{3} \right) &= \frac{1}{3}v_{f}\left( \left\{ 1,2,3 \right\},\left( x_{1},x_{2},x_{3} \right) \right)\\
	&= \frac{1}{3}(R - 3C + 3rC - qC)x_{1}x_{2}x_{3} + \frac{1}{3}(R + rC - 2C)\left( x_{1}x_{2} + x_{1}x_{3} + x_{2}x_{3} \right) + \frac{1}{3}(R - C)(x_{1} + x_{2} + x_{3}),\\
	a_{2}^{v_{f}}\left( x_{1},x_{2},x_{3} \right) &= \frac{1}{3}v_{f}\left( \left\{ 1,2,3 \right\},\left( x_{1},x_{2},x_{3} \right) \right)\\
	&= \frac{1}{3}(R - 3C + 3rC - qC)x_{1}x_{2}x_{3} + \frac{1}{3}(R + rC - 2C)\left( x_{1}x_{2} + x_{1}x_{3} + x_{2}x_{3} \right) + \frac{1}{3}(R - C)(x_{1} + x_{2} + x_{3}),\\
	a_{3}^{v_{f}}\left( x_{1},x_{2},x_{3} \right)& = \frac{1}{3}v_{f}\left( \left\{ 1,2,3 \right\},\left( x_{1},x_{2},x_{3} \right) \right)\\
	&= \frac{1}{3}(R - 3C + 3rC - qC)x_{1}x_{2}x_{3} + \frac{1}{3}(R + rC - 2C)\left( x_{1}x_{2} + x_{1}x_{3} + x_{2}x_{3} \right) + \frac{1}{3}(R - C)(x_{1} + x_{2} + x_{3}).
	\end{split}
\end{align}

Thus, the first stage of the biform game is the non-cooperative game
$\left( N,(a_{i}^{v_{f}}),(\mathrm{\Delta}_{i}) \right)$ derived from
the biform game, where the payoff function for department $i$ is the
allocation function $a_{i}^{v_{f}}$ in Eq. \eqref{H2}.

Taking the partial derivative of $a_{1}^{v_{f}}$ with respect to
$x_{1}$, we obtain
\begin{align}\label{H3}
	\begin{split}
	\frac{\partial a_{1}^{v_{f}}}{\partial x_{1}} &= \frac{1}{3}(R - 3C + 3rC - qC)x_{2}x_{3} - \frac{1}{3}(R - 2C + rC)\left( x_{2} + x_{3} \right) + \frac{1}{3}(R - C)\\
	&\geq \frac{1}{3}\left[(R - C)x_{2}x_{3} - (R - C)\left( x_{2} + x_{3} \right) + (R - C) + (2rC - 2C)x_{2}x_{3} + (rC - qC)x_{2}x_{3} + (C - rC)\left( x_{2} + x_{3} \right)\right]\\
	&\geq \frac{1}{3}(R - C)\left( 1 - x_{2})(1 - x_{3} \right) + (1 - r)C\left[ x_{2}\left( 1 - x_{3} \right) + x_{3}\left( 1 - x_{2} \right)\right] + (r - q)Cx_{2}x_{3} > 0.
	\end{split}
\end{align}

Similarly, we can obtain
$\frac{\partial a_{1}^{v_{f}}}{\partial x_{2}} > 0$ and
$\frac{\partial a_{1}^{v_{f}}}{\partial x_{3}} > 0$. Therefore,
$(1,1,1)$ is the unique maximum point of
$$\frac{1}{3}v_{f}\left( \left\{ 1,2,3 \right\},\left( x_{1},x_{2},x_{3} \right) \right),$$
then $(1,1,1)$ is the unique Nash equilibrium of the non-cooperative
game \eqref{H1}, and also the unique solution of the biform game. It follows
that if the biform game allocates the cooperative benefits equally in
the cooperative stage, $(1,1,1)$ becomes the unique solution of the
biform game derived from \eqref{H1}. The corresponding utility value is

\[
a_{1}^{v_{f}}(1,1,1) = \frac{1}{3}v_{f}\left( \left\{ 1,2,3 \right\},(1,1,1) \right) = R - qC,
\]

\[
a_{2}^{v_{f}}(1,1,1) = \frac{1}{3}v_{f}\left( \left\{ 1,2,3 \right\},(1,1,1) \right) = R - qC,
\]

\[
a_{3}^{v_{f}}(1,1,1) = \frac{1}{3}v_{f}\left( \left\{ 1,2,3 \right\},(1,1,1) \right) = R - qC.
\]

The issue of cooperative regulation among three departments, through
cooperation between departments and the rational allocation of
cooperative utility, makes it a Nash equilibrium in the non-cooperative
stage for each department to fully engage and collaborate in regulation,
achieving the Pareto optimal solution with the maximum overall utility.
This approach helps to escape the inefficient social dilemma and avoid
the ``anti-commons tragedy'' in the exercise of public power.
\subsection{Summary of This Section}
In this section, we have taken three food safety regulatory departments as an example to construct a strategic-form game model. Through the analysis of non-cooperative games, we have explored the social dilemma that is difficult to avoid in cooperative regulation, namely the "anti-commons tragedy" in the exercise of public power. Given that competition and cooperation are ubiquitous in the field of regulation, transforming the problem into a biform game is of significant importance. We have made an attempt to study the issue of cooperative regulation using biform games.

Firstly, from the perspective of non-cooperative games, that is, departments compete with each other and seek to maximize their own departmental utility, regulatory departments do not consider cooperation even if they participate in cooperative regulation. Under such circumstances, the choice of each regulatory department to fully refrain from participating in cooperative regulation constitutes a Nash equilibrium in the non-cooperative game. In other words, the optimal choice for regulatory departments is to fully refrain from participating in cooperative regulation. This conclusion reveals a social dilemma that cooperative regulation often falls into, which is the ¡°anti-commons tragedy¡± in the exercise of public power.

In the context of cooperative regulation, if participants fully adopt a non-cooperative approach, the Nash equilibrium reveals that regulatory bodies are likely to fall into a dilemma characterized by low overall efficiency. However, whether it is the primary government and higher-level regulatory authorities, or the various regulatory departments themselves, if they can clearly recognize that this approach will lead to a dilemma and also understand that cooperation can break the deadlock, they naturally should not refuse to avoid the tragedy through cooperation. The flaw of the cooperative game solution is the lack of self-enforcement; cooperation often requires prior agreement. In terms of cooperative regulation, without prior agreement, regulatory departments will choose to betray in order to seek greater departmental benefits. The agreement here can be a contract that the participating departments must follow, or it can be the rules and requirements set by the higher-level department.

Regarding the issue of cooperative regulation, during the cooperation phase of the biform game, the cooperative benefits are distributed according to the average utility. The solution to the biform game is Pareto optimal, achieving the maximization of overall benefits, escaping the social dilemma of not fulfilling the responsibility of cooperative regulation, and solving the ¡°anti-commons tragedy.¡± This distribution method differs from the strategy game (3.1) where the utilities of each department are compared. The former involves the average distribution of benefits without cost sharing, while the latter involves the simultaneous average distribution of both benefits and costs, that is, the so-called average distribution according to utility. Cooperation among regulatory departments involves, on the one hand, transferable utility and, on the other hand, cost sharing.

The conclusions derived from the model's solution and analysis reveal, from the perspective of game theory, the inherent inevitability of the ¡°anti-commons tragedy¡± in the exercise of public power. They also highlight the new pattern characterized by the coexistence of cooperation and competition. Moreover, the impact of parameter values in the model on the results provides the following strategic recommendations for food quality and safety regulation:

1. Given that the pursuit of self-interest by each department is inevitable, and considering that non-cooperation often leads to the social dilemma of regulatory inaction, while cooperation can enhance collaborative effects, the government or higher-level regulatory authorities should establish rational assessment and incentive mechanisms to promote cooperation among departments.

2. Based on the marginal utility functions of the regulatory departments, such as the positive relationship between  and  given by Eq. \eqref{H3}, the government and higher-level departments should significantly increase the utility weight of cooperative regulation in the assessment and evaluation of regulatory departments. This will enhance the enthusiasm of each department to participate in cooperative regulation.

3. The different outcomes produced by the utility function of the strategy game \eqref{H1} and the allocation function during the cooperation phase indicate that to promote active participation in cooperative regulation, the key to breaking the deadlock is to differentiate the extent of each regulatory department's involvement in cooperative regulation. Therefore, the government or higher-level regulatory authorities must make the assessment of cooperative regulation work an important part of the overall departmental assessment. They must clearly evaluate the degree of involvement and the contributions made by each department in cooperative regulation.

\section{Biform Game Analysis of Green Technology Investment in a Market Competition Model}
\subsection{Problem Formulation}

Enhancing food quality and safety is a crucial objective within the realm of green technology. Green technology encompasses innovative systems designed to minimize resource utilization, mitigate environmental pollution, and foster ecological equilibrium. This approach permeates all stages of the product lifecycle, including design, manufacturing, consumption, recycling, and reuse.

Green technology not only conserves energy and minimizes pollution but also enhances food safety and quality. By incorporating these innovations into enterprise operations and production management, a comprehensive and systematic approach to food safety can be established. Moreover, the utilization of green synthesis techniques in food packaging material preparation not only improves food preservation but also diminishes the risk of harmful substance release compared to conventional packaging. This approach also contributes to environmental protection through self-degradation processes, thereby reducing pollution.

The adoption of green technology is crucial for enhancing the market competitiveness of food companies. Nevertheless, its implementation in ensuring food safety encounters challenges. A primary obstacle is the discrepancy between the high product premium and consumers' limited willingness to pay. The substantial investment needed for research, development, and implementation of green technology acts as a deterrent to its widespread adoption. Furthermore, the lack of standardized protocols for endorsing green products, coupled with low consumer recognition and acceptance of green synthesis methods, hinders their proliferation in the market. Efforts to enhance public awareness and approval of green products are imperative for overcoming these barriers.

To promote the widespread adoption of green technologies in the food industry, governments and relevant authorities have implemented a series of measures. These include: strengthening public education campaigns to enhance consumer awareness and trust in green technologies for food safety production; increasing financial support and policy incentives for green technology research and development, and encouraging collaboration between enterprises and research institutions to overcome technical challenges; accelerating the improvement of legal frameworks and standards to establish clear market access criteria for eco-friendly synthetic products. With these initiatives underway, investment in green technologies has gained growing attention from both businesses and society. However, it must be acknowledged that corporate enthusiasm and proactive engagement in green technology investments remain insufficient. To truly motivate enterprises, we must fully leverage market forces ¨C stimulating internal drive through corporate objectives and understanding market competition dynamics to identify inherent operational patterns. This rationale underpins our section's focus on analyzing green technology investments through market competition models.

This section still uses the dual-game method in Section \ref{Sec1} to study the issue of corporate green technology investment in market competition. It mainly focuses on two types of market competition models: The first type is the price competition model, specifically the Bertrand duopoly price competition model. The specific approach is to incorporate green technology investment items into the classic Bertrand model, establish a Bertrand model that includes green technology investment items, introduce cooperative pricing to construct a dual-game model, and interpret the famous Bertrand paradox through the comparison and analysis of non-cooperative game solutions and dual-game solutions. Meanwhile, it also proposes a specific path to resolve the price war dilemma and achieve win-win cooperation. For the second type, which targets the food supply chain, the Cournot model with green technology investment items is established, and cost sharing (or a cost-sharing mechanism) is introduced to construct a dual-type game model. Through discussion and analysis of the dual-type game solutions, the results show that enterprises in the supply chain can collaborate to increase corporate profits, and this outcome is fully consistent with raising the level of corporate green technology investment. It should be noted that in the above two types of models, improving the level of green technology investment is a requirement for maximizing corporate profits, and this is also the endogenous driving force for corporate green technology investment.

Given that the content of this section covers two types of models, we divide it into two parts for elaboration.

\subsection{Biform Game Analysis Based on the Bertrand Model}\label{}
\subsubsection{Problem Description}
The Cournot and Bertrand models are fundamental game theory frameworks used to analyze market competition in oligopolies, each representing the production and pricing strategies of firms within an oligopoly. The Nash equilibrium of the game reflects the underlying principles and inevitable patterns of the market.

The Bertrand model is widely used to study the competitive decision-making behaviors of firms across various industries. For example,  Ye et al. \cite{Ye2008}, Yang et al.  \cite{Yang2017}, and Du Chuanzhong et al. \cite{Du2017} have explored the mutual influence and outcomes of decisions such as market pricing and promotional subsidies. Wang Aihu et al. \cite{Wang2020} applied a non-cooperative-cooperative game framework to study the optimal decision-making of firms and governments in a mixed duopoly market. Zhang Wei et al. \cite{Zhang2016}, Chen Qi and Wang Qiang \cite{Chen2012}, Kong Yue et al. \cite{Kong2021}, and Meng Zhan et al. \cite{Meng2023} have used the Cournot and Bertrand models to analyze corporate research and development investment, especially within the context of corporate competition mechanisms and government-business interactions in green technological innovation, including the impact and driving forces of government subsidies.

In the first part of this section, we base our study on the Bertrand model, using a duopoly of food producers as an example to analyze the pricing competition behavior of firms. The focus is primarily on the strategic choice behavior of food companies regarding their investment in green technologies, particularly by incorporating the factor of increased green technology investment.

Considering the reality of both competition and cooperation in the market, this section will use the dual-type game model from Section \ref{Sec1} to examine this issue. In fact, as long as there is a market, price acts as the 'invisible hand,' driving market economic activities. The modern market economy differs from a laissez-faire economy: in addition to the market's strong self-regulation function, there are also macroeconomic controls and even interventions. In terms of corporate price competition, aside from the dilemmas and deadlocks revealed by the Nash equilibrium in price wars, there are also cooperative, win-win strategies or tacit understandings aimed at survival and development. Reality shows that firms do not always strictly follow the Nash equilibrium of the Bertrand model; the 'killing a thousand, losing eight hundred' decision in price wars is often avoided by both parties. Competition is undoubtedly the inevitable choice of individual rationality, but avoiding a lose-lose situation is also a unique wisdom of humanity.

In the Bertrand price competition model, we will take two oligopolistic food producers of similar products as an example. By introducing green technology investment as a strategic factor, we will establish a strategic game model, and will then consider cooperative pricing to derive the corresponding dual-type game model from this game problem. This study aims to explore the strategic choices made by firms in competition and cooperation. The main research topics include: How should pricing be considered? How should the level of green technology investment be determined? Furthermore, the study will examine the impact of their cooperation and the interactive effects of their strategic choices, as well as analyze the influence of parameter changes in the model on the outcomes. Finally, it will explore the role that governments or regulatory bodies can play in promoting these effects.

\subsubsection{Assumptions and Modeling}
Consider two competing producers $i,\left( i=1,2 \right)$ in a market, producing homogeneous products. Let $q_1$ and $q_2$ represent the quantities produced by producers 1 and 2, respectively, and $p_1$ and $p_2$ represent the prices of their products. Let $D$ denote the market's potential demand, and $\alpha$ be the price coefficient, which measures the sensitivity of demand to price changes. Assume that both firms have no fixed costs, and their marginal production costs are $c$. According to the Bertrand model, for producers 1 and 2, the firm with the lower price will capture the entire market, while the firm with the higher price will gain no market share at all.

We introduce green technology investment into the demand function, where producers 1 and 2 invest in green technology to increase their market share. Let $\theta \in \left[ 0,1 \right]$ represent the level of green technology investment. Considering constraints such as funding and technological capabilities, let $A$ denote the maximum achievable level of green technology investment for each producer $P$, where $A\theta$ represents the level of green technology investment by the producer, and let $\alpha$ be the coefficient of the impact of green technology innovation on production output.

The profit functions for producers 1 and 2 are respectively:
\begin{equation}\label{eq:5}
	\pi _i\left( p_1,\mathrm{}p_2,\theta _1,\theta _2 \right) =\left\{ \begin{matrix}
		\left( p_i-c \right) \left[ a-bp_i+\lambda A\theta _i \right] -\mu \left( A\theta _i \right) ^2,&		p_i<p_j,\\
		\frac{1}{2}\left( p_i-c \right) \left[ a-bp_i+\lambda (A\theta _1+A\theta _2) \right] -\mu \left( A\theta _i \right) ^2,
		&		p_i=p_j,\\
		0,&		p_i>p_j.\\
	\end{matrix} \right. \,\, i=1,2, i\ne j,
\end{equation}

\textbf{Assumption 4.1}: Assume that if the product is sold at cost, there will definitely be demand; otherwise, there is no need for production, i.e., $a>bc$.

\textbf{Assumption 4.2}: The food pricing should not be lower than the cost, i.e., $p\ge c$. At the same time, a minimum demand quantity $a_0\left( 0<a_0\le a \right) $ should be ensured to meet $a-bp\ge a_0$, otherwise, food enterprises cannot proceed with production and sales.

\textbf{Note 4.1}: From $a-bp\ge a_0$, we can derive $p\le \frac{1}{b}\left( a-a_0 \right)$. Additionally, $\frac{1}{b}\left( a-a_0 \right) \ge p\ge c$ can be inferred, i.e., $p\in [c,\frac{1}{b}\left( a-a_0 \right) ]$, and let $H=[c,\frac{1}{b}\left( a-a_0 \right) ]$.

If the two producers adopt non-cooperative strategies and compete for market share through price competition, then \eqref{eq:5} constitutes a non-cooperative game. Let $X_1=X_2=H\times[0,1]$; then the non-cooperative game \eqref{eq:5} is denoted as $\left( \left\{ 1,2 \right\} ,(X_1,X_2) ,(\pi _1,\pi _2)\right)$. It is easy to see that its Nash equilibrium is $(c,c)$, meaning both producers will set their prices equal to the marginal cost, i.e., $p_1=p_2=c$, and in this case, both producers earn zero profit.

In real-world situations, such a mutually detrimental outcome does not always occur, which is the famous Bertrand paradox. There are several explanations for this phenomenon, including: 1. Production capacity limitations: The Bertrand model assumes that firms can supply the market infinitely, but in reality, firms have limited production capacities and cannot meet the entire market demand. Therefore, firms that raise prices will capture market share based on the allocation principle, which is the market demand minus the output of other firms, and will earn corresponding profits. 2. Dynamic decision-making process: The game between firms is not a one-time event. After losing market share, firms can adjust their prices to regain market share. 3. Product differentiation: The assumption of product homogeneity does not fully align with reality. In practice, products especially when combined with services often have differences, even among similar products. 4. Limited impact of price differences: A small price advantage does not always translate into an absolute market advantage. The effect of price differences on consumers can be delayed, and consumer preferences and consumption habits may reduce the impact of price differences between products from different firms.

These explanations clarify the reasons behind the Bertrand paradox, based on the differences between reality and theoretical assumptions. In fact, there is another reason that cannot be ignored: the Bertrand model assumes that producers cannot collude at all, but in reality, producers often reach some form of cooperation or an unwritten understanding. In other words, since the economic agent's complete rationality is an absolute pursuit of profit, why, knowing that non-cooperation leads to mutual destruction, would they not aim for higher super-rationality in pursuit of greater profit? For this reason, we now consider a situation where competition and cooperation coexist and discuss the problem as a dual-type game. In the non-cooperative phase, the two producers determine their levels of investment in green technology to maximize profits. Based on different levels of green technology investment, the two producers cooperate during pricing and jointly set a reasonable product price.

According to the analysis of the Bertrand model, producers avoid the consequences of a price war by adopting cooperative pricing. Both parties seek a mutually optimal price, $p_1=p_2=p$, in the market and then distribute the benefits of cooperation. Each party determines the level of investment in green technology based on their own circumstances.

In the cooperative pricing phase, both parties determine the optimal price for different levels of green technology investment. According to the dual-type game model in section 2.2, we can derive the alliance function $v\left( p_1,\mathrm{}p_2,\theta _1,\theta _2 \right)$ after cooperative pricing as follows:
\begin{align}
	\nonumber
	\begin{split}
			&\mathrm{}v\left( \emptyset ,\left( p_1,\mathrm{}p_2,\theta _1,\theta _2 \right) \right) =0,i=1,2,\\
			&	\mathrm{}v\left( \left\{ i \right\} ,\left( p_1,\mathrm{}p_2,\theta _1,\theta _2 \right) \right) =\pi _i\left( p_1,\mathrm{}p_2,\theta _1,\theta _2 \right) ,i=1,2,\\
			&v\left( \left\{ 1,2 \right\} ,\left( p_1,\mathrm{}p_2,\theta _1,\theta _2 \right) \right) =\pi _1\left( p_1,\mathrm{}p_2,\theta _1,\theta _2 \right) +\pi _2\left( p_1,\mathrm{}p_2,\theta _1,\theta _2 \right),\\
			&	\mathrm{(}p_1,\mathrm{}p_2)\in H^C\subset H\times H,\\
			&H^C=\left\{ \left( \hat{p},\hat{p} \right) :v\left( \left\{ 1,2 \right\} ,\left( \hat{p},\hat{p},\theta _1,\theta _2 \right) \right) =\max_{(p,p)\in I\times I} \!\:v\left( \left\{ 1,2 \right\} ,\left( p,p,\theta _1,\theta _2 \right) \right) \right\} .
	\end{split}
\end{align}
The above game can also be simplified as:
\begin{align}\label{eq:6}
	&v\left( \emptyset ,( p_1,p_2,\theta _1,\theta _2 ) \right) =0,\quad i=1,2; \notag \\
	&v\left( \{ i \} ,( \hat{p},\hat{p},\theta _1,\theta _2 ) \right) =\pi _i( \hat{p},\hat{p},\theta _1,\theta _2 ) \notag \\
	&\quad =\frac{1}{2}( \hat{p}-c ) ( a-b\hat{p}+\lambda (A\theta _1+A\theta _2) ) -\mu (A\theta _i)^2,\quad i=1,2, \\
	&v\left( \{ 1,2 \} ,( \hat{p},\hat{p},\theta _1,\theta _2 ) \right) = \pi _1( \hat{p},\hat{p},\theta _1,\theta _2 ) +\pi _2( \hat{p},\hat{p},\theta _1,\theta _2 ) \notag \\
	&\quad = ( \hat{p}-c ) ( a-b\hat{p}+\lambda (A\theta _1+A\theta _2) ) -\mu (A\theta _1)^2-\mu (A\theta _2)^2,\ (\hat{p},\hat{p}) \in H^c. \notag
\end{align}
\subsubsection{Solution and Analysis}
Next, we solve for the solution of the dual-type game. First, we present the cooperation set $H^c$ in model \eqref{eq:6}.

For each $\theta _1,\theta _2\in [0,1]$, when the two manufacturers cooperate on pricing, the cooperation set $I^C$ is the price that maximizes the total profit, which corresponds to solving for the optimal solution $\left( \hat{p},\hat{p} \right)$ of $\mathrm{}v\left( \left\{ 1,2 \right\} ,\left( p_1,\mathrm{}p_2,\theta _1,\theta _2 \right) \right)$.Because
\begin{align*}
	v(1,2,(p,p,\theta _1,\theta _2)) &= \pi _1(p,p,\theta _1,\theta _2) + \pi _2(p,p,\theta _1,\theta _2) \\
	&= \frac{1}{2}(p-c)(a-bp+\lambda A\theta _1) - \mu (A\theta _1)^2 + \frac{1}{2}(p-c)(a-bp+\lambda A\theta _2) - \mu (A\theta _2)^2 \\
	&= (p-c)(a-bp) + (p-c)\lambda (A\theta _1 + A\theta _2) - (\mu (A\theta _1)^2 + \mu (A\theta _2)^2).
\end{align*}

Differentiating $\mathrm{}v\left( \left\{ 1,2 \right\} ,\left( p,p,\theta _1,\theta _2 \right) \right)$ with respect to $p$, we can obtain
\begin{align*}
	\frac{dv}{dp}=a-2bp+bc+\lambda \left( A\theta _1+A\theta _2 \right).
\end{align*}
Set the derivative equal to 0, and we can obtain
\begin{align*}
	\hat{p}=\frac{a+bc+\lambda (A\theta _1+A\theta _2)}{2b}.
\end{align*}
Thus, $H^c=\{\hat{p}\}$, at this point
\begin{align*}
	v(1,2,(\hat{p},\hat{p},\theta _1,\theta _2))&=(\hat{p}-c)[(a-b\hat{p})+\lambda (A\theta _1+A\theta _2)]-(\mu (A\theta _1)^2+\mu (A\theta _2)^2)
	\\
	&=(\frac{a-bc+\lambda (A\theta _1+A\theta _2)}{2})(\frac{a-bc+\lambda (A\theta _1+A\theta _2)}{2b})\mathrm{}-\left( \mu \left( A\theta _1 \right) ^2+\mu \left( A\theta _2 \right) ^2 \right) 
	\\
	&=\frac{[a-bc+\lambda \left( A\theta _1+A\theta _2 \right) ]^2}{4b}\mathrm{}-\left( \mu \left( A\theta _1 \right) ^2+\mu \left( A\theta _2 \right) ^2 \right). 
\end{align*}

Under the premise of cooperative pricing, the two producers each determine their green technology investment levels to maximize their own profits. This is the non-cooperative phase of a Biform game. The payoffs of the non-cooperative game come from the distribution of the cooperative surplus, also known as the non-cooperative game derived from the Biform game, denoted as $(\left\{ 1,2 \right\} ,[0,1]^2,\left( a_{1}^{v},a_{2}^{v} \right) )$. The payoff function $a_{1}^{v}(\theta _1,\theta _2),a_{2}^{v}(\theta _1,\theta _2)$ is the allocation function of the Biform game $(\left\{ 1,2 \right\} ,,\mathrm{}v(\bullet ,\left( \theta _1,\theta _2 \right) ),[0,1]^2)$,  which is the solution to the cooperative game $(\left\{ 1,2 \right\} ,\mathrm{}v(\bullet ,(\theta _1,\theta _2)).$

We choose the allocation function in two ways: one is based on marginalism, and the other is based on egalitarianism, both of which we will discuss.

First, according to the marginal
\begin{align*}
	\varphi _1\left( \theta _1,\theta _2\mathrm{} \right) &=\frac{1}{2}\left( \hat{p}-c \right) \left( \mathrm{}a-b\hat{p}+\lambda \left( A\theta _1+A\theta _2 \right) \right) -\mu (A\theta _1)^2
	\\
	&=\frac{[a-bc+\lambda \left( A\theta _1+A\theta _2 \right) ]^2}{8b}-\mu (A\theta _1)^2,
\end{align*}
\begin{align*}
	\varphi _2\left( \theta _1,\theta _2\mathrm{} \right) &=\frac{1}{2}\left( \hat{p}-c \right) \left( \mathrm{}a-b\hat{p}+\lambda \left( A\theta _1+A\theta _2 \right) \right) -\mu \left( A\theta _1 \right) ^2
	\\
	\mathrm{}&=\frac{[a-bc+\lambda \left( A\theta _1+A\theta _2 \right) ]^2}{8b}-\mu (A\theta _1)^2.
\end{align*}

The above allocation method is the most direct way to distribute based on the size of contributions, where each party receives their own contribution (their respective profit) from the total profit.

The partial derivatives of $\varphi _1\left( \theta _1,\theta _2\mathrm{} \right) ,\mathrm{}\varphi _2\left( \theta _1,\theta _2\mathrm{} \right) $ with respect to $\theta _1$ and $\theta _2$ are as follows:
\begin{align*}
	\frac{\partial \varphi _1}{\partial \theta _1}=\frac{\lambda A[a-bc+\lambda \left( A\theta _1+A\theta _2 \right) ]}{4b}-2\mu A^2\theta _1,
\end{align*}
\begin{align*}
	\mathrm{}\frac{\partial \varphi _2}{\partial \theta _2}=\frac{\lambda A[a-bc+\lambda \left( A\theta _1+A\theta _2 \right) ]}{4b}-2\mu A^2\theta _2.
\end{align*}
Let $\frac{\partial \varphi _1}{\partial \theta _1}=\mathrm{}\frac{\partial \varphi _2}{\partial \theta _2}=0$, we get
\begin{align*}
	\hat{\theta}_1=\hat{\theta}_2=\frac{\lambda (a-bc)}{2A(4\mu b-\lambda ^2)}.
\end{align*}
Considering the sign of $(4\mu b-\lambda ^2)$, two cases need to be discussed.

\textbf{(B1)} when $\mu >\frac{\lambda ^2}{4b}$ is the case, that is, when the cost of green technology investment is relatively high, if $\theta _2=\hat{\theta}_2=\frac{\lambda (a-bc)}{2A(4\mu b-\lambda ^2)}$, then
\begin{align*}
	\left. \frac{\partial \varphi _1}{\partial \theta _1} \right|_{\theta _2=\hat{\theta}_2}=\frac{\lambda A\left[ a-bc+\lambda \left( A\theta _1+A\hat{\theta}_2 \right) \right]}{4b}-2\mu A^2\theta _1
	\\
	=\frac{\lambda A\left[ a-bc+\lambda \left( A\theta _1+A\times \frac{\lambda \left( a-bc \right)}{2A(4\mu b-\lambda ^2)} \right) \right] -8\mu bA^2\theta _1}{4b}
	\\
	=\frac{A}{4b}[(\mathrm{}\lambda ^2-8\mu b)A\theta _1+\frac{8\mu b\lambda \left( a-bc \right) -\lambda ^3\left( a-bc \right)}{2\left( 4\mu b-\lambda ^2 \right)}]
	\\
	\mathrm{}=\frac{A}{4b}[\frac{\lambda \left( a-bc \right) \left( 8\mu b-\lambda ^2 \right)}{2\left( 4\mu b-\lambda ^2 \right)}-(8\mu b-\lambda ^2)A\theta _1].
\end{align*}
Then, when $\theta _1<\hat{\theta}_2=\frac{\lambda \left( a-bc \right)}{2A\left( 4\mu b-\lambda ^2 \right)}$, $\left. \frac{\partial \varphi _1}{\partial \theta _1} \right|_{\theta _2=\hat{\theta}_2}>0$ holds; and when $\theta _1>\hat{\theta}_2=\frac{\lambda \left( a-bc \right)}{2A\left( 4\mu b-\lambda ^2 \right)}$, $\left. \frac{\partial \varphi _1}{\partial \theta _1} \right|_{\theta _2=\hat{\theta}_2}<0$ holds.
Similarly, it can be obtained that $\left. \frac{\partial \varphi _2}{\partial \theta _2} \right|_{\theta _1=\hat{\theta}_1}$ satisfies the same property.

From this, we know that when $\hat{\theta}_1=\hat{\theta}_2=\lambda (a-bc)/2A(4\mu b-\lambda ^2)$, $\varphi _1\left( \theta _1,\hat{\theta}_2 \right) ,\varphi _1\left( \hat{\theta}_1,\theta _2 \right) $ attain their maximum values at $\hat{\theta}_1,\hat{\theta}_2$ respectively, that is
\begin{align*}
	\varphi _1\left( \hat{\theta}_1,\hat{\theta}_2 \right) \ge \varphi _1\left( \theta _1,\hat{\theta}_2 \right) , \forall \theta _1 \in [0,1],
\end{align*}
\begin{align*}
	\varphi _2\left( \hat{\theta}_1,\hat{\theta}_2 \right) \ge \varphi _2\left( \hat{\theta}_1,\theta _2 \right) , \forall \theta _2 \in [0,1].
\end{align*}

That is to say, $\left( \hat{\theta}_1,\hat{\theta}_2\mathrm{} \right)$ is the Nash equilibrium of the non-cooperative stage, and thus serves as the solution to the Biform game problem (2.4.2). At this point, the profits of the manufacturers are respectively
\begin{align*}
	\mathrm{}\varphi _1\left( \hat{\theta}_1,\hat{\theta}_2\mathrm{} \right) &=\varphi _2\left( \hat{\theta}_1,\hat{\theta}_2\mathrm{} \right) 
	\\
	\mathrm{}&=\frac{\left[ a-bc+\lambda \left( A\hat{\theta}_1+A\hat{\theta}_2 \right) \right] ^2}{8b}-\mu (A\theta _1)^2=\left( a-bc \right) ^2\frac{\left[ 1+\frac{\lambda ^2}{\left( 4\mu b-\lambda ^2 \right)} \right] ^2}{8b}-\frac{\mu \lambda ^2\left( a-bc \right) ^2}{4\left( 4\mu b-\lambda ^2 \right) ^2}
	\\
	&=\frac{\mu \left( a-bc \right) ^2\left( 8\mu b-\lambda ^2 \right)}{4\left( 4\mu b-\lambda ^2 \right) ^2}>0.
\end{align*}

\textbf{(B2)} when $\mu \le \frac{\lambda ^2}{4b}$, that is, in the case where the investment cost of green technology is relatively low, we have
\begin{align*}
	\frac{\partial \varphi _1}{\partial \theta _1}&=\frac{\lambda A\left[ a-bc+\lambda \left( A\theta _1+A\theta _2 \right) \right]}{4b}-2\mu A^2\theta _1
	\\
	\mathrm{}&=\frac{\lambda \left( a-bc \right) +2(\lambda ^2-4\mu b)A\theta _1+\lambda ^2(A\theta _2-A\theta _1)}{4b},
\end{align*}
\begin{align*}
	\mathrm{}\frac{\partial \varphi _2}{\partial \theta _2}&=\frac{\lambda A\left[ a-bc+\lambda \left( A\theta _1+A\theta _2 \right) \right]}{4b}-2\mu A^2\theta _2
	\\
	&\mathrm{}=\frac{\lambda \left( a-bc \right) +2(\lambda ^2-4\mu b)A\theta _1+\lambda ^2(A\theta _1-A\theta _2)}{4b}.
\end{align*}

For each $\left(\theta _1,\theta _2 \right) \in [0,1]^2$, if $\theta _1<\theta _2$, then we have $\frac{\partial \varphi _1}{\partial \theta _1}>0$. Taking $\tau _1=\frac{\theta _1+\theta _2}{2}$, then we have $\varphi _1\left(\tau _1,\theta _2\mathrm{} \right) >\varphi _1\left(\theta _1,\theta _2\mathrm{} \right) $. That is to say, $\theta _1$ cannot be the maximum point of $\varphi _1\left(\bullet ,\theta _2\mathrm{} \right) $. By the same logic, if $\theta _2<\theta _1$, then $\theta _2$ cannot be the maximum point of $\varphi _1\left(\theta _1,\bullet \mathrm{} \right) $. Therefore, the points that make $\varphi _1\left(\bullet ,\theta _2\mathrm{} \right)$ and $\varphi _1\left(\theta _1,\bullet \mathrm{} \right) $ attain their maximum values respectively can only be the points of $\theta _1=\theta _2$. Note that the points of $\theta _1=\theta _2$ satisfy both $\frac{\partial \varphi _1}{\partial \theta _1}>0$ and $\frac{\partial \varphi _2}{\partial \theta _2}>0$. Therefore, the points that make $\varphi _1\left(\bullet ,\theta _2\mathrm{} \right) $ and $\varphi _1\left(\theta _1,\bullet \mathrm{} \right) $ attain their maximum values respectively are $\theta _1=\theta _2=1\text{}$.

From this, it can be obtained that $\left( \theta _{1}^{*},\theta _{2}^{*} \right) =(1,1)$ constitutes the Nash equilibrium of the non-cooperative game, that is

\begin{align*}
	\varphi _1(1,1) \ge \psi _1\left( \theta _1,1 \right) , \quad \forall \theta _1 \in [0,1],
\end{align*}
\begin{align*}
	\varphi _2(1,1) \ge \psi _2\left( 1,\theta _2 \right) , \quad \forall \theta _2 \in [0,1].
\end{align*}
At this point, the profits of the manufacturers are
\begin{align*}
	\varphi _1\left( 1,1 \right) &=\mathrm{}\varphi _2\left( 1,1 \right) =\mathrm{}\frac{\left[ a-bc+\lambda \left( A\theta _{1}^{*}+A\theta _{2}^{*}\mathrm{} \right) \right] ^2}{8b}-\mu \left( A\theta _{1}^{*})^2 \right] 
	\\
	&\mathrm{}=\mathrm{}\frac{1}{8b}\left[ \left( a-bc+2\lambda A \right) ^2-8\mu bA^2 \right] 
	\\
	&=\frac{1}{8b}[(a-bc)^2+4\lambda A\left( a-bc \right) +2(2\lambda ^2-4\mu b)A^2].
\end{align*}

Next, we discuss the case where the allocation function follows egalitarianism. Under egalitarian allocation, we directly set the allocation function as
\begin{align*}
	\mathrm{}\psi _1\left( \theta _1,\theta _2\mathrm{} \right) &=\frac{1}{2}v\left( \left\{ 1,2 \right\} ,\left( \theta _1,\theta _2\mathrm{} \right) \right) 
	\\
	&\mathrm{}=\frac{1}{2}[\left( \hat{p}-c \right) \left( \mathrm{}a-b\hat{p}+\lambda \left( A\theta _1+A\theta _2 \right) \right) -\mu (A\theta _1)^2-\mu \left( A\theta _1 \right) ^2]
	\\
	&=\frac{[a-bc+\lambda \left( A\theta _1+A\theta _2 \right) ]^2}{8b}-\frac{1}{2}[\mu \left( A\theta _1)^2+\mu \left( A\theta _1 \right) ^2 \right],
\end{align*}
\begin{align*}
	\psi _2\left( \theta _1,\theta _2\mathrm{} \right) &=\frac{1}{2}v\left( \left\{ 1,2 \right\} ,\left( \theta _1,\theta _2\mathrm{} \right) \right) 
	\\
	&\mathrm{}=\frac{1}{2}[\left( \hat{p}-c \right) \left( \mathrm{}a-b\hat{p}+\lambda \left( A\theta _1+A\theta _2 \right) \right) -\mu (A\theta _1)^2-\mu \left( A\theta _1 \right) ^2]
	\\
	&=\frac{[a-bc+\lambda \left( A\theta _1+A\theta _2 \right) ]^2}{8b}-\frac{1}{2}[\mu \left( A\theta _1)^2+\mu \left( A\theta _1 \right) ^2 \right] .
\end{align*}

Taking the partial derivatives of $\psi _1\left( \theta _1,\theta _2\mathrm{} \right) ,\mathrm{}\psi _2\left( \theta _1,\theta _2\mathrm{} \right)$ with respect to $\theta _1$ and $\theta _2$ respectively, we can obtain
\begin{align*}
	\frac{\partial \psi _1}{\partial \theta _1}=\frac{\lambda A[a-bc+\lambda \left( A\theta _1+A\theta _2 \right) ]}{4b}-\mu A^2\theta _1,
\end{align*}
\begin{align*}
	\mathrm{}\frac{\partial \psi _2}{\partial \theta _2}=\frac{\lambda A[a-bc+\lambda \left( A\theta _1+A\theta _2 \right) ]}{4b}-\mu A^2\theta _2.
\end{align*}
Set $\frac{\partial \psi _1}{\partial \theta _1}=\mathrm{}\frac{\partial \psi _2}{\partial \theta _2}=0$, and we can obtain
\begin{align*}
	\theta _{1}^{*}=\theta _{2}^{*}=\frac{\lambda (a-bc)}{A(4\mu b-2\lambda ^2)}.
\end{align*}
Similar to the case where the allocation function is $\varphi _1,\varphi _2$, we also discuss it in two cases.

\textbf{(C1)} When $\mu >\frac{2\lambda ^2}{4b}$, that is, in the case where the investment cost of green technology is relatively high, if $\theta _2=\theta _{2}^{*}=\frac{\lambda (a-bc)}{A(4\mu b-2\lambda ^2)}$, we have
\begin{align*}
	\left. \frac{\partial \psi _1}{\partial \theta _1} \right|_{\theta _2=\hat{\theta}_2}=\frac{\lambda A\left[ a-bc+\lambda \left( A\theta _1+A\theta _{2}^{*} \right) \right]}{4b}-\mu A^2\theta _1
	\\
	=\frac{\lambda A\left[ a-bc+\lambda \left( A\theta _1+A\times \frac{\lambda \left( a-bc \right)}{A(4\mu b-2\lambda ^2)} \right) \right] -4\mu bA^2\theta _1}{4b}
	\\
	=\frac{A}{4b}[\lambda \left( a-bc \right) +(\mathrm{}\lambda ^2-4\mu b)A\theta _1+\frac{\lambda ^3\left( a-bc \right)}{\left( 4\mu b-2\lambda ^2 \right)}]
	\\
	\mathrm{}=\frac{A}{4b}[\frac{\lambda \left( a-bc \right) \left( 4\mu b-\lambda ^2 \right)}{\left( 4\mu b-2\lambda ^2 \right)}-(4\mu b-\lambda ^2)A\theta _1].
\end{align*}
Therefore, when $\theta _1<\theta _{2}^{*}=\frac{\lambda (a-bc)}{A(4\mu b-2\lambda ^2)}$, $\left. \frac{\partial \psi _1}{\partial \theta _1} \right|_{\theta _2=\hat{\theta}_2}>0$; when $\theta _1>\theta _{2}^{*}=\frac{\lambda (a-bc)}{A(4\mu b-2\lambda ^2)}$, $\left. \frac{\partial \psi _1}{\partial \theta _1} \right|_{\theta _2=\hat{\theta}_2}<0.$.
Therefore, when $\theta _1<\theta _{2}^{*}=\frac{\lambda (a-bc)}{A(4\mu b-2\lambda ^2)}$, $\left. \frac{\partial \psi _2}{\partial \theta _2} \right|_{\theta _1=\hat{\theta}_1}>0$; when $\theta _1>\theta _{2}^{*}=\frac{\lambda (a-bc)}{A(4\mu b-2\lambda ^2)}$, $\left. \frac{\partial \psi _2}{\partial \theta _2} \right|_{\theta _1=\hat{\theta}_1}<0$.

From this, we know that when $\theta _{1}^{*}=\theta _{2}^{*}=\frac{\lambda (a-bc)}{A(4\mu b-2\lambda ^2)}$, $\psi _1\left( \theta _1,\theta _{2}^{*}\mathrm{} \right) ,\mathrm{}\psi _2\left( \theta _{1}^{*},\theta _2\mathrm{} \right) $ attain their maximum values at $\theta _{1}^{*},\theta _{2}^{*}$ respectively, that is
\begin{align*}
	\psi _1\left( \theta _{1}^{*},\theta _{2}^{*} \right) \ge \psi _1\left( \theta _1,\theta _{2}^{*} \right) , \quad \forall \theta _1 \in [0,1],
\end{align*}
\begin{align*}
	\psi _2\left( \theta _{1}^{*},\theta _{2}^{*} \right) \ge \psi _2\left( \theta _{1}^{*},\theta _2 \right) , \quad \forall \theta _2 \in [0,1].
\end{align*}
That is to say, $\left( \theta _{1}^{*},\theta _{2}^{*}\mathrm{} \right)$ is the Nash equilibrium of the non-cooperative stage, and thus serves as the solution to the Biform game problem \eqref{eq:6}. At this point, the profits of the manufacturers are respectively
\begin{align*}
	\mathrm{}\psi _1\left( \theta _{1}^{*},\theta _{2}^{*}\mathrm{} \right) &=\psi _2\left( \theta _{1}^{*},\theta _{2}^{*}\mathrm{} \right) 
	\\
	\mathrm{}&=\mathrm{}\frac{[a-bc+\lambda \left( A\theta _{1}^{*}+A\theta _{2}^{*}\mathrm{} \right) ]^2}{8b}-\frac{1}{2}[\mu \left( A\theta _{1}^{*})^2+\mu \left( A\theta _{2}^{*} \right) ^2 \right] 
	\\
&	=\left( a-bc \right) ^2\times \frac{\left[ 1+\lambda \times 2A\times \frac{\lambda}{A\left( 4\mu b-2\lambda ^2 \right)} \right] ^2}{8b}-\mu \times A^2\times \frac{\lambda ^2\left( a-bc \right) ^2}{A^2\left( 4\mu b-2\lambda ^2 \right) ^2}
	\\
&	=\mu \left( a-bc \right) ^2[\frac{2\mu b}{\left( 4\mu b-2\lambda ^2 \right) ^2}-\frac{\lambda ^2}{\left( 4\mu b-2\lambda ^2 \right) ^2}]=\frac{\mu \left( a-bc \right) ^2}{2\left( 4\mu b-2\lambda ^2 \right)}>0.
\end{align*}

\textbf{(C2)} When $\mu \le \frac{2\lambda ^2}{4b}$, that is, in the case where the investment cost of green technology is relatively low, we have
\begin{align*}
	\frac{\partial \psi _1}{\partial \theta _1}&=\frac{\lambda A\left[ a-bc+\lambda \left( A\theta _1+A\theta _2 \right) \right]}{4b}-\mu A^2\theta _1
	\\
	\mathrm{}&=\frac{\lambda \left( a-bc \right) +(2\lambda ^2-4\mu b)A\theta _1+\lambda ^2(A\theta _2-A\theta _1)}{4b},
\end{align*}
\begin{align*}
	\frac{\partial \psi _2}{\partial \theta _2}&=\frac{\lambda A\left[ a-bc+\lambda \left( A\theta _1+A\theta _2 \right) \right]}{4b}-\mu A^2\theta _2
	\\
	\mathrm{}&=\frac{\lambda \left( a-bc \right) +(2\lambda ^2-4\mu b)A\theta _1+\lambda ^2(A\theta _1-A\theta _2)}{4b}.
\end{align*}
For each $\left(\theta _1,\theta _2 \right) \in [0,1]^2$, if $\theta _1<\theta _2$, then we have $\mathrm{}\frac{\partial \psi _1}{\partial \theta _1}>0$. Taking $\tau _1=\frac{\theta _1+\theta _2}{2}$, then we have $\mathrm{}{\psi _1}_1\left( \tau _1,\theta _2\mathrm{} \right) >{\psi _1}_1\left( \theta _1,\theta _2\mathrm{} \right)$. That is to say, $\theta _1$ cannot be the maximum point of $\psi _1\left( \bullet ,\theta _2\mathrm{} \right) $. By the same logic, if $\theta _2<\theta _1$, then $\theta _2$ cannot be the maximum point of $\psi _2\left( \theta _1,\bullet \mathrm{} \right)  $. Therefore, the points that make $\psi _1\left( \bullet ,\theta _2\mathrm{} \right)$ and $\psi _2\left( \theta _1,\bullet \mathrm{} \right)  $ attain their maximum values respectively can only be the points of $\theta _1=\theta _2$. Note that the points of $\theta _1=\theta _2$ satisfy both $\frac{\partial \varphi _1}{\partial \theta _1}>0$ and $\frac{\partial \varphi _2}{\partial \theta _2}>0$. Therefore, the points that make $\psi _1\left( \bullet ,\theta _2\mathrm{} \right)$ and $\psi _2\left( \theta _1,\bullet \mathrm{} \right)  $ attain their maximum values respectively are $\theta _1=\theta _2=1\text{}$.
From this, it can be obtained that $\left( \theta _{1}^{*},\theta _{2}^{*} \right) =(1,1)$ constitutes the Nash equilibrium of the non-cooperative game, that is
\begin{align*}
	\psi _1(1,1) \ge \psi _1(\theta _1,1), \quad \forall \theta _1 \in [0,1],
\end{align*}
\begin{align*}
	\psi _2(1,1) \ge \psi _2(1,\theta _2), \quad \forall \theta _2 \in [0,1].
\end{align*}
At this point, the profits of the manufacturers are
\begin{align*}
	\psi _1\left( 1,1 \right) &=\mathrm{}\psi _2\left( 1,1 \right) =\mathrm{}\frac{\left[ a-bc+\lambda \left( A\theta _{1}^{*}+A\theta _{2}^{*}\mathrm{} \right) \right] ^2}{8b}-\frac{1}{2}[\mu \left( A\theta _{1}^{*})^2+\mu \left( A\theta _{2}^{*} \right) ^2 \right] 
	\\
	&=\mathrm{}\frac{1}{8b}\left[ \left( a-bc+2\lambda A \right) ^2-8\mu bA^2 \right] 
	\\
	&=\frac{1}{8b}[\left( a-bc)^2+4\lambda A\left( a-bc \right) +2\left( 2\lambda ^2-4\mu b \right) A^2 \right].
\end{align*}

Finally, we analyze the differences in green technology investment levels. First, cooperative pricing increases manufacturers' profits, while engaging in price wars will result in zero profits for both parties. Under the circumstances of complete non-cooperation, green technology is out of the question, and the green technology investment level is zero. Through cooperative pricing, not only do profits increase, but the green technology investment level is also improved.

Additionally, the two different allocation methods also lead to different outcomes. We compare the green technology investment levels under the two allocation methods. Since $\mathrm{}\theta _{1}^{*}=\mathrm{}\theta _{2}^{*},\hat{\theta}_1=\hat{\theta}_2$, we only need to compare $\mathrm{}\theta _{1}^{*}$ with $\hat{\theta}_1$. We discuss it in two cases.

\textbf{(D1)} when $\mu >\frac{2\lambda ^2}{4b}$, then we have $\mu > \lambda^2/4b$. From the previous analysis, we have
\begin{align*}
	\mathrm{  }\psi _1\left( \theta _{1}^{*},\theta _{2}^{*}\mathrm{  } \right) -\varphi _1\left( \hat{\theta}_1,\hat{\theta}_2\mathrm{  } \right) &=\frac{\mu \left( a-bc \right) ^2}{2\left( 4\mu b-2\lambda ^2 \right)}-\frac{\mu \left( a-bc \right) ^2\left( 8\mu b-\lambda ^2 \right)}{4\left( 4\mu b-\lambda ^2 \right) ^2}
	\\
	&\mathrm{  }=\frac{\mu \left( a-bc \right) ^2}{4\left( 4\mu b-2\lambda ^2 \right) \left( 4\mu b-\lambda ^2 \right) ^2}\left[2\left( 4\mu b-\lambda ^2 \right) ^2-\left( 8\mu b-\lambda ^2 \right) \left( 4\mu b-2\lambda ^2 \right) \right]
	\\
	&\mathrm{  }=\frac{\mu \left( a-bc \right) ^2}{4\left( 4\mu b-2\lambda ^2 \right) \left( 4\mu b-\lambda ^2 \right) ^2}\left[ \left( 32\mu ^2b^2-16\mu b\lambda ^2+2\lambda ^4 \right)  - \left( 32\mu ^2b^2-20\mu b\lambda ^2+2\lambda ^4 \right) \right]
	\\
	&=\frac{4\mu ^2b\lambda ^2\left( a-bc \right) ^2}{4\left( 4\mu b-2\lambda ^2 \right) \left( 4\mu b-\lambda ^2 \right) ^2}>0.
\end{align*}
That is $\mathrm{}\psi _1\left( \theta _{1}^{*},\theta _{2}^{*}\mathrm{} \right) >\varphi _1\left( \hat{\theta}_1,\hat{\theta}_2\mathrm{} \right) $. Therefore, compared with the marginalist allocation method, the egalitarian allocation method will promote the improvement of green technology investment level and at the same time increase manufacturers' profits.

\textbf{(D2)} when $\mu \le \frac{2\lambda ^2}{4b}$, then we have $\mathrm{}\theta _{1}^{*}=1$, and obviously we have $\mathrm{}\theta _{1}^{*}\ge \hat{\theta}_1$. For further analysis, if $\mu \le \frac{\lambda ^2}{4b}$, then $\hat{\theta}_1=\mathrm{}\theta _{1}^{*}=1$; at this point, the profits of both manufacturers are
\begin{align*}
	\psi _1\left( 1,1 \right) =\varphi _1\left( 1,1 \right) =\frac{1}{8b}[\left( a-bc)^2+4\lambda A\left( a-bc \right) +2\left( 2\lambda ^2-4\mu b \right) A^2 \right]. 
\end{align*}
If $\frac{\lambda ^2}{4b}<\mu \le \frac{2\lambda ^2}{4b}$, then $\mathrm{}\theta _{1}^{*}=1\ge \hat{\theta}_1=\frac{\lambda \left( a-bc \right)}{2A\left( 4\mu b-\lambda ^2 \right)}$. Additionally
\begin{align*}
	\psi _1\left( \theta _{1}^{*},\theta _{2}^{*} \right) - \varphi _1\left( \hat{\theta}_1,\hat{\theta}_2 \right) 
	&= \psi _1\left( 1,1 \right) - \varphi _1\left( \hat{\theta}_1,\hat{\theta}_2 \right) \\
	&= \frac{1}{8b} \left[ (a-bc)^2 + 4\lambda A (a-bc) + 2(2\lambda^2 - 4\mu b) A^2 \right] \\
	&\quad - \frac{\mu (a-bc)^2 (8\mu b - 2\lambda^2) + \lambda^2}{4(4\mu b - \lambda^2)^2} \\
	&= \frac{1}{8b(4\mu b - \lambda^2)} \left[ (a-bc)^2 (4\mu b - \lambda^2) 
	+ 4\lambda A (a-bc) (4\mu b - \lambda^2) \right. \\
	&\quad \left. + 2(2\lambda^2 - 4\mu b)(4\mu b - \lambda^2) A^2 - 4\mu b (a-bc)^2 \right] \\
	&\quad + \frac{\lambda^2 \mu (a-bc)^2}{4(4\mu b - \lambda^2)^2} \\
	&= \frac{1}{8b(4\mu b - \lambda^2)} \left[ 4\lambda A (a-bc)(4\mu b - \lambda^2) \right. \\
	&\quad \left. + 2(2\lambda^2 - 4\mu b)(4\mu b - \lambda^2) A^2 \right] \\
	&\quad + \frac{\lambda^2 (a-bc)^2 (\lambda^2 - 2\mu b)}{8b(4\mu b - \lambda^2)^2} > 0.
\end{align*}

That is, $\mathrm{}\psi _1\left( \theta _{1}^{*},\theta _{2}^{*}\mathrm{} \right) >\varphi _1\left( \hat{\theta}_1,\hat{\theta}_2\mathrm{} \right)$. Therefore, in either case, there is both $\theta _{1}^{*}=\theta _{2}^{*}\ge \hat{\theta}_1=\hat{\theta}_2$ and $\mathrm{}\psi _1\left( \theta _{1}^{*},\theta _{2}^{*}\mathrm{} \right) \ge \varphi _1\left( \hat{\theta}_1,\hat{\theta}_2\mathrm{} \right)$; and except in the case of the cost coefficients $\mu \le \frac{\lambda ^2}{4b}$ and $\frac{\lambda \left( a-bc \right)}{2A\left( 4\mu b-\lambda ^2 \right)}=1$, there is both $\theta _{1}^{*}=\theta _{2}^{*}>\hat{\theta}_1=\hat{\theta}_2$ and $\mathrm{}\psi _1\left( \theta _{1}^{*},\theta _{2}^{*}\mathrm{} \right) >\varphi _1\left( \hat{\theta}_1,\hat{\theta}_2\mathrm{} \right)$.

To sum up, regardless of the value of the green technology investment coefficient $\mu$, when the two manufacturers adopt cooperative pricing, both their profits and green technology investment levels are higher than those in the case of complete non-cooperation. Meanwhile, the two different allocation methods of cooperative benefits also lead to different outcomes; whether in terms of profits or green technology investment levels, the egalitarian allocation method is superior to the marginalist one by comparison.

Additionally, we can also discuss the differences in prices.
\begin{align*}
	\mathrm{}\hat{p}-\bar{p}&=[\frac{a+bc}{2b}+\frac{\left( a-bc \right)}{2b\left( 4\mu b-1 \right)}]
	\\
	&-[\frac{a+\mathrm{}\left( b+\lambda \right) c}{2b+\lambda}+\frac{2\left( a-bc \right) \left( b+\lambda \right) ^2}{\left( 2b+\lambda \right) \left[ \mu \left( 2b+3\lambda \right) \left( 2b+\lambda \right) ^2-2\left( b+\lambda \right) ^2 \right]}]
	\\
	&=\frac{4\mu a+4\mu bc-2c}{2\left( 4\mu b-1 \right)}
	-\frac{\mu [a+\mathrm{}\left( b+\lambda \right) c]\left( 2b+3\lambda \right) \left( 2b+\lambda \right) ^2-2\left( b+\lambda \right) ^2\left( 2b+\lambda \right) c}{(2b+\lambda )\left[ \mu \left( 2b+3\lambda \right) \left( 2b+\lambda \right) ^2-2\left( b+\lambda \right) ^2 \right]}
	\\
	&=\frac{\mu \left( 2b+3\lambda \right) \left( 2b+\lambda \right) [\left( 2b+\lambda \right) \left( 2\mu a+2\mu bc-c \right) -\left( 4\mu b-1 \right) \left( a+\mathrm{}\left( b+\lambda \right) c \right) ]}{\left( 4\mu b-1 \right) \left[ \mu \left( 2b+3\lambda \right) \left( 2b+\lambda \right) ^2-2\left( b+\lambda \right) ^2 \right]}
	\\
	&+\frac{2\left( b+\lambda \right) ^2[4\mu bc-\left( 2\mu a+2\mu bc \right) ]}{\left( 4\mu b-1 \right) \left[ \mu \left( 2b+3\lambda \right) \left( 2b+\lambda \right) ^2-2\left( b+\lambda \right) ^2 \right]}
	\\
	&=\frac{\mu \left( a-bc \right) \left[ \left( 2b+3\lambda \right) \left( 2b+\lambda \right) \left( 1+2\mu \lambda \right) -4\left( b+\lambda \right) ^2 \right]}{\left( 4\mu b-1 \right) \left[ \mu \left( 2b+3\lambda \right) \left( 2b+\lambda \right) ^2-2\left( b+\lambda \right) ^2 \right]}
	\\
	&=\frac{\mu \lambda \left( a-bc \right) \left[ 8\mu \lambda ^2+\left( 16\mu b-1 \right) \lambda +8ub \right]}{\left( 4\mu b-1 \right) \left[ \mu \left( 2b+3\lambda \right) \left( 2b+\lambda \right) ^2-2\left( b+\lambda \right) ^2 \right]}>0\mathrm{}
\end{align*}

Therefore, when $\mu >\frac{1}{4b}$, we have $\hat{p}>\bar{p}$. That is to say, when the green investment cost coefficient exceeds a certain level, cooperative pricing enables green investment and profits to increase, and at the same time, product prices also rise.
\subsection{Based on competition within the supply chain and cost sharing: Biform game analysis}
\subsubsection{Problem Description}
In the first part of this section, based on the Bertrand model, we explore how manufacturers of similar products formulate reasonable prices and green technology investments in market competition, and achieve profit maximization by virtue of price advantages and food safety quality.

In the second part of this section, based on the Cournot model, we use the Biform game method to study the strategic selection behavior of enterprises' green technology investment within the supply chain from the perspective of competition and cost sharing in the supply chain, and thereby discuss the incentive measures to improve enterprises' green technology investment levels.

Regarding green technology investment in the supply chain, many scholars have conducted research on the development of green supply chains from different perspectives, mainly including research on enterprise carbon emission decisions (see Yang Huixiao and Ou Jinwen \cite{Yang2020}, Wang Na and Zhang Yulin \cite{Wang2021}, Zhang Zhipeng \cite{Zhang2022}, Liu Mingwu et al. \cite{Liu2022}), as well as game analysis and evolutionary dynamics research on enterprises' investment in product greenness and government promotion (see Shao Bilin and Hu Linglin \cite{Shao2021}, Zhu Qinghua and Dou Yijie \cite{Zhu2007}, \cite{Zhu2011}, Jiang Shiying and Li Suicheng \cite{Jiang2015}, Bai Chunguang and Tang Jiafu \cite{Bai2017}, Xu Geni \cite{Xu2020}). Considering the coexistence of competition and cooperation within the supply chain, the Biform game has become an effective method for researching green technology investment in the supply chain (see Nan Jiangxia et al. \cite{Nan2021}, Li Mengqi et al. \cite{Li2023a}, Li Dengfeng et al. \cite{Li2023b}, Liang Kairong and Li Dengfeng \cite{Liang2023}).

For the research on green technology investment in the food supply chain, we will apply the Biform game model proposed in 2.2 to discuss the competitive and cooperative behaviors among raw material suppliers, food manufacturers, and distributors. The main contents include: by adding a green technology investment term to the demand function of the Cournot model, we construct a strategic game model for competition among enterprises within the supply chain; then, we establish a Biform game model derived from the strategic game through cost sharing. In this model, the first stage is a non-cooperative game, where enterprises in the supply chain determine the manufacturers' green technology investment levels and the final retail prices from the perspective of maximizing their own profits. The second stage is a cooperative game, where enterprises in the chain determine the cost-sharing and total profit allocation methods based on different green technology investment levels. Finally, the solution to the Biform game is obtained based on the non-cooperative game corresponding to the allocation function. (2) Through the analysis of the solution to the Biform game, the corresponding conclusions are drawn, including: determining the allocation function of enterprises in the chain based on cost sharing can achieve the Pareto optimality of the total profit of the supply chain; by analyzing the relationships between various parameters in the model, green technology investment levels, and profits, it is found that cooperation not only increases profits but also promotes enterprises in the chain to improve green technology investment levels under certain conditions. The research results reveal the inherent laws of food enterprises increasing the application of green technology and also provide decision-making references for the government and regulatory authorities to promote food enterprises to increase green technology investment.
\subsubsection{Assumptions and Modeling}
In response to the concept of sustainable development, consumers are paying increasing attention to the green and environmentally friendly attributes of food. Food enterprises have begun to increase their investment in green technology, and the food industry from raw material supply to food processing, packaging, and then to product sales widely adopts green technology to enhance product competitiveness. We consider a food industry supply chain, which consists of one supplier $S$, one manufacturer $P$, and one distributor $R$, denoted as $N=\left\{ S,P,R \right\}$. Within the supply chain, the supplier procures raw materials at a unit cost of $c$ and supplies them to the manufacturer in bulk at a unit price of m; the manufacturer sells (products) to the downstream distributor at a wholesale price of $w$; and the distributor sells to retailers and consumers at a wholesale price of $p$.

Let the demand function of this product be
\begin{align*}
	Q=a-bp+A\theta,
\end{align*}
where $a$ represents the potential market demand, and $b$ is the price coefficient, which measures the sensitivity of demand quantity to price changes.

\textbf{Assumption 4.3}: Manufacturer $P$ invests in green technology to increase the product's market share, where $\theta \in [0,1]$ denotes the green technology investment level. Considering constraints such as capital and technical capabilities, we assume the maximum achievable green technology investment level of manufacturer $P$ is $A$, and $A\theta$ represents the green technology investment level of manufacturer $P$.

\textbf{Assumption 4.4}: $a>bc$ If the product is sold at cost, there is definitely demand for it; otherwise, there is no need for production¡ªthat is, $a>bc$.

\textbf{Assumption 4.5}: The food pricing is not lower than the cost, i.e., $p \ge c$. Meanwhile, a minimum demand quantity $a_0$ (with $0 < a_0 \le a$) should also be guaranteed to meet $a - bp \ge a_0$; otherwise, food enterprises cannot carry out production and sales.

\textbf{Note 4.5} From $a-bp\ge a_0$, we can obtain $p\le \frac{1}{b}\left( a-a_0 \right) $. Additionally, we can also deduce $\frac{1}{b}\left( a-a_0 \right) \ge p\ge c$, i.e., $p\in [c,\frac{1}{b}\left( a-a_0 \right) ]$, which is denoted as $I=[c,\frac{1}{b}\left( a-a_0 \right) ]$.

\textbf{Assumption 4.6}: When the supplier, manufacturer, and retailer do not cooperate, the logistics cost will increase. For simplicity, it is assumed that the logistics cost is zero when the three parties cooperate. Let $h_w,h_P>0_w$ respectively represent the logistics costs incurred by the manufacturer and the retailer, which respectively satisfy $h_w<w-m$ and $h_P<p-w$.

Without considering cooperation, the green technology investment is borne entirely by Manufacturer $P$, and the profits of the supplier, manufacturer, and retailer are respectively expressed as follows:
\begin{equation}\label{eq:9}
	\begin{cases}
		\pi _S\left( p,w,m,\theta \right) =\left( m-c \right) \left( a-bp+A\theta \right),\\
		\pi _P\left( p,w,m,\theta \right) =[w-m-h_w]\left( a-bp+A\theta \right) -\mu (A\theta )^2,\\
		\pi _R\left( p,w,m,\theta \right) =[p-w-h_p]\left( a-bp+A\theta \right).\\
	\end{cases}\,\,  \theta \in [0,1],p\in I,
\end{equation}

For the convenience of discussion, we express the unit revenue increments of the supplier, manufacturer, and retailer as proportions of $p-c$. Let $m-c=\beta _1\left( p-c \right) $ and $w-m=\beta _2\left( p-c \right)$; then $p-w=p-c+c-m+m-w=\left( p-c \right) -\beta _1(p-c)-\beta _2\left( p-c \right) =(1-\beta _1-\beta _2)\left( p-c \right)$. Let $h_w=l_1\left( p-c \right) ,\mathrm{}h_P=l_2(p-c)$ as well. Here, $0<l_1<\beta _2,0<l_2<1-\beta _1-\beta _2$.
Thus, the profits of the supplier, manufacturer, and retailer can also be expressed as follows respectively:
\begin{equation}\label{eq:10}
	\begin{cases}
		\pi _S\left( p,\beta _1,\beta _2,\theta \right) =\beta _1\left( p-c \right) \left( a-bp+A\theta \right),\\
		\pi _P\left( p,\beta _1,\beta _2,\theta \right) =\left( \beta _2-l_1 \right) \left( p-c \right) \left( a-bp+A\theta \right) -\mu (A\theta )^2,\\
		\pi _R\left( p,\beta _1,\beta _2,\theta \right) =\left( 1-\beta _1-\beta _2-l_2 \right) \left( p-c \right) \left( a-bp+A\theta \right).\\
	\end{cases}\,\, \theta \in [0,1],\mathrm{}\beta _1,\beta _2\in \left[ 0,1 \right] ,p\in I,\mathrm{}
\end{equation}
where $\beta _1,\beta _2,p$ correspond to the selling prices of the products of the supplier, manufacturer, and retailer respectively, all of which are abbreviated as price.
Let $\pi \left( p,\beta _1,\beta _2,\theta \right) =(\pi _S\left( p,\beta _1,\beta _2,\theta \right) ,\pi _P\left( p,\beta _1,\beta _2,\theta \right) ,\pi _R\left( p,\beta _1,\beta _2,\theta \right) )$ and $X=[0,1]\times \left[ 0,1 \right] ^2\times I$ hold. At this point, the supplier has no strategic choice in selecting $\beta _1$, and its strategy set is $[0,1]$; the manufacturer selects $\beta _2$ and the green technology investment level $\theta$, with its strategy set being $\left[ 0,1 \right] ^2=[0,1]\times [0,1]$; the retailer determines the retail price $p$, and its strategy choice set is $I$. At this time, \eqref{eq:10} constitutes a strategic form game, denoted as $(\left\{ S,P,R \right\} ,X,\pi )$.

From the profit functions of supply chain members, it can be seen that there is not only competition over the intermediate prices $m$ and $w$ among the supplier, manufacturer, and distributor, but also their respective demands for the retail price $p $exists. We assume that the intermediate prices have been agreed upon; the issue to be considered is how the three enterprises in the chain set the price of retail products from the perspective of maximizing their own interests, and what the manufacturer's green technology investment level is under different pricing levels. Regarding green technology investment, there exists cooperation driven by its cost sharing. Therefore, if we consider how to set prices and determine the green technology investment level in stages, this issue constitutes a Biform game. The supplier, manufacturer, and distributor all participate in both the non-cooperative game and the cooperative game at the same time. The profit function of the non-cooperative game part is affected by the research and development cost sharing ratio and the green technology level; based on different retail product sales prices, the supplier, manufacturer, and distributor will also jointly determine the green technology investment level through cooperation.

Next, we regard competition and cooperation within the supply chain as a Biform game.

\subsection{The solution to the Biform game}
If the three supply chain members do not cooperate on green technology investment, the profits of the supply chain members are as shown in \eqref{eq:10}. At this time, each member considers its own profit maximization.

For each set of prices $\beta _1,\beta _2\in \left[ 0,1 \right] ,\mathrm{}p\in I,\mathrm{}\pi _P\left( p,\beta _1,\beta _2,\theta \right) $, taking the derivative with respect to $\theta$, we can obtain
\begin{align*}
	\frac{d\pi _P}{d\theta}=\left( \beta _2-l_1 \right) \left( p-c \right) A-2\mu A^2\theta ,  
\end{align*}
Let $\frac{d\pi _p}{d\theta}=0$ hold; we can obtain $\theta =\frac{\left( \beta _2-l_1 \right) \left( p-c \right)}{2\mu A}$, which is denoted as $\hat{\theta}=\frac{\left( \beta _2-l_1 \right) \left( p-c \right)}{2\mu A}$. Since $\frac{d^2\pi _p}{d\theta ^2}=-2\mu A^2<0$, then
\begin{align*}
	\pi _P\left( p,\beta _1,\beta _2 \right) &=\pi _P\left( p,\beta _1,\beta _2,\hat{\theta} \right) =\pi _P\left( p,\beta _1,\beta _2,\hat{\theta} \right) 
	\\
	&=\max_{\theta \epsilon \left[ 0,1 \right]} \!\:\left\{ \left( \beta _2-l_1 \right) \left( p-c \right) \left( a-bp+A\theta \right) -\mu \left( A\theta \right) ^2 \right\} 
	\\
	&=\left( \beta _2-l_1 \right) \left( p-c \right) \left( a-bp+\frac{\left( \beta _2-l_1 \right) \left( p-c \right)}{2\mu} \right) -\mu A^2\left( \frac{\left( \beta _2-l_1 \right) \left( p-c \right)}{2\mu A} \right) ^2
	\\
	&=\left( \beta _2-l_1 \right) \left( p-c \right) \left( a-bp \right) +\frac{\left( \beta _2-l_1 \right) ^2\left( p-c \right) ^2}{4\mu},
\end{align*}

\begin{align*}
	\pi _S\left( p,\beta _1,\beta _2 \right) &=\pi _S\left( p,\beta _1,\beta _2,\hat{\theta} \right) =\pi _S\left( p,\beta _1,\beta _2,\hat{\theta} \right) 
	\\
	&=\beta _1\left( p-c \right) \left( a-bp \right) +\frac{\beta _1\left( \beta _2-l_1 \right) \left( p-c \right) ^2}{2\mu},
\end{align*}
\begin{align*}
	\pi _R\left( p,\beta _1,\beta _2 \right) &=\pi _R\left( p,\beta _1,\beta _2,\hat{\theta} \right) =\pi _R\left( p,\beta _1,\beta _2,\hat{\theta} \right) 
	\\
	&=\left( 1-\beta _1-\beta _2-l_2 \right) \left( p-c \right) \left( a-bp \right) +\frac{\left( 1-\beta _1-\beta _2-l_2 \right) \left( \beta _2-l_1 \right) \left( p-c \right) ^2}{2\mu}.
\end{align*}
From this, we can obtain that
\begin{align*}
	&\pi _S\left( p,\beta _1,\beta _2 \right) +\pi _P\left( p,\beta _1,\beta _2 \right) +\pi _R\left( p,\beta _1,\beta _2 \right) 
	\\
	&=\beta _1\left( p-c \right) \left( a-bp \right) +\frac{\beta _1\left( \beta _2-l_1 \right) \left( p-c \right) ^2}{2\mu}
	\\
	&+\left( \beta _2-l_1 \right) \left( p-c \right) \left( a-bp \right) +\frac{\left( \beta _2-l_1 \right) ^2\left( p-c \right) ^2}{4\mu}
	\\
	&+\left( 1-\beta _1-\beta _2-l_2 \right) \left( p-c \right) \left( a-bp \right) +\frac{\left( 1-\beta _1-\beta _2-l_2 \right) \left( \beta _2-l_1 \right) \left( p-c \right) ^2}{2\mu}
	\\
	&=\left( 1-l_1-l_2 \right) \left( p-c \right) \left( a-bp \right) +\frac{\left( 1-\beta _2-l_2 \right) \left( \beta _2-l_1 \right) \left( p-c \right) ^2}{2\mu}+\frac{\left( \beta _2-l_1 \right) ^2\left( p-c \right) ^2}{4\mu}.
\end{align*}
If the three enterprises cooperate in green technology investment, the cooperation method is to jointly share the costs of green technology investment. It is assumed that the cooperation among the members within the chain also eliminates the logistics costs $l_1,l_2$, which is also the coordination effect brought by cooperation. At this time, the total profit of the supply chain is
\begin{align*}
	v\left( \left\{ S,P,R \right\} ,\left( p,\beta _1,\beta _2,\theta \right) \right) =\left( p-c \right) \left( a-bp+A\theta \right) -\mu (A\theta )^2.
\end{align*}
And it is not difficult to obtain that the coalition function of the cooperative stage in the strategic Biform game is:

\begin{flalign}\label{H4}
	& v\left( \left\{ \emptyset \right\} ,\left( p,\beta _1,\beta _2,\theta \right) \right) =0; & \nonumber \\
	& v\left( \left\{ S \right\} ,\left( p,\beta _1,\beta _2,\theta \right) \right) =\pi _S\left( p,\beta _1,\beta _2,\theta \right) =\beta _1\left( p-c \right) \left( a-bp+A\theta \right) ; & \nonumber \\
	& v\left( \left\{ P \right\} ,\left( p,\beta _1,\beta _2,\theta \right) \right) =\pi _p\left( p,\beta _1,\beta _2,\theta \right) =\left( \beta _2-l_1 \right) \left( p-c \right) \left( a-bp+A\theta \right) -\mu \left( A\theta \right) ^2; & \nonumber \\
	& v\left( \left\{ R \right\} ,\left( p,\beta _1,\beta _2,\theta \right) \right) =\pi _R\left( p,\beta _1,\beta _2,\theta \right) =\left( 1-\beta _1-\beta _2-l_2 \right) \left( p-c \right) \left( a-bp+A\theta \right) ; & \nonumber \\
	& v\left( \left\{ S,P \right\} ,\left( p,\beta _1,\beta _2,\theta \right) \right) =\pi _S\left( p,\beta _1,\beta _2,\theta \right) +\pi _p\left( p,\beta _1,\beta _2,\theta \right) & \nonumber \\
	& \quad = \beta _1\left( p-c \right) \left( a-bp+A\theta \right) +\left( \beta _2-l_1 \right) \left( p-c \right) \left( a-bp+A\theta \right) -\mu \left( A\theta \right) ^2; & \nonumber \\
	& v\left( \left\{ S,R \right\} ,\left( p,\beta _1,\beta _2,\theta \right) \right) =\pi _S\left( p,\beta _1,\beta _2,\theta \right) +\pi _R\left( p,\beta _1,\beta _2,\theta \right) & \nonumber \\
	& \quad = \beta _1\left( p-c \right) \left( a-bp+A\theta \right) +\left( 1-\beta _1-\beta _2-l_2 \right) \left( p-c \right) \left( a-bp+A\theta \right) ; & \nonumber \\
	& v\left( \left\{ P,R \right\} ,\left( p,\beta _1,\beta _2,\theta \right) \right) =\pi _P\left( p,\beta _1,\beta _2,\theta \right) +\pi _R\left( p,\beta _1,\beta _2,\theta \right) & \nonumber \\
	& \quad = \left( \beta _2-l_1 \right) \left( p-c \right) \left( a-bp+A\theta \right) -\mu \left( A\theta \right) ^2+\left( 1-\beta _1-\beta _2-l_2 \right) \left( p-c \right) \left( a-bp+A\theta \right) ; & \nonumber \\
	& v\left( \left\{ S,P,R \right\} ,\left( p,\beta _1,\beta _2,\theta \right) \right) =\left( p-c \right) \left( a-bp+A\theta \right) -\mu \left( A\theta \right) ^2,
\end{flalign}
where $\theta \in [0,1],\ \beta _1,\beta _2\in \left[ 0,1 \right] ,\ p\in I$. \eqref{H4} is a Biform game derived from the strategic game \eqref{eq:10}, and is denoted as $(\left\{ S,P,R \right\} ,v,X)$.

For each set of prices $\beta _1,\beta _2\in \left[ 0,1 \right] ,p\in I$, if the three enterprises within the chain adopt a cooperative approach to green technology investment i.e., share the cost of the manufacturer's green technology investment then when considering the maximum profit of the supply chain, we take the derivative of $v\left( \left\{ S,P,R \right\} ,\left( p,\beta _1,\beta _2,\theta \right) \right) $ with respect to $\theta$, and can obtain
\begin{align*}
	\frac{dv}{d\theta}=\left( p-c \right) A-2\mu A^2\theta .
\end{align*}
Let $\frac{\partial v}{\partial \theta}=\left( p-c \right) A-2\mu A^2\theta =0$ hold; we can obtain $\theta =\frac{p-c}{2\mu A}$, which is denoted as $\theta ^*=\frac{p-c}{2\mu A}$. From the second derivative $\frac{d^2v}{d\theta ^2}=-2\mu A^2<0$, we can conclude that the maximum profit of the supply chain is
\begin{align*}
	v\left( \left\{ S,P,R \right\} ,\left( p,\mathrm{}\beta _1,\beta _2 \right) \right) &=v\left( \left\{ S,P,R \right\} ,\left( p,\mathrm{}\beta _1,\beta _2,\theta ^* \right) \right) 
	\\
	&=\max_{\theta} \!\:\left( p-c \right) \left( a-bp+A\theta \right) -\mu \left( A\theta \right) ^2
	\\
	&=\left( p-c \right) \left( a-bp+A\theta ^* \right) -\mu \left( A\theta ^* \right) ^2
	\\
	&=\left( p-c \right) \left( a-bp+\frac{\left( p-c \right)}{2\mu} \right) -\mu A^2\left( \frac{\left( p-c \right)}{2\mu A} \right) ^2
	\\
	&=\left( p-c \right) \left( a-bp \right) +\frac{\left( p-c \right) ^2}{4\mu}.
\end{align*}

At this time, the total profit of the supply chain is not only affected by the unit product value-added, but also by the price elasticity coefficient $b$ and the green technology investment cost coefficient $\mu $, both of which exert an inverse impact. Specifically, the more intense the market competition, the greater the price elasticity, and the lower the profit; meanwhile, the higher the green technology investment cost coefficient, the lower the profit.

Next, we make a comparison between the green technology investment levels $\hat{\theta}$ and $\theta ^*$ under the two scenarios of non-cooperation and cooperation among supply chain members. It should be noted that
\begin{align*}
	\theta ^*-\hat{\theta}=\frac{\left( p-c \right)}{2\mu A}-\frac{\left( \beta _2-l_1 \right) \left( p-c \right)}{2\mu A}=\frac{\left( 1-\beta _2+l_1 \right) \left( p-c \right)}{2\mu A}>0.
\end{align*}
From this, we can conclude that under the cost-sharing scenario, the green technology investment level is higher than that under non-cooperation.

Additionally, let us also compare the total profits of the supply chain under these two scenarios; we have
\begin{align*}
	&v\left( \left\{ S,P,R \right\} ,\left( p,\beta _1,\beta _2 \right) \right) -\left[ \pi _S\left( p,\beta _1,\beta _2 \right) +\pi _P\left( p,\beta _1,\beta _2 \right) +\pi _R\left( p,\beta _1,\beta _2 \right) \right] 
	\\
	&=v\left( \left\{ S,P,R \right\} ,\left( p,\beta _1,\beta _2\theta ^* \right) \right) -\left[ \pi _S\left( p,\beta _1,\beta _2,\hat{\theta} \right) +\pi _P\left( p,\beta _1,\beta _2,\hat{\theta} \right) +\pi _R\left( p,\beta _1,\beta _2,\hat{\theta} \right) \right] 
	\\
	&=\left( p-c \right) \left( a-bp \right) +\frac{\left( p-c \right) ^2}{4\mu}-[\left( 1-l_1-l_2 \right) \left( p-c \right) \left( a-bp \right) 
	\\
	&-\frac{\left( 1-\beta _2-l_2 \right) \left( \beta _2-l_1 \right) \left( p-c \right) ^2}{2\mu}-\frac{\left( \beta _2-l_1 \right) ^2\left( p-c \right) ^2}{4\mu}]
	\\
	&=(l_1+l_2)\left( p-c \right) \left( a-bp \right) +\frac{\left( p-c \right) ^2}{4\mu}\left[ \left( 1-\beta _2 \right) ^2-{l_1}^2+2\beta _2l_2+2l_1-2l_1l_2 \right].
\end{align*}

Additionally, it should be noted that $\beta _2-l_1>0$ i.e., $\beta _2>l_1$ and $l_1<1$¡ªand then we have
\begin{align*}
	&v\left( \left\{ S,P,R \right\} ,\left( p,\beta _1,\beta _2 \right) \right) -\left[ \pi _S\left( p,\beta _1,\beta _2 \right) +\pi _P\left( p,\beta _1,\beta _2 \right) +\pi _R\left( p,\beta _1,\beta _2 \right) \right] 
	\\
	&\ge (l_1+l_2)\left( p-c \right) \left( a-bp \right) +\frac{\left( p-c \right) ^2}{4\mu}[(\beta _1+l_2)^2-{l_1}^2+2l_1l_2+2l_1-2l_1l_2]
	\\
	&>\frac{\left( p-c \right) ^2}{4\mu}\left[ \left( \beta _1+l_2 \right) ^2+2l_1-{l_1}^2 \right] >0.
\end{align*}
From this, we can conclude that under the cost-sharing scenario, the total profit of the supply chain is also higher than that under the non-cooperation scenario.

For each price combination $\beta _1,\beta _2\in \left[ 0,1 \right] ,p\in I$, after determining the green technology investment level through cost sharing, the supplier, manufacturer, and retailer each set the sales price $\beta _1,\beta _2,p$, which constitutes a non-cooperative game and also forms the first stage of the Biform game $(\left\{ S,P,R \right\} ,v,X)$. The payoff function of this non-cooperative game comes from the distribution of the total profit of the supply chain in the cooperative stage. At this time, the distribution function of the cooperative stage is crucial to the solution of the Biform game.

Next, we determine the distribution function by means of the cost sharing ratio and the value of $\beta _1,\beta _2,p$, so that the Nash equilibrium solution of the first stage becomes the Pareto optimal solution that maximizes the total profit.

Suppose the green degree cost sharing ratios of the supplier, manufacturer, and retailer are $\lambda _S,\lambda _P$ and $1-\lambda _S-\lambda _P$ respectively. Considering that the cost of green technology investment matches the benefit i.e., the sharing ratios are $\lambda _S=\beta _1,\lambda _P=\beta _2$ and $1-\lambda _S-\lambda _P=1-\beta _1-\beta _2$ respectively the distribution corresponding to the solution to the cooperative game is exactly the Shapley value. The distribution of total profit is based on contribution (the profit each obtains); in other words, the cost therein is shared according to the proportion of the corresponding income distribution. The distribution function is:
\begin{align*}
	\varphi _S\left( p,\beta _1,\beta _2 \right) &=\varphi _S\left( p,\beta _1,\beta _2,\theta ^* \right) 
	\\
	&=\beta _1\left( p-c \right) [\left( a-bp \right) +A\theta ^*]-\beta _1\mu (A\theta ^*)^2
	\\
	&=\beta _1\left( p-c \right) [\left( a-bp \right) +\frac{p-c}{2\mu}]-\beta _1\mu A^2(\frac{p-c}{2\mu A})^2
	\\
	&=\beta _1[\left( p-c \right) \left( a-bp \right) +\frac{\left( p-c \right) ^2}{4\mu}]=\beta _1v\left( \left\{ S,P,R \right\} ,\left( p,\beta _1,\beta _2,\theta ^* \right) \right) ;
\end{align*}
\begin{align*}
	\varphi _P\left( p,\beta _1,\beta _2 \right) &=\varphi _P\left( p,\beta _1,\beta _2,\theta ^* \right) 
	\\
	&=\beta _2\left( p-c \right) [\left( a-bp \right) +A\theta ^*]-\beta _2\mu (A\theta ^*)^2
	\\
	&=\beta _2\left( p-c \right) [a-bp+A\times \frac{p-c}{2\mu A}]-\frac{\beta _2\left( p-c \right) ^2}{4\mu}
	\\
	&=\beta _2[\left( p-c \right) \left( a-bp \right) +\frac{\left( p-c \right) ^2}{4\mu}]=\beta _2v\left( \left\{ S,P,R \right\} ,\left( p,\beta _1,\beta _2,\theta ^* \right) \right) ;
\end{align*}
\begin{align*}
	\varphi _R\left( p,\beta _1,\beta _2 \right) &=\varphi _R\left( p,\beta _1,\beta _2,\theta ^* \right) 
	\\
	&=\left( 1-\beta _1-\beta _2 \right) \left( p-c \right) [\left( a-bp \right) +A\theta ^*]-\left( 1-\beta _1-\beta _2 \right) \mu (A\theta ^*)^2
	\\
	&=\left( 1-\beta _1-\beta _2 \right) \left( p-c \right) \left[ a-bp+A\times \frac{p-c}{2\mu A} \right] -\frac{\left( 1-\beta _1-\beta _2 \right) \left( p-c \right) ^2}{4\mu}
	\\
	&=\left( 1-\beta _1-\beta _2 \right) \left[ \left( p-c \right) \left( a-bp \right) +\frac{\left( p-c \right) ^2}{4\mu} \right] 
	\\
	&=\left( 1-\beta _1-\beta _2 \right) v\left( \left\{ S,P,R \right\} ,\left( p,\beta _1,\beta _2,\theta ^* \right) \right). 
\end{align*}
Thus, the three enterprises within the chain each set price $p,\beta _1,\beta _2$ to maximize their respective profits, which constitutes the non-cooperative game $(\left\{ S,P,R \right\} ,X,(\varphi _S,\varphi _P,\varphi _R))$ in the first stage of the Biform game, and its payoff functions are respectively:
\begin{equation}\label{eq:11}
	\begin{cases}
		\varphi _S\left( p,\beta _1,\beta _2 \right) =\mathrm{}=\beta _1v\left( \left\{ S,P,R \right\} ,\left( p,\beta _1,\beta _2,\theta ^* \right) \right),\\
		\varphi _P\left( p,\beta _1,\beta _2 \right) =\mathrm{}=\beta _2v\left( \left\{ S,P,R \right\} ,\left( p,\beta _1,\beta _2,\theta ^* \right) \right),\\
		\varphi _R\left( p,\beta _1,\beta _2 \right) =\mathrm{}=\left( 1-\beta _1-\beta _2 \right) v\left( \left\{ S,P,R \right\} ,\left( p,\beta _1,\beta _2,\theta ^* \right) \right).\\
	\end{cases}
\end{equation}
Eq.  \eqref{eq:11} is the non-cooperative game derived from the Biform game $(\left\{ S,P,R \right\} ,v,X)$. To find the solution to the Biform game $(\left\{ S,P,R \right\} ,v,X)$ means to find the solution to the non-cooperative game $(\left\{ S,P,R \right\} ,(\varphi _S,\varphi _P,\varphi _R),X)$.
First, find the maximum point of total profit $v\left( \left\{ S,P,R \right\} ,\left( p,\beta _1,\beta _2,\theta ^* \right) \right) $. From
\begin{align*}
	\frac{\partial v}{\partial p}=\left( a-bp \right) -b\left( p-c \right) +\frac{p-c}{2\mu}=\left( a+bc \right) -\left( 2b-\frac{1}{2\mu} \right) p-\frac{c}{2\mu}.
\end{align*}
Let $\frac{\partial v}{\partial p}=0$ hold, then we solve for $p=\frac{2\mu \left( a+bc \right) -c}{4\mu b-1}$ and denote it as $p^*=\frac{2\mu \left( a+bc \right) -c}{4\mu b-1}$; and then from $\frac{\partial ^2v}{\partial p^2}=-\frac{1}{2\mu}\left( 4\mu b-1 \right)$.
Considering the positive or negative nature of $4\mu b-1$, we discuss it under two scenarios.

\textbf{(E1)} When $\mu >\frac{1}{4b}$ holds, we have $\frac{\partial ^2v}{\partial p^2}<0$; then $p^*=\frac{2\mu \left( a+bc \right) -c}{4b\mu -1}$ is the maximum point of $v\left( \left\{ S,P,R \right\} ,\left( p,\beta _1,\beta _2,\theta ^* \right) \right) $, and the maximum profit at this time is
\begin{align*}
	v^*&=\mathrm{}v\left( \left\{ S,P,R \right\} ,\left( p^*,\beta _1,\beta _2,\theta ^* \right) \right) 
	\\
	&=\left( p^*-c \right) \left( a-bp^* \right) +\frac{\left( p^*-c \right) ^2}{4\mu}
	\\
	&=\left( \frac{2\mu \left( a+bc \right) -c}{4b\mu -1}-c \right) \left( a-b\times \frac{2\mu \left( a+bc \right) -c}{4b\mu -1} \right) +\frac{\left( \frac{2\mu \left( a+bc \right) -c}{4b\mu -1}-c \right) ^2}{4\mu}
	\\
	&=\left( \frac{2\mu a-2bc\mu}{4\mu b-1} \right) \left( \frac{2ab\mu -2\mu b^2c-a+bc}{4\mu b-1} \right) +\frac{\left( \frac{2\mu a-2bc\mu}{4\mu b-1} \right) ^2}{4\mu}=\frac{\mu \left( a-bc \right) ^2}{(4\mu b-1)}.
\end{align*}
It is not difficult to verify that $(p,\beta _1,\beta _2)$ is the solution to the non-cooperative game \eqref{eq:11} if and only if $p=p^*$ and $0\le \beta _1+\beta _2\le 1$. Denote the price combination satisfying this condition as $(p^*,{\beta _1}^*,{\beta _2}^*)$; then $(p^*,{\beta _1}^*,{\beta _2}^*)$ is the solution to the Biform game $(\left\{ S,P,R \right\} ,X,v)$.

To further analyze the relationships between the profit $v^*$ and price $p^*$corresponding to solution $(p^*,{\beta _1}^*,{\beta _2}^*)$, and parameters $\mu$ and $b$.

Taking the partial derivative of $v^*=\frac{\mu \left( a-bc \right) ^2}{(4\mu b-1)}$ with respect to $\mu$, we can obtain
\begin{align*}
	\frac{\partial v^*}{\partial \mu}=\frac{\left( a-bc \right) ^2\left( 4\mu b-1 \right) -4\mu b\left( a-bc \right) ^2}{\left( 4\mu b-1 \right) ^2}=-\frac{\left( a-bc \right) ^2}{\left( 4\mu b-1 \right) ^2}<0.
\end{align*}
Therefore, the total profit of the supply chain is decreasing with respect to the green technology cost coefficient; in other words, the lower the green technology investment cost coefficient, the higher the total profit.

Taking the partial derivative of $p^*=\frac{2\mu \left( a+bc \right) -c}{4b\mu -1}$ with respect to $\mu$ and $b$, we can obtain
\begin{align*}
	\frac{\partial p^*}{\partial \mu}=\frac{-2\left( a-bc \right)}{\left( 4\mu b-1 \right) ^2}<0,
\end{align*}
\begin{align*}
	\frac{\partial p^*}{\partial b}=\frac{2\mu c\left( 4\mu b-1 \right) -4\mu (2\mu \left( a+bc \right) -c)}{\left( 4\mu b-1 \right) ^2}=-\frac{2\mu \left[ 4a\mu -c \right]}{\left( 4\mu b-1 \right) ^2}<-\frac{2c\mu \left[ 4b\mu -1 \right]}{\left( 4\mu b-1 \right) ^2}<0.
\end{align*}

Therefore, the sales price $p^*$ is decreasing with respect to the green technology investment cost coefficient $\mu$ in other words, the lower the green technology investment cost coefficient $\mu$, the lower the optimal sales price $p^*$ corresponding to maximum profit. Similarly, the sales price $p^*$ is also decreasing with respect to the price elasticity coefficient $b$, which means that the more intense the price competition, the lower the optimal sales price $p^*$ and the larger the price elasticity coefficient $b$ is, the more intense the price competition becomes.

\textbf{(E2)} When $\mu <\frac{1}{4b}$ holds, we have $\frac{\partial ^2v}{\partial p^2}>0$;when $p=c$,have
\begin{align*}
	\frac{\partial v}{\partial p}|_{p=c}=\left( a+bc \right) -\left( 2b-\frac{1}{2\mu} \right) c-\frac{c}{2\mu}=a-bc>0.
\end{align*}
Then, when $p\ge c$, $v$ is an increasing function with respect to $p$.

From assumption 4.5, $p\in I$. Therefore, $v$ attains its maximum value at $p^{**}=\frac{1}{b}\left( a-a_0 \right)$. At this point, the maximum profit is
\begin{align*}
	v^{**}&=v\left( \left\{ S,P,R \right\} ,\left( p^{**},\beta _1,\beta _2,\theta ^* \right) \right) =\left( p^{**}-c \right) \left( a-bp^{**} \right) +\frac{\left( p^{**}-c \right) ^2}{4\mu}
	\\
	&=\frac{1}{b}\left( \left( a-a_0 \right) -bc \right) a_0+\frac{\left( \frac{1}{b}\left( a-a_0 \right) -c \right) ^2}{4\mu}].
\end{align*}

Similar to \textbf{(E1)}, it is not difficult to verify that $(p,\beta _1,\beta _2)$ is the solution to the non-cooperative game \eqref{eq:11} if and only if $p=p^{**}$ and $0\le \beta _1+\beta _2\le 1$. Denote the price combination satisfying this condition as $(p^{**},{\beta _1}^{**},{\beta _2}^{**})$; then $(p^{**},{\beta _1}^{**},{\beta _2}^{**})$ is the solution to the Biform game $(\left\{ S,P,R \right\} ,X,v)$. Similarly, we can analyze the relationships between the profit $v^{**}$ and price $p^{**}$ that correspond to $(p^{**},{\beta _1}^{**},{\beta _2}^{**})$ and the parameters $\mu$ and $b$. The results show that when $\mu <\frac{1}{4b}$, $v^{**}$ is also decreasing with respect to $\mu$¡ªmeaning the lower the green technology cost coefficient, the higher the maximum profit of the supply chain. At the same time, the larger the price elasticity $b$, i.e., the more intense the price competition, the lower the optimal price $p^{**}$ corresponding to maximum profit.

It should be noted that the conditions for ${\beta _1}^{*},{\beta _2}^{*}$ in the first case and ${\beta _1}^{**},{\beta _2}^{**}$ in the second case to be Nash equilibria are $0\le {\beta _1}^*+{\beta _2}^*\le 1$ (for the first case) and the corresponding condition for the second case respectively. Taking ${\beta _1}^{*}$ and ${\beta _2}^{*}$ as an example, this condition means that ${\beta _1}^{*},{\beta _2}^{*}$ and the retailer¡¯s allocation coefficient $1-{\beta _1}^*-{\beta _2}^*$ add up to 1; i.e., the three enterprises in the chain exactly distribute the total profit of the supply chain in full. This solution is similar to the Nash equilibrium of the cake-cutting game. Regarding the selection of ${\beta _1}^{*},{\beta _2}^{*}$ or ${\beta _1}^{**},{\beta _2}^{**}$, it is theoretically a problem worthy of further research. In practical applications, it can be determined by the market and each enterprise based on the actual situation of specific issues.
\subsection{Summary of This Section}
In this section, the Biform game analysis method established in Section \ref{Sec1} is applied to study the issue of green technology investment by food enterprises in market competition. The research is mainly conducted based on two models: the Bertrand model and the supply chain Cournot model.

The first part is based on the Bertrand model, i.e., the duopoly price competition model. The method involves introducing a green technology investment term into the classic Bertrand model, then interpreting the famous Bertrand paradox through cooperative pricing, exploring the logical basis for achieving win-win cooperation, and resolving the duopoly price war dilemma in the Bertrand model. By converting the strategic game into a Biform game and adopting two distribution methods marginalism and equalitarianism for the distribution of cooperative benefits, the payoff functions of the non-cooperative game are derived, and the solution to the Biform game is obtained i.e., the price and green technology investment level of the duopoly market price competition model. A further analysis is also conducted on the relationship between profits and relevant parameters.

The research results show that: (1) Cooperation among manufacturers increases the level of green technology investment and also raises the total profit of the supply chain. In other words, cooperation not only avoids the mutually destructive outcome of price wars and increases manufacturers¡¯ profits, but also encourages manufacturers to raise the level of green technology investment and improve food safety quality. Therefore, governments and regulatory authorities should introduce relevant policies and incentive measures to promote cooperation among manufacturers, further encourage industrial cooperation, and increase green technology investment. (2) The smaller the green technology cost coefficient, the higher the level of green technology, and the greater the total profit of manufacturers. Therefore, to promote enterprises to increase green technology investment, the government can implement subsidy policies for green technology investment to reduce enterprises¡¯ investment costs; alternatively, it can establish green technology research platforms, actively introduce relevant policies to guide the participation of research institutions and manufacturers, and encourage research and development cooperation between manufacturing enterprises and research institutions or among manufacturing enterprises thereby reducing the research and development costs of green technology.

The second part is based on the food supply chain model. A green technology investment term is introduced into this model, and cooperation is achieved by means of cost sharing of green technology this not only increases the level of green technology investment by enterprises within the chain, but also boosts their profits. This part mainly focuses on a food industry supply chain consisting of one supplier, one manufacturer, and one retailer. Considering the competition between the supply chains products and similar or substitute products in actual operations, both the selling price and green technology investment exert an impact on product demand, and also affect the respective profits of enterprises. Thus, to maximize profits, supply chain members compete with each other over the selling prices of their respective products and have respective demands for the manufacturer's green technology investment leading to a non-cooperative scenario for this issue. In addition, members within the chain can also cooperate by sharing the manufacturer's green technology investment costs, which means the issue also contains cooperative elements. In fact, the analysis results show that cooperation can increase the total profit of the supply chain. Therefore, for the supply chain Cournot model, we apply the Biform game model from Section \ref{Sec1} to the study of this issue: we first establish a strategic game model, and then derive a Biform game from it. During the cooperation phase, through the supplier and retailer's cost sharing of the manufacturer's green technology investment, the level of green technology investment is increased, the total profit of the supply chain is raised, and Pareto optimality in the profits of members within the chain is achieved. On this basis, we analyzed the relationships between the total profit of the supply chain, the optimal price, and relevant parameters.

The research results show that: (1) Cooperation among supply chain members increases the level of green technology investment in the supply chain and also raises the total profit of the supply chain; only when supply chain members share the green technology costs according to the level of revenue increment can the profits of supply chain members achieve Pareto optimality, i.e., the maximization of the total profit of the supply chain. Therefore, governments and regulatory authorities should introduce relevant policies and incentive measures to promote in-depth cooperation among members of the industrial chain, and upstream and downstream members of the food industry chain should jointly bear the green technology costs this not only helps increase enterprise profits and fiscal revenue, but also increases enterprises investment in green technology and improves the level of food safety and quality; (2) The smaller the green technology cost coefficient, the higher the level of green technology, and the greater the total profit of the supply chain. Therefore, to promote enterprises to increase green technology investment, the government can implement subsidy policies for green technology investment to reduce enterprises investment costs, or establish green technology research and development platforms, actively introduce relevant policies to guide the participation of research institutions and food manufacturing enterprises, and encourage research and development cooperation between manufacturing enterprises and research institutions or among manufacturing enterprises thereby reducing the research and development costs of green technology; (3) The lower the green technology cost coefficient, the lower the optimal sales price of the supply chain; at the same time, the higher the price elasticity coefficient, the lower the optimal retail price of the supply chain. Therefore, reducing the green technology investment coefficient of the supply chain, stimulating market vitality and maintaining the intensity of product competition can promote the reduction of the supply chain's retail price, thereby allowing consumers to enjoy more benefits.

In summary, the two models mentioned above provide strong theoretical analysis and methodological support for studying the issue of green technology investment by food enterprises in market competition, especially the issue of the integration of competition and cooperation among food enterprises. The state has introduced relevant policies to encourage the construction of supply chain innovation networks led by enterprises and featuring cooperation among industry, universities, research institutions and users, build innovation service platforms in cross-boundary and interdisciplinary fields, and provide services such as technology research and development, brand development, market expansion, standardization services, and inspection, testing and certification (Notice of the General Office of the State Council on Actively Promoting the Innovation and Application of Supply Chains (Guobanfa [2017] No. 84)), which includes supporting the establishment of supply chain research institutes and encouraging qualified localities to build supply chain science, technology and innovation research and development centers. These measures will promote cooperation among supply chain members, reduce the cost of green technology investment, which is not only conducive to increasing enterprise profits, but also brings benefits to consumers in terms of prices and improves consumer satisfaction.




\end{document}